\documentclass[twocolumn]{aastex63}
\usepackage{amsmath}
\usepackage{amssymb}
\usepackage{euscript}
\usepackage{algorithm2e,algorithmic}
\usepackage{bbold}
\usepackage{paralist}
\usepackage{hyperref}
\usepackage{booktabs}
\usepackage{rotating}
\usepackage{lineno}
\hypersetup{         
    unicode=true,          
    colorlinks=true,       
    linkcolor=red,          
    citecolor=blue,        
    filecolor=magenta,      
    urlcolor=red           
    }

\def\gtorder{\mathrel{\raise.3ex\hbox{$>$}\mkern-14mu
             \lower0.6ex\hbox{$\sim$}}}
\def\ltorder{\mathrel{\raise.3ex\hbox{$<$}\mkern-14mu
             \lower0.6ex\hbox{$\sim$}}}

\def\gtorder{\mathrel{\raise.3ex\hbox{$>$}\mkern-14mu
             \lower0.6ex\hbox{$\sim$}}}
\def\ltorder{\mathrel{\raise.3ex\hbox{$<$}\mkern-14mu
             \lower0.6ex\hbox{$\sim$}}}

\newcommand{\package}[1]{\texttt{#1}}
\newcommand{\galex}{\textit{GALEX}}
\newcommand{\swift}{\textit{Swift}}
\newcommand{\wise}{\textit{WISE}}

\newcommand{\kms}{km\,s$^{-1}$}

\newcommand{\msun}{$M_{\odot}$}

\newcommand{\Nifs}{$^{56} \rm Ni$}

\newcommand{\CIV}{\ion{C}{4}}

\newcommand{\oqm }{SN\,2022oqm}

\newcommand{\OIV}{\ion{O}{4}}
\newcommand{\OV}{\ion{O}{5}}
\shortauthors{Irani et al.}

\newcommand{\referee}{}
\newcommand{\secondref}{}

\graphicspath{{./}{./}}

\begin{document}

\title{SN\,2022oqm -- a Ca-rich Explosion of a Compact Progenitor Embedded in \ion{C}{0}/\ion{O}{0} Circumstellar Material}

\author[0000-0002-7996-8780]{Ido ~Irani}
\affiliation{Department of Particle Physics and Astrophysics,
             Weizmann Institute of Science,
             234 Herzl St, 7610001 Rehovot, Israel}
\email{idoirani@gmail.com}

\author[0000-0003-0853-6427]{Ping Chen}

\affiliation{Department of Particle Physics and Astrophysics,
             Weizmann Institute of Science,
             234 Herzl St, 7610001 Rehovot, Israel}

\author{Jonathan Morag}
\affiliation{Department of Particle Physics and Astrophysics,
             Weizmann Institute of Science,
             234 Herzl St, 7610001 Rehovot, Israel}

\author[0000-0001-6797-1889]{Steve ~Schulze}
\affiliation{Department of Physics, 
             The Oskar Klein Center, Stockholm University, 
             AlbaNova, SE-10691 Stockholm, Sweden}

\author[0000-0002-3653-5598]{Avishay ~Gal-Yam}
\affiliation{Department of Particle Physics and Astrophysics,
             Weizmann Institute of Science,
             234 Herzl St, 7610001 Rehovot, Israel}

\author[0000-0002-4667-6730]{Nora L. Strotjohann}
\affiliation{Department of Particle Physics and Astrophysics,
             Weizmann Institute of Science,
             234 Herzl St, 7610001 Rehovot, Israel}

\author[0000-0002-0301-8017]{Ofer Yaron}
\affiliation{Department of Particle Physics and Astrophysics,
             Weizmann Institute of Science,
             234 Herzl St, 7610001 Rehovot, Israel}

\author[0000-0001-8985-2493]{Erez A. Zimmerman}
\affiliation{Department of Particle Physics and Astrophysics,
             Weizmann Institute of Science,
             234 Herzl St, 7610001 Rehovot, Israel}
                            
\author{Amir Sharon}
\affiliation{Department of Particle Physics and Astrophysics,
             Weizmann Institute of Science,
             234 Herzl St, 7610001 Rehovot, Israel}

\author{Daniel A. Perley}
\affiliation{Astrophysics Research Institute, Liverpool John Moores University, IC2 Liverpool Science Park, 146 Brownlow Hill, Liverpool L3 5RF, UK}

\author[0000-0003-1546-6615]{J.~Sollerman}
\affiliation{Department of Astronomy, 
             The Oskar Klein Center, Stockholm University, 
             AlbaNova, SE-10691 Stockholm, Sweden}
    
\author[0000-0002-2810-8764]{Aaron Tohuvavohu}
\affiliation{David A. Dunlap Department of Astronomy and Astrophysics, University of Toronto, Toronto, ON, Canada}
                    
\author{Kaustav K. Das}
\affiliation{Division of Physics, Mathematics and Astronomy, California Institute of Technology, Pasadena, CA 91125, USA}

\author[0000-0002-5619-4938]{Mansi M. Kasliwal}
\affiliation{Division of Physics, Mathematics and Astronomy, California Institute of Technology, Pasadena, CA 91125, USA}

\author{Rachel Bruch}
\affiliation{Department of Particle Physics and Astrophysics,
             Weizmann Institute of Science,
             234 Herzl St, 7610001 Rehovot, Israel}
          
\author[0000-0001-5955-2502]{Thomas~G.~Brink}
\affiliation{Department of Astronomy, University of California, Berkeley, CA 94720-3411, USA}
\affiliation{Wood Specialist in Astronomy}

\author{WeiKang~Zheng}
\affiliation{Department of Astronomy, University of California, Berkeley, CA 94720-3411, USA}
\affiliation{Eustace Specialist in Astronomy}

\author[0000-0003-3460-0103]{Alexei~V.~Filippenko}
\affiliation{Department of Astronomy, University of California, Berkeley, CA 94720-3411, USA}

\author[0000-0002-1092-6806]{Kishore C. Patra}
\affiliation{Department of Astronomy, University of California, Berkeley, CA 94720-3411, USA}
\affiliation{Nagaraj-Noll-Otellini Graduate Fellow}

\author[0000-0002-4951-8762]{Sergiy~S. Vasylyev}
\affiliation{Department of Astronomy, University of California, Berkeley, CA 94720-3411, USA}
\affiliation{Steven Nelson Graduate Fellow}

\author[0000-0002-6535-8500]{Yi Yang}
\affiliation{Department of Astronomy, University of California, Berkeley, CA 94720-3411, USA}
\affiliation{Bengier-Winslow-Robertson Postdoctoral Fellow}
          
\author[0000-0002-3168-0139]{Matthew J. Graham}
\affiliation{Division of Physics, Mathematics and Astronomy, California Institute of Technology, Pasadena, CA 91125, USA}

\author[0000-0002-7777-216X]{Joshua~S.~Bloom}
\affiliation{Department of Astronomy, University of California, Berkeley, CA 94720-3411, USA}
\affiliation{Lawrence Berkeley National Laboratory, 1 Cyclotron Road, MS 50B-4206, Berkeley, CA 94720, USA}

\author[0000-0001-6876-8284]{Paolo Mazzali}
\affiliation{Astrophysics Research Institute, Liverpool John Moores University, IC2 Liverpool Science Park, 146 Brownlow Hill, Liverpool L3 5RF, UK}
\affiliation{Max-Planck Institute for Astrophysics, Garching, Germany}

\author[0000-0003-1227-3738]{Josiah Purdum}
\affiliation{Division of Physics, Mathematics and Astronomy, California Institute of Technology, Pasadena, CA 91125, USA}

\author[0000-0003-2451-5482]{Russ R. Laher}
\affiliation{IPAC, California Institute of Technology, 1200 E. California Boulevard, Pasadena, CA 91125, USA}

\author[0000-0002-9998-6732]{Avery Wold}
\affiliation{IPAC, California Institute of Technology, 1200 E. California Boulevard, Pasadena, CA 91125, USA}

\author[0000-0003-4531-1745]{Yashvi Sharma}
\affiliation{Division of Physics, Mathematics and Astronomy, California Institute of Technology, Pasadena, CA 91125, USA}

\author[0000-0003-0629-5746]{Leander Lacroix}
\affiliation{LPNHE, Sorbonne-Université, Paris}

\author[0000-0002-7226-0659]{Michael S. Medford}
\affiliation{Department of Astronomy, University of California, Berkeley, CA 94720-3411, USA}
\affiliation{Lawrence Berkeley National Laboratory, 1 Cyclotron Road, MS 50B-4206, Berkeley, CA 94720, USA}

\newcommand{\missing}[1]{\textcolor{red}{\textbf{#1}}}

\begin{abstract}
We present the discovery and analysis of SN\,2022oqm, a Type Ic supernova (SN) detected $<1$\,day after explosion. The SN rises to a blue and short-lived (2\,days) initial peak. Early-time spectral observations of SN\,2022oqm show a hot (40,000\,K) continuum with high-ionization C and O absorption features at velocities of 4000\,km\,s$^{-1}$, while its photospheric radius expands at 20,000\,\kms, indicating a pre-existing distribution of expanding C/O material. After $\sim2.5$\,days, both the spectrum and light curves evolve into those of a typical SN Ic, with line velocities of $\sim10,000$\,km\,s$^{-1}$, in agreement with the photospheric radius evolution. The optical light curves reach a second peak at $t\approx15$\,days. By $t=60$\,days, the spectrum of \oqm\ becomes nearly nebular, displaying strong \ion{Ca}{2} and [\ion{Ca}{2}] emission with no detectable [\ion{O}{1}], marking this event as Ca-rich. The early behavior can be explained by $10^{-3}$\,\msun\ of optically thin circumstellar material (CSM) surrounding either (1) a massive compact progenitor such as a Wolf-Rayet star, (2) a massive stripped progenitor with an extended envelope, or (3) a binary system with a white dwarf. We propose that the early-time light curve is powered by both interaction of the ejecta with the optically thin CSM and shock cooling (in the massive-star scenario). The observations can be explained by CSM that is optically thick to X-ray photons, is optically thick in the lines as seen in the spectra, and is optically thin to visible-light continuum photons that come either from downscattered X-rays or from the shock-heated ejecta. Calculations show that this scenario is self-consistent.\\
\newpage
\end{abstract}

\section{Introduction}
\label{sec:introduction}
While the light curves of Type Ia supernovae (SNe~Ia) are well explained by the radioactive decay of \Nifs, many core-collapse SNe (CCSNe) require an additional powering mechanism for their early-time light curves \citep[For reviews, see][and references therein]{Maguire2017, Arcavi2017, Pian2017}. In the absence of circumstellar material (CSM), the early ultraviolet (UV) through optical light curves of SNe are expected to be the result of shock breakout from the stellar surface \citep{Matzner_1999}, or due to the subsequent cooling of the shocked material \citep[for a review, see][]{Waxman2017,levinson_physics_2020}. 

If CSM is present, an early-time UV-optical component can be explained by the breakout of a radiation-mediated shock (RMS) from the CSM \citep{Campana2006,Waxman2007,Ofek2010,Ofek2014a,Waxman2017,forster_delay_2018}, or possibly due to interaction of the ejecta with confined CSM ejected shortly ($\sim1$\,yr) prior to explosion \citep{Murase2014,maeda_final_2021,maeda_properties_2022}.  In contrast, interaction resulting from progenitors with typical Wolf-Rayet (W-R) stellar winds is not expected to contribute to the optical light curve in SNe~Ib/c \citep{Chevalier2006}. 

The early-time light curves and spectra of SNe are sensitive to the properties of the progenitor star. If shock cooling is the dominant source of energy, the early light curves will be sensitive to the progenitor radius and mass, as well as to the slope of the outer density profile \citep{Nakar2010, Rabinak2011, Piro2015, sapir2016, Piro2021, Morag2022}. If there is confined CSM around the progenitor star, the result of elevated mass loss in the months prior to explosion \citep[e.g.,][]{Ofek2014c,strotjohann2021}, early-time SN spectra can show narrow high-ionization features \citep{galyam2014,Khazov2016,Yaron2017,Bruch2021, JacobsonGalan2022b}. 

The past decade has seen a rapid increase in the early detection of SNe by high-cadence wide-field surveys such as the Palomar Transient Factory (PTF; \citealt{law2009,kulkarni2013}), which detected several such SNe \citep{galyam2011,arcavi2011,nugent2011,galyam2014, Benami2014, Khazov2016, Yaron2017}. Since then, the Astroid-Terrestrial impact Last Alert System (ATLAS; \citealt{Tonry2018}), the Zwicky Transient Facility (ZTF; \citealt{Bellm2019b,Graham2019}), the Distance Less than 40\,Mpc Survey \citep[DLT40;][]{Tartaglia2018}, and most recently the Young Supernovae Experiment (YSE;  \citealt{Jones2021}) have been conducting 1--3\,day cadence wide-field surveys and regularly detect SNe \citep[e.g.][]{Ho2019,Soumagnac2020,Bruch2021, galyam2022,Perley2022,Terreran2022,JacobsonGalan2022b,Tinyanont2022,Hosseinzadeh2022} and fast-transients \citep[e.g.][]{perley2018,ho2020a,Perley2021,Ofek2021} shortly after explosion. Consequently, the study of the early emission from SN explosions is at the forefront of current efforts in the field \citep{Modjaz2019}.

While the early evolution of SNe~II and IIb is relatively well studied \citep{Bersten2012,galyam2014,rubin_2016,Garnavich_2016,Rubin_2017,Arcavi_2017b,Khazov2016,Bersten2018,Prentice2020b,Bruch2021,Ganot_2022,Martinez2022,Medler2022}, multiband observations in both UV and visible light have been obtained during the first few days for only a handful of stripped-envelope SNe \citep[SESNe; but see][]{Taddia2015}. Occasionally, a coincident gamma-ray burst (GRB) or X-ray flash (XRF) resulted in intense UV-optical follow-up observations \citep{Campana2006, Soderberg2008}. Several other well-studied normal and peculiar SNe~Ic \citep{De2018,Horesh2020}, broad-lined SNe~Ic (Ic-BL; reported by \citealt{Ho2019,ho2020b}), and SNe~Icn \citep{galyam2022,Perley2022,Pellegrino2022,Gagliano2022} have been found, showing diverse properties. In some of these cases, a short-lived early blue peak has been observed, possibly consistent with the shock cooling of a low-mass envelope, shock-breakout from a confined shell of CSM, or the subsequent cooling of the shocked material. For example, the SN~Ic iPTF15dtg \citep{Taddia2016b} had an early blue peak associated with the cooling envelope of a massive star, as $M_{\rm ej} \approx 10$\,\msun\ were ejected in the explosion, while other events, such as PTF11mnb, developed a longer double-peaked structure \citep{Taddia2018a}. 

Owing to the absence of He and H in their spectra, SN~Ic progenitors have been suggested to lose their envelope prior to explosion, either due to stellar winds \citep[][and references therein]{Filippenko1997} or through binary interaction \citep{Podsiadlowski1992,Yoon2010,Smith2014b}. The presence and distribution of CSM around SESN progenitors, as well as the measurement of progenitor properties from the shock-cooling peak of SNe,  can provide vital clues to better understand the yet unknown details of the evolution of massive progenitors of SNe and their explosion mechanism.

Here, we report the early-time detection and extensive follow-up observations of SN\,2022oqm, a relatively normal SN~Ic with an early-time UV peak as well as high-ionization and short-lived C/O lines with  4000--5000\,\kms\ velocities, likely originating in an optically thin CSM surrounding the expanding ejecta. In $\S$~\ref{sec:discovery} we report the discovery of the SN. We describe in $\S$~\ref{sec:observations} the multiwavelength monitoring campaign of  \oqm\ and its host galaxy. Section~\ref{sec:analysis} presents an analysis of the spectral and photometric evolution, and we derive basic properties of the explosion such as its blackbody evolution, ejected (and \Nifs) mass, and the  host-galaxy properties. In $\S$~\ref{sec:discussion},
 we discuss our findings and propose that the early-time light curve is explained by an initial CSM interaction possibly followed by a brief period of shock-cooling emission. We present possible interpretations for the origin of \oqm\ in $\S$~\ref{sec:Interpretation}, and we summarize our findings in $\S$~\ref{sec:summary}. 
 
 Throughout the paper we use a $\Lambda$CDM cosmological model with H$_{0} = 67.4$\,\kms\,Mpc$^{-1}$, $\Omega_{M} = 0.315$, and $\Omega_{\Lambda} = 0.685$ \citep{Planck2018a}. 


\section{Discovery}
\label{sec:discovery}

\subsection{Supernova Discovery}
\label{discovery}

\begin{figure}[t]
\centering
\includegraphics[width=\columnwidth]{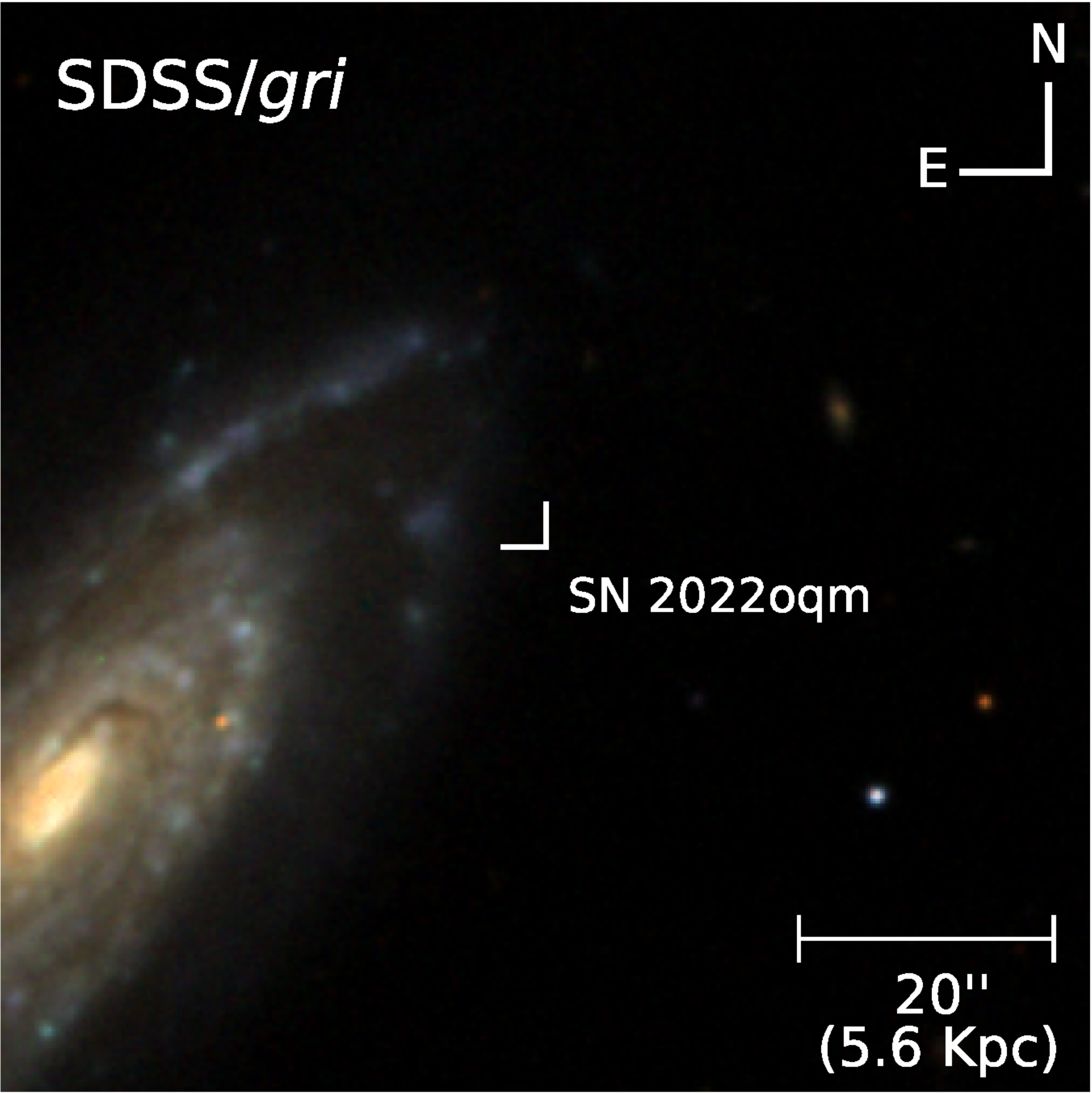} \\
\caption{The location of SN\,2022oqm marked by a white arrow on a false-color SDSS $gri$ image of the host galaxy NGC~5875. The SDSS $gri$ images were combined using the methods of \cite{Lupton2004}.}
\label{fig:discovery}
\end{figure}


 \oqm\ was first detected by the ZTF survey \citep{Bellm2019b,Graham2019} at  $\alpha = 15^{\rm hr}09^{\rm m}08.22^{\rm s}$, $\delta = +52^\circ 32' 05\farcs28$ (J2000.0). It was observed on 2022 July 11 at 04:40 (UTC dates are used throughout this paper; JD = 2,459,771.695) with a $g$-band magnitude of $17.32 \pm 0.04$, following a nondetection one day prior (JD = 2,459,770.764) with a 5$\sigma$ limit of $g = 19.94\,$mag, indicating a rise of $\gtrsim 2.6$\,mag in just one day.

The SN was internally designated ZTF22aasxgjp and was reported to the Transient Name Server\footnote{\url{https://www.wis-tns.org/}} by a ZTF duty astronomer \citep{TNSdiscovery}.
It was discovered in NGC 5875 with a redshift of $z=0.0113$ {\referee \citep{Albareti2017}};  its location is shown in Fig. \ref{fig:discovery}.

We adopt a Hubble-flow distance of $d=58\pm4.1$\,Mpc provided by the NASA Extragalactic Database {\secondref calculator, embedded in the NGC 5875 object page} (NED)\footnote{\hyperlink{https://ned.ipac.caltech.edu/}{https://ned.ipac.caltech.edu/}} and corrected for Virgo, Great Attractor, and Shapley supercluster infall {\referee \citep{Mould2000}}, corresponding to a distance modulus of $33.82\pm0.15$\,mag. This implies that the absolute magnitude of \oqm\ at discovery was $M_g = -16.6 \pm 0.16$\,mag {\secondref (corrected for Galactic reddening; see $\S$~\ref{extinction} )}.


Rapid spectroscopic and photometric observations were obtained shortly thereafter, following the methodology of \citet{galyam2011}. Within the first 8\,hr, we obtained optical $ugri$ photometry and a low-resolution spectrum with the Spectral Energy Distribution Machine (SEDM; \citealt{benami2012}; \citealt{Blagorodnova2018a}), a spectrum with the Gemini Multi-Object Spectrograph (GMOS; \citeauthor{gemini} \citeyear{gemini}), and UV photometry using the UV/Optical Telescope (UVOT; \citealt{Gehrels2004}) at the Neil Gehrels {\it Swift} Observatory. The SEDM photometry showed \oqm\ to be blue ($g-r=-0.29$\,mag) and rapidly rising, with an additional rise of $\sim 0.3$\,mag over 3\,hr (i.e., rising with a rate of 2.4\,mag\,day$^{-1}$). SEDM spectroscopy (resolution $\mathcal{R} \approx 100$) and GMOS ($\mathcal{R} \approx 1500$) spectra revealed highly ionized \CIV, \OV, and \OIV\ features with velocities of $\sim4000\,$\kms. UV photometry indicated that \oqm\ was bright, with $M_{\text{UVM2}}=-17.79\pm0.15\,$mag. Attempts to obtain UV spectra of \oqm\  using the {\it Hubble Space Telescope (HST)} and {\it Swift} UVOT were unsuccessful owing to technical reasons ({\it HST}) and lack of sufficient signal ({\it Swift}).

\subsection{Explosion-Time Estimate}
\label{time}
Typically, in order to establish an explosion time using a well-sampled light curve,  the flux in a given band can be extrapolated to zero assuming a power-law rise \citep[e.g., ][]{Bruch2021, Soumagnac2020}. 

A potentially superior alternative is possible for objects with spectral energy distributions (SEDs) that are well fit by a blackbody. In such cases, the light-curve behavior in a given band is determined by the radius and temperature evolution. A physically motivated model for the rise can thus be acquired from assuming a power-law behavior for the temperature and radius,

\begin{equation}
\label{eq:bb_pl}
T_{\rm eff}=T_{0}(t-t_{0})^{\alpha}\ {\rm and\ } R_{\rm BB}=R_{0}(t-t_{0})^{\beta}\, ,
\end{equation}
\noindent
{\referee where $T_{0}$ and $R_{0}$ are respectively the temperature and radius at day 1, and $\alpha$ and $\beta$ are their corresponding power-law slopes.} Using this model, we constrain the rise using all multiband information during the first 2\,days. The advantage of this method is its sensitivity to the decline of the UV bands as well as the rise of the optical bands, and the ability to simultaneously utilize all available photometric bands. We discuss the fitting process in detail in $\S$~\ref{subsec:blackbody}, and adopt an estimate for the explosion date (JD) of $t_{\rm exp} = 2,459,771.2\pm0.2$. Times $t$ reported hereafter are relative to this date. This estimate is consistent with an explosion time measured using a power-law extrapolation of the $g$-band flux to zero. 

\subsection{Extinction}
\label{extinction}
We correct for foreground Galactic reddening using the \cite{schlafly2011} recalibration of the \cite{schlegel1998} extinction maps. At the location of \oqm, these imply a reddening of $E(B-V) = 0.017$\,mag, which we correct assuming a \cite{Cardelli1989} extinction law with $R_{V}=3.1$. To estimate the host-galaxy extinction, we apply the methods of \cite{stritzinger2018b}, who found that the variance of the optical colors of SNe~Ic is minimal 10\,days after maximum brightness. We determine the peak times and colors by fitting a low-order polynomial to a range of 10\,days around the respective peak time and 10\,days after. We evaluate the $g-r$ and $g-i$ color 10\,days after the $g$-band peak, as well as the $g-r$ and $r-i$ colors 10\,days after the $r$-band peak, and find good agreement with a negligible ($E(B-V)<0.02$\,mag) amount of host extinction. This conclusion is consistent with the location of the SN at a large offset from its host galaxy (Fig.~\ref{fig:discovery}), and with the absence of  narrow Na\,I\,D doublet in absorption. This line is correlated with dust extinction and reddening \citep{poznanski2012}, but was not detected in any of our high signal-to-noise-ratio (S/N) spectra. Hence, we do not apply any host extinction correction to our data.

\section{Observations}
\label{sec:observations}

All observations are made public via WISeREP \citep{yaron2012}.

\subsection{Spectroscopy}
\label{subsec:spectroscopy}

We obtained {\referee 36} epochs of spectroscopy between $t=0.6$\,days and $t=60.1$\,days. The details of the observations and reductions for the telescopes used are provided below.


\begin{itemize}
    \item 8\,m Gemini North telescope on Maunakea; 2 epochs.
    GMOS was used to obtain the data. 
    For each spectrum, four 900\,s exposures were obtained {\referee in the long-slit mode} with the B600 grating {\referee ($\mathcal{R}\approx1500$)}, and two different central wavelengths of 5200\,\AA\ and 5250\,AA\ were adopted to cover the chip gap. The data were reduced using the Gemini IRAF package v. 1.14.\footnote{\href{http://www.gemini.edu/observing/phase-iii/understanding-and-processing-data/data-processing-software/gemini-iraf-general}{http://www.gemini.edu/observing/phase-iii/understanding-and-processing-data/data-processing-software/gemini-iraf-general}}. {\referee The slit was oriented at or near the parallactic angle to minimize slit losses caused by atmospheric dispersion \citep{Filippenko1982}}.

    \item 1.5\,m telescope at Palomar Observatory (P60); 16 epochs.
    Data were acquired with the integral field unit (IFU; {\referee $\mathcal{R}\approx 100$}) SEDM, and reduced using the automatic SEDM pipeline \citep{rigault2019, Kim2022}. {\referee The slit was oriented at or near the parallactic angle.} 

    \item 2.56\,m Nordic Optical Telescope (NOT) at the Observatorio del Roque de los Muchachos on La Palma (Spain); 7 epochs between July 11 and September 8.\footnote{Program ID 64-501; PI J. Sollerman}.
    Low-resolution spectra were obtained with the Alhambra Faint Object Spectrograph and Camera (ALFOSC)\footnote{\href{http://www.not.iac.es/instruments/alfosc}{{http://www.not.iac.es/instruments/alfosc}}}  with a 1\farcs0 wide slit and grism \#4 {\referee ($\mathcal{R}$ = 360)}, providing a wavelength coverage of 3500--9000\,\AA. The data were reduced using standard methods with the data-reduction pipelines PyNOT\footnote{\href{https://github.com/jkrogager/PyNOT}{https://github.com/jkrogager/PyNOT}} v. 1.0.1 and PypeIt v. 1.8.1 \citep{Prochaska2020a}. {\referee The slit was oriented at or near the parallactic angle.}

    \item 3\,m Shane telescope at Lick Observatory; 7 epochs. We used the Kast double spectrograph \citep{Miller1988} configured with the 2\farcs0 wide slit, the 600/4310 grism, and the 300/7500 grating to obtain a series of 7 optical spectra.  This configuration resulted in a spectral resolution of a $\sim 5$\,\AA\ on the blue side ($\sim 3630$--5680\,\AA) and $\sim 12$\,\AA\ on the red side ($\sim 5450$--10,740\,\AA), corresponding to a resolving power of $\mathcal{R} \approx 800$ across the observed band. All spectra were obtained at an airmass less than 1.6 and with the slit was oriented at or near the parallactic angle. Data were reduced (including removal of telluric features) following the approach described by \citet{Silverman2012}. At each epoch 3 red side exposures were taken to minimize the effects of cosmic rays. A single blue side exposure was taken with an additional 60\,s in order to synchronize its readout time with the final red exposure.
    
    \item 5\,m Hale telescope at Palomar Observatory (P200); 2 epochs. We used the Double Beam Spectrograph (DBSP; \citealt{Oke1982}). The data were reduced following standard procedures using the P200/DBSP pipeline described by \cite{Roberson2022}. {\referee The 600/4000 and 316/7500 gratings were used in the blue and red arms, respectively, corresponding to a resolving power of $\mathcal{R}\approx 1000$ over the observed 3200--10,000\,\AA\ bandpass.} {\referee The slit was oriented at or near the parallactic angle.}
    
    \item 2.0\,m Liverpool Telescope \citep[LT;][]{Steele2004}; one epoch. The Spectrograph for the Rapid Acquisition of Transients (SPRAT; \citealt{Piascik2014}) was used to obtain data, which were reduced using the LT pipeline \citep{Barnsley2016}. {\referee The blue optimized mode was used, with a central spectral resolution of $\mathcal{R}=350$. {\referee The slit was oriented at or near the parallactic angle.}}
    
    \item {\referee 10\,m Keck I telescope at the W. M. Keck Observatory; 1 epoch.  The Low-Resolution Imaging Spectrometer
\citep[LRIS; ][]{Oke1995} was used to acquire a single long-slit spectrum using the 1\farcs0 wide slit oriented at the parallactic angle.  The 600/4000 grism and 400/8500 grating were used for the blue and red arms, respectively.  This configuration resulted in spectral resolutions of $\sim 5$\,\AA\ on the blue side ($\sim 3165$--5643\,\AA) and $\sim 9$\,\AA\ on the red side ($\sim 5359$--10,256\,\AA), and a resolving power of $\mathcal{R}\approx900$. Data were reduced using the \package{LPipe} automaed pipeline \citep{Perley2019}}. 
    
\end{itemize}

\begin{figure*}[t]
\centering
\includegraphics[width=\textwidth]{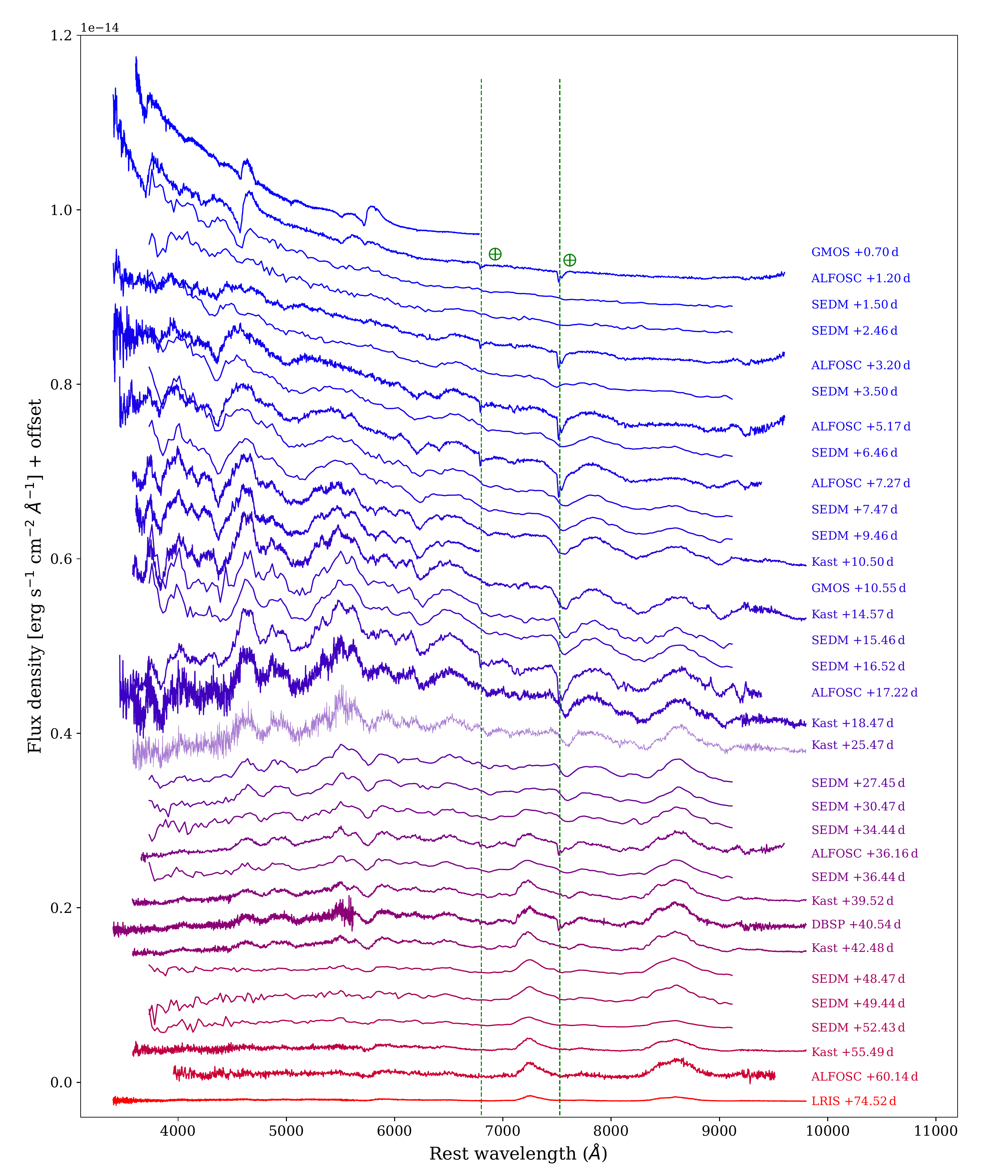} \\
\caption{Spectral evolution of SN\,2022oqm. The phases are reported relative to the estimated explosion time of JD = 2,459,771.217. The locations of telluric features are marked with vertical green dashed lines.}
\label{fig:spec_main}
\end{figure*}

The details of the spectroscopic observations are listed in Table~\ref{tab:spec}, and the spectra are shown in Fig.~\ref{fig:spec_main}. All spectra have been calibrated to the Galactic-extinction-corrected ZTF $g$, $r$, and $i$ photometry by scaling the reduced spectrum with a linear function to match the flux obtained from photometry. 

\begin{deluxetable}{lcccc}
\tablecaption{Summary of Spectroscopic Observations of SN 2022oqm \label{tab:spec}} 
\tablecolumns{5}
\tabletypesize{\footnotesize}  
\tablehead{
\colhead{{\referee Start Date (UTC)}} &
\colhead{{\referee Phase (d)}} &
\colhead{Telescope} &
\colhead{Spectrograph} &
\colhead{Exp (s)}
}
\startdata 
2022-07-11.31 &  0.60 &       P60 &       SEDM &    1800 \\
2022-07-11.42 &  0.70 &  Gemini-N &       GMOS &    3600 \\
2022-07-11.91 &  1.20 &       NOT &     ALFOSC &    1800 \\
2022-07-12.21 &  1.50 &       P60 &       SEDM &    1800 \\
2022-07-12.98 &  2.26 &        LT &      SPRAT &     750 \\
2022-07-13.18 &  2.46 &       P60 &       SEDM &    1800 \\
2022-07-13.92 &  3.20 &       NOT &     ALFOSC &    1800 \\
2022-07-14.22 &  3.50 &       P60 &       SEDM &    1800 \\
2022-07-15.89 &  5.17 &       NOT &     ALFOSC &     600 \\
2022-07-17.18 &  6.46 &       P60 &       SEDM &    1800 \\
2022-07-17.99 &  7.27 &       NOT &     ALFOSC &    1800 \\
2022-07-18.18 &  7.47 &       P60 &       SEDM &    1800 \\
2022-07-20.18 &  9.46 &       P60 &       SEDM &    1800 \\
2022-07-21.23 & 10.52 &     Shane &       Kast &    2160/2100 \\
2022-07-21.27 & 10.56 &  Gemini-N &       GMOS &    1600 \\
2022-07-22.32 & 11.60 &      P200 &       DBSP &     300 \\
2022-07-25.30 & 14.59 &     Shane &       Kast &    1860/1800 \\
2022-07-26.17 & 15.46 &       P60 &       SEDM &    1800 \\
2022-07-27.24 & 16.52 &       P60 &       SEDM &    1800 \\
2022-07-27.93 & 17.22 &       NOT &     ALFOSC &    1200 \\
2022-07-29.20 & 18.48 &     Shane &       Kast &    1860/2100 \\
2022-08-05.19 & 25.47 &     Shane &       Kast &    2460/2400 \\
2022-08-07.16 & 27.45 &       P60 &       SEDM &    1800 \\
2022-08-10.19 & 30.47 &       P60 &       SEDM &    1800 \\
2022-08-14.16 & 34.44 &       P60 &       SEDM &    1800 \\
2022-08-15.88 & 36.16 &       NOT &     ALFOSC &    2400 \\
2022-08-19.16 & 36.44 &       P60 &       SEDM &    1800 \\
2022-08-19.24 & 36.52 &     Shane &       Kast &    3360/3300 \\
2022-08-20.26 & 40.54 &      P200 &       DBSP &     600 \\
2022-08-22.20 & 42.49 &     Shane &       Kast &    3360/3300 \\
2022-08-28.18 & 48.47 &       P60 &       SEDM &    2250 \\
2022-08-29.16 & 49.44 &       P60 &       SEDM &    2250 \\
2022-09-01.14 & 52.43 &       P60 &       SEDM &    2250 \\
2022-09-04.21 & 55.49 &     Shane &       Kast &    3360/3300 \\
2022-09-08.85 & 60.14 &       NOT &     ALFOSC &    3600 \\
{\referee 2022-09-23.24} & {\referee 74.53} &      {\referee Keck} &       {\referee LRIS} &    {\referee 1200}\\
\enddata
\end{deluxetable}

\subsection{Photometry}
\label{subsec:opt-photometry}

ZTF photometry in the \textit{gri} bands was acquired using the ZTF camera \citep{Dekany2020} mounted on the 48\,inch (1.2\,m) Samuel Oschin Telescope at Palomar Observatory (P48). 
These data were processed using the ZTF Science Data System \citep[ZSDS;][]{Masci2019}. 
Light curves were obtained using the 
ZTF forced-photometry service\footnote{See ztf\_forced\_photometry.pdf under \url{https://irsa.ipac.caltech.edu/data/ZTF/docs}} on difference images produced using the optimal image subtraction algorithm of \citet{Zackay2016} at the position of the SN, calculated from the median ZTF alert locations to lie at $\alpha = 15^{\rm hr}09^{\rm m}08.213^{\rm s}$, $\delta = +52^\circ 32' 05\farcs17$ (J2000.0). We removed images that have flagged difference images, bad pixels close to the SN position, a large standard deviation in the background region, or a seeing of more than 4\arcsec. We performed a baseline correction to ensure the mean of the pre-SN flux is zero. We report detections above a $3\sigma$ threshold, and $5\sigma$ nondetections. These data are provided in Table \ref{tab:phot}.

In addition to the ZTF photometry, we triggered an extensive photometric follow-up campaign using the following telescopes.
\begin{itemize}

\item The UV-Optical Telescope (UVOT) onboard the \textit{Neil Gehrels Swift Observatory} \citep{Gehrels2004,Roming2005}. The images were reduced using the \swift\ \package{HEAsoft}\footnote{\url{https://heasarc.gsfc.nasa.gov/docs/software/heasoft/}
v. 6.26.1.} toolset. Individual exposures comprising a single epoch were summed using  \package{uvotimsum}. Source counts were then extracted using \package{uvotsource} from the summed images using
a 5{\arcsec} circular aperture. The background was estimated from several larger regions surrounding the host galaxy. These counts were then converted to fluxes using the photometric zero points of \cite{Breeveld2011} with the latest calibration files from September 2020. We did not attempt to subtract the host flux at the location of the SN. This is justified, as the field was observed before the SN exploded in the $UVW1$ and $U$ bands, revealing no underlying sources. This is also corroborated by archival Legacy Survey images \citep{Dey2019} from the Beijing-Arizona Sky Survey fields \citep{Zou2017}, and by deep PS1 imaging \citep{Flewelling2020}. We used the {\it Swift} pre-SN images in the $UVW1$ and $U$ bands, and the surrounding host flux in the $UVW2$, $UVM2$, $B$, and $V$ bands, and estimate the host contribution to the SN flux as negligible in all bands for all of the epochs presented in this paper. 

\item 
The Optical Imager (IO:O) at the 2.0\,m robotic LT the Observatorio del Roque de los Muchachos. We used the $u$, $g$, $r$, $i$, and $z$ filters.  Images were reduced using the IO:O automatic pipeline; image subtraction versus Pan-STARRS ($g$, $r$, $i$, $z$) or SDSS ($u$) reference imaging was performed with a custom IDL routine.  Aperture photometry was conducted on the subtracted image using SDSS secondary standards.

\item The Rainbow Camera \citep{Blagorodnova2018a} on the Palomar 60\,inch (1.52\,m) telescope (P60; \citealt{Cenko2006}). Reductions were performed using the automatic pipeline described by \cite{fremling2016}. 

\item The 0.75\,m Katzman Automatic Imaging Telescope (KAIT) and the 1.0\,m Nickel telescope at Lick Observatory. The data were reduced using a custom pipeline\footnote{https://github.com/benstahl92/LOSSPhotPypeline} presented by \citet{stahl2019}. No image-subtraction procedure was applied (see above for {\it Swift}), and the Pan-STARRS1\footnote{http://archive.stsci.edu/panstarrs/search.php} catalog was used for calibration. Point-spread-function (PSF) photometry was obtained using DAOPHOT (Stetson 1987) from the IDL Astronomy User’s Library\footnote{http://idlastro.gsfc.nasa.gov/}. Apparent magnitudes were all measured in the KAIT4/Nickel2 natural system, and then transformed back to the standard system using local calibrators and color terms for KAIT4 and Nickel2 \citep{stahl2019}.

\item The 6.5\,m MMT equipped with the Magellan infrared spectrograph (MMIRS) at the Fred Lawrence Whipple Observatory. We acquired 1 epoch of $JHK_{s}$ imaging with 90$\arcsec$ dithers between different exposures. The images were reduced by customized scripts within IRAF which include dark subtraction, sky subtraction, and coaddition of multiple exposures. Instrumental magnitudes of all stars in the imaging field with S/N $>5$  were obtained with PSF photometry using IRAF task {\it daophot}, and the zero point was obtained by calibrating the instrumental magnitudes to the 2MASS catalog \citep{Skrutskie2006}. 

\end{itemize}

\begin{figure}[t]
\centering
\includegraphics[width=\columnwidth]{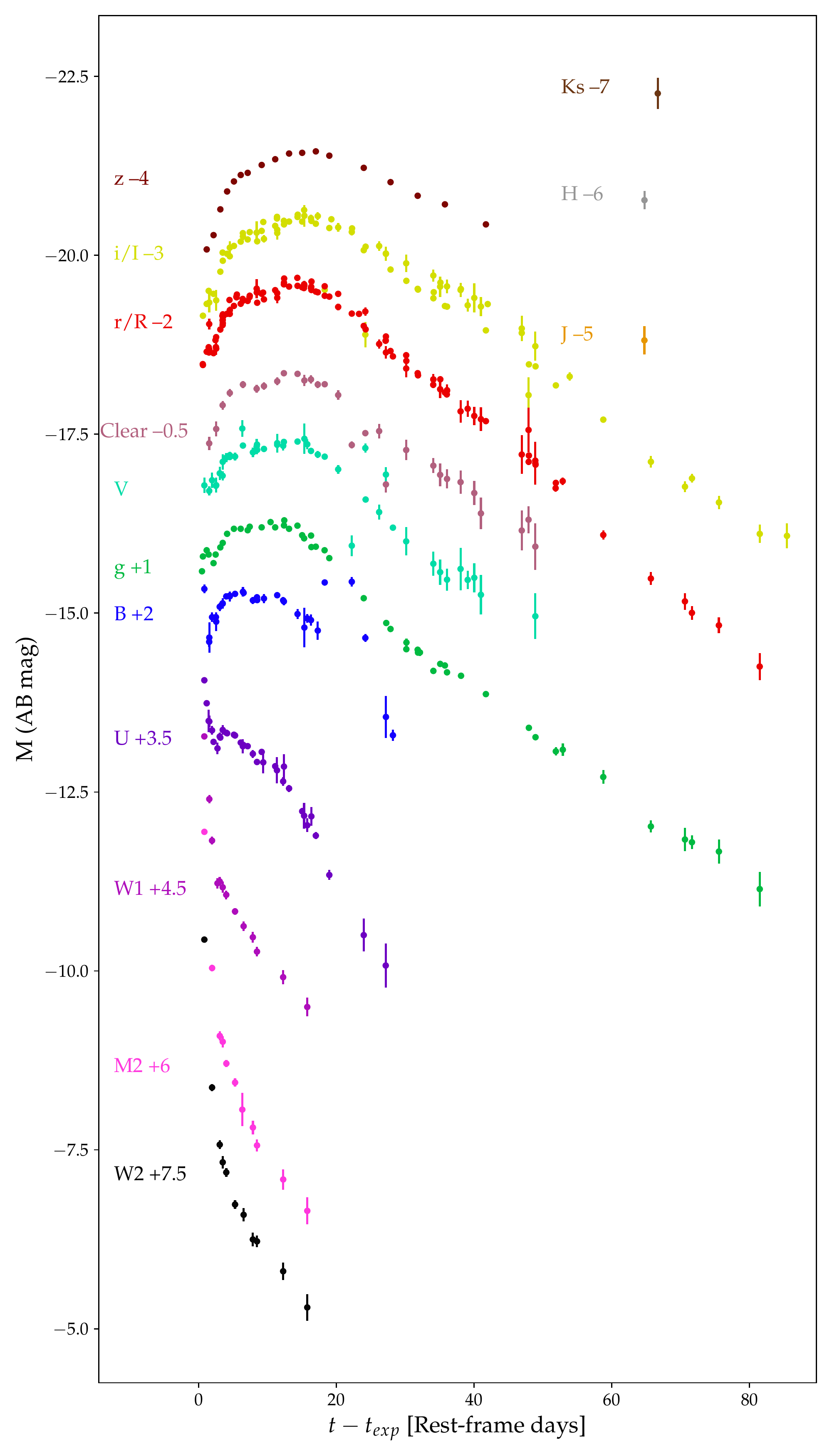} \\
\caption{The UV-optical and NIR light curves of SN\,2022oqm. Note the early blue peak, most notable in the $u$ and $g$ bands.}
\label{fig:lc_main}
\end{figure}

The resulting light curves appear in Fig.~\ref{fig:lc_main}. Some cross-instrument differences between the \swift/UVOT and the KAIT and Nickel $V$-band photometry remain, so we apply a $-0.18$ mag offset for clarity to the  \swift/UVOT $V$-band photometry to align these data in all figures where the light curves appear. The offset is not applied in our analysis, and applying it does would change any of the results. 

\begin{deluxetable*}{ccllc}
\label{tab:phot}
\centering
\tablecaption{Log of photometric observations (truncated)}
\tablewidth{24pt} 
\tablehead{\colhead{JD} & \colhead{$t$ [rest-frame days]} & \colhead{Instrument} & \colhead{Filter} & \colhead{AB Magnitude}} 
\tabletypesize{\scriptsize} 
\startdata
2459771.7 & 0.47 & P48/ZTF & \textit{g} & $17.23 \pm 0.02$ \\
2459771.81 & 1.47 & P60/SEDM & \textit{r} & $17.34 \pm 0.02$ \\
2459771.83 & 1.51 & P60/SEDM & \textit{g} & $17.03 \pm 0.03$ \\
2459771.84 & 1.53 & P60/SEDM & \textit{r} & $17.35 \pm 0.02$ \\
2459771.84 & 2.53 & P60/SEDM & \textit{i} & $17.66 \pm 0.03$ \\
2459772.04 & 3.44 & Swift/UVOT & \textit{W1} & $16.04 \pm 0.04$ \\
2459772.04 & 3.51 & Swift/UVOT & \textit{U} & $16.26 \pm 0.04$ \\
2459772.04 & 3.58 & Swift/UVOT & \textit{B} & $16.47 \pm 0.06$ \\
2459772.04 & 4.47 & Swift/UVOT & \textit{W2} & $15.88 \pm 0.03$ \\
2459772.05 & 5.46 & Swift/UVOT & \textit{V} & $16.85 \pm 0.11$ \\
2459772.05 & 5.5 & Swift/UVOT & \textit{M2} & $15.88 \pm 0.03$ \\
2459772.39 & 5.54 & LT/IO:O & \textit{g} & $16.91 \pm 0.08$ \\
2459772.39 & 8.43 & LT/IO:O & \textit{r} & $17.17 \pm 0.05$ \\
2459772.39 & 8.51 & LT/IO:O & \textit{i} & $17.51 \pm 0.07$ \\
2459772.39 & 10.43 & LT/IO:O & \textit{u} & $16.62 \pm 0.06$ \\
2459772.39 & 11.37 & LT/IO:O & \textit{z} & $17.78 \pm 0.09$ \\
\enddata

\tablenotetext{a}{All measurements are reported in the AB system and are corrected for Galactic line-of-sight reddening.}
\tablenotetext{b}{The full table is made available electronically on WISeREP.}
\vspace{-0.15cm}

\end{deluxetable*}

\subsection{X-ray Follow-up Observations}
\label{subsec:xray}

While monitoring SN\,2022oqm with UVOT, \swift\ also observed the field between 0.3 and 10\,keV with its onboard X-ray telescope (XRT) in photon-counting mode \citep{Burrows2005a}. We analyzed these data with the online tools provided by the UK \swift\ team\footnote{\href{https://www.swift.ac.uk/user_objects/}{https://www.swift.ac.uk/user\_objects}} that use the methods described by \citet{Evans2007a, Evans2009a} and the software package \package{HEASoft} v. 6.29.

\oqm\ evaded detection at all epochs ($N=16$, between $t=0.8$ and $t=16$\,days). The median $3\sigma$ count (ct)-rate limit of each observing block is 0.006\,ct\,s$^{-1}$ (0.3--10\,keV). Stacking all data lowers the upper limit to $0.0004\,{\rm ct\,s}^{-1}$. Assuming a Galactic neutral hydrogen column density of $n(H)= 1.73\times10^{20}\,{\rm cm}^{-2}$ \citep{HI4PI2016a} and a power-law spectrum with a photon index of 2, the count rates correspond to an unabsorbed flux limit of $2.2\times10^{-13}$ (the median luminosity of the unbinned data) and $1.5\times10^{-14}\,{\rm erg\,cm}^{-2}\,{\rm s}^{-1}$ (binned over all epochs) in the 0.3--10\,keV bandpass. At the distance of \oqm\ this corresponds to luminosity $L_{X} < 8.7\times10^{40}\,{\rm erg\,s}^{-1}$ (unbinned) and $L_{X} < 6.2\times10^{39}\,{\rm erg\,s}^{-1}$ (binned) in the range 0.3--10\,keV.

\subsection{Search for Prediscovery Emission}
\label{subsec:precursor}

\begin{figure}[t]
\centering
\includegraphics[width=0.45\textwidth]{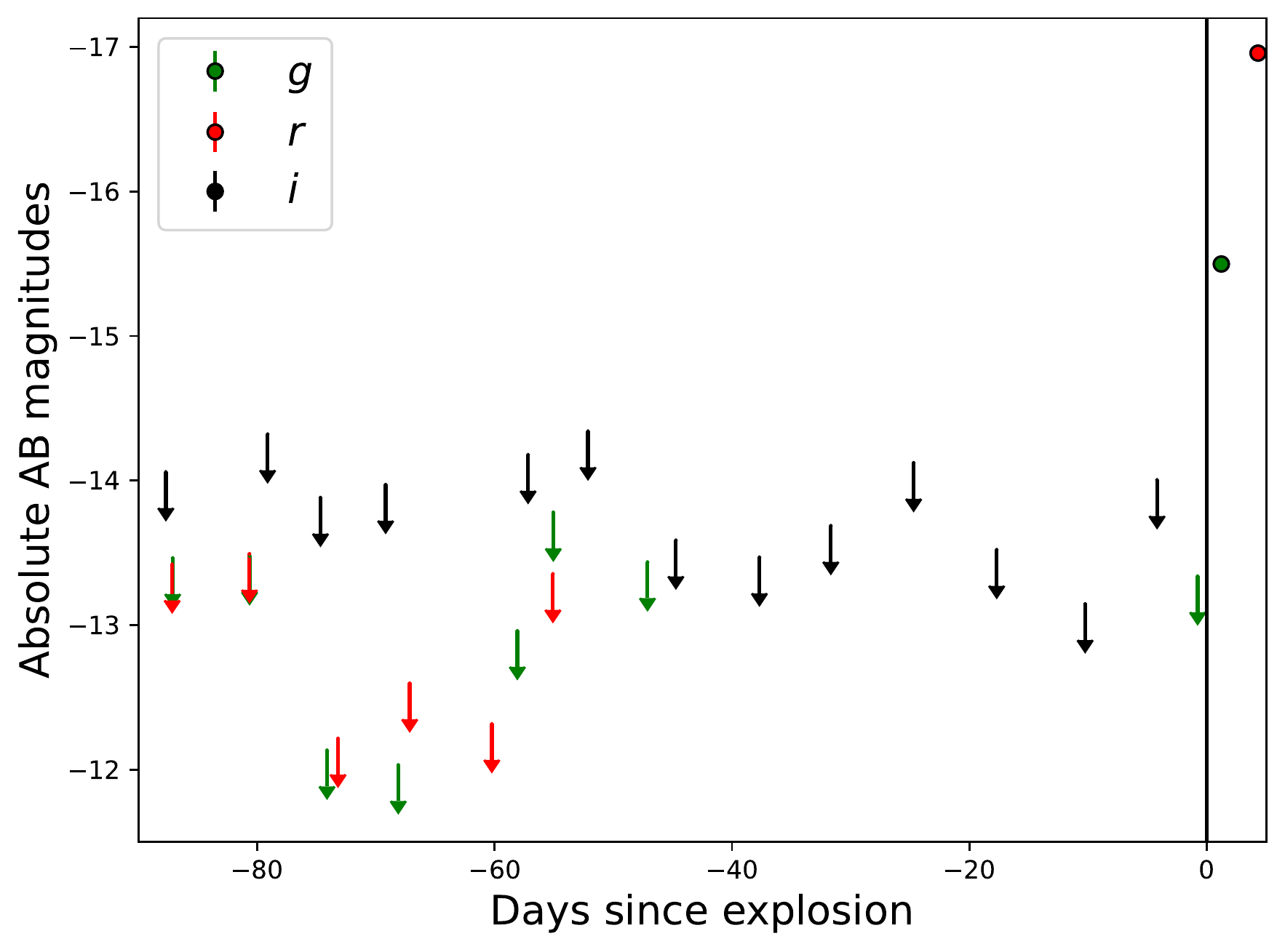} \\
\caption{Light curve in 7\,day bins for the last 100\,days before the SN explosion. Arrows mark $5\sigma$ upper limits in the ZTF $g$, $r$, and $i$ bands. No prediscovery emission is detected at the SN site throughout the ZTF survey.}
\label{fig:presn_lc_zoom}
\end{figure}

Many SNe with spectroscopic and photometric signatures of CSM show prediscovery emission in the weeks, months, or years before the explosion. While in most cases prediscovery emission has been detected for SNe~IIn \citep[e.g.,][]{Fraser2013,Mauerhan2013,Ofek2013,Pastorello_2013,Margutti_2013,strotjohann2021}, precursors have also been detected for SNe~Ibn \citep[][]{Pastorello2007,Foley2007}, SNe~Ic-BL \citep[][]{corsi2014,Ho2019}, and possibly for a SN~IIb \citep[][]{Strotjohann2015}. Here, we check for prediscovery emission for \oqm. 

The ZTF survey first started monitoring the position of SN\,2022oqm 4.3\,yr before the SN explosion and we obtain a forced-photometry light curve for all difference images available at IPAC\footnote{\url{https://irsa.ipac.caltech.edu/applications/ztf/}} following the methods described by \citet{strotjohann2021}. We discard 5.8\% of the observations because they have either flagged difference images, bad pixels close to the SN position, a large standard deviation in the background region, or a seeing disk $> 4''$. After these quality cuts, we are left with a total of 2329 pre-SN observations during 658 different nights. We perform a baseline correction and verify that the error bars are large enough to account for random scatter before the SN explosion. Next, we bin the light curve using a variety of bin sizes (1, 3, 7, 15, 30, and 90-day-long bins) owing to the unknown outburst duration and search the unbinned and binned light curves for $5\sigma$ detections before the SN explosion. 

We do not detect any pre-SN outbursts and here present limits for 7-day-long bins. We correct for the Galactic foreground extinction of $E(B-V)=0.017$\,mag and adopt a distance modulus of 33.8\,mag. The median limiting magnitude is $-12.9$\,mag in the $g$ and $r$ bands, and $-13.9$\,mag in the $i$ band. ZTF $i$-band observations are generally less constraining owing to the reduced sensitivity of the CCD and because fewer observations are obtained in this band. Eruptions that are brighter than $-13$\,mag in the $r$ band and last for at least a week can be excluded during 78 weeks (84 weeks for the $g$ band); this corresponds to 35\% of the time during the 4.3\,yr before the explosion.
In the last 3\,months before the SN explosion, the position was mostly observed in the $i$ band and the absolute magnitude limits in this time window are shown in Fig.~\ref{fig:presn_lc_zoom}. During this time we can exclude week-long precursor eruptions that are brighter than $-14$ mag in the $i$ band. {\referee This rules our bright precursors, in the range observed for strongly interacting SNe~IIn {\secondref and SNe~Ibn}, which typically reach $M_{r}\approx -14$\,mag, and can occasionally reach $-17$\,mag \citep[][and references therein]{strotjohann2021}.}


\section{Analysis}
\label{sec:analysis}

\subsection{Spectral Analysis}
\label{subsec:spec_analysis}

\begin{figure*}[t]
\centering
\includegraphics[width=\textwidth]{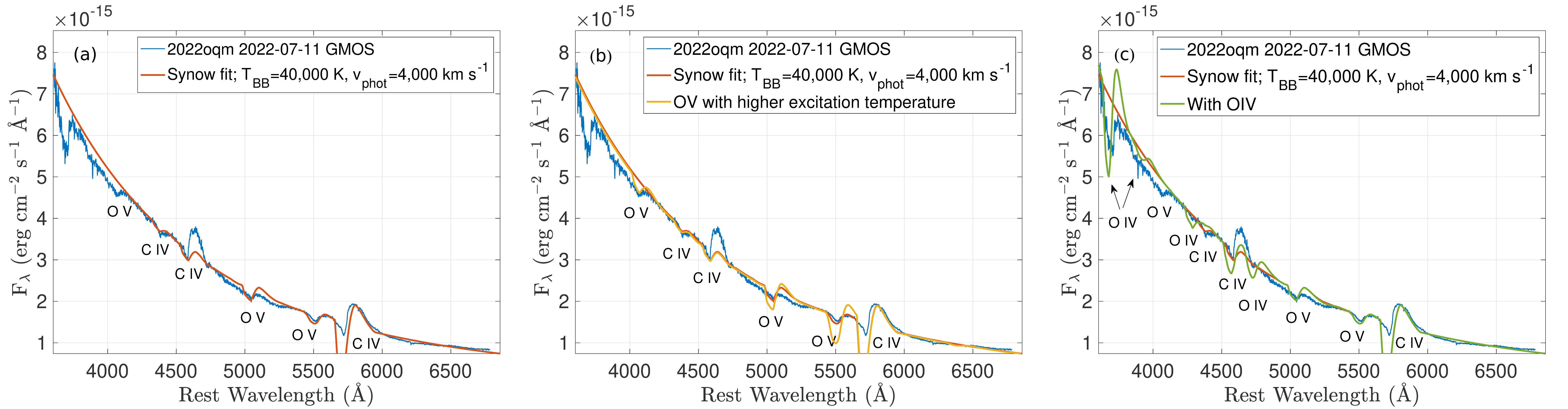} 
\caption{\package{SYNOW} fits to the {\referee first high-spectral resolution} spectrum (GMOS; $t=0.7$\,d). All fits assume a 40,000\,K continuum and an expansion velocity of 4000\,\kms. (a) \ion{C}{4} and \ion{O}{5} using the default parameters of \package{SYNOW}. (b) The same as (a) but with a higher specific excitation temperature for \ion{O}{5}. (c) The same as (a), but also including \ion{O}{4}.}
\label{fig:synow}
\end{figure*}

We use the parameterized supernova synthetic-spectrum \package{SYNOW} code \citep{branch2005} in order to interpret the $t = 0.7$\,d GMOS spectrum, {\referee chosen since it is the earliest high-resolution spectrum}. Using this approach, a blackbody is first fit to the continuum. A spherical expansion velocity is assumed and various ions are added in order to match the lines. Owing to the simplifying underlying assumptions of the \package{SYNOW} approach (e.g., spherical, homologous expansion, and resonant-scattering line formation above a sharp blackbody-spectrum-emitting photosphere), this modeling can only be used to identify and verify the prominent line features, but not to assess physical parameters such as elemental abundances or relative mass fractions. We therefore also avoid performing any fine tuning of the different ion parameters. 

We display three possible fits in Fig.~\ref{fig:synow}, with an increasing amount of lines matched by the fit. The fits are obtained for an expansion velocity of 4000\,\kms\ and for a blackbody temperature of 40,000 $\pm$ 10,000\,K. In the first panel, we acquire a good match for all features $\gtrsim 4300$\,\AA\ by high-ionization lines of pure carbon and oxygen (\ion{C}{4} and \ion{O}{5}). In the second panel, the overplotted yellow fit shows that the dip around 4070\,\AA\ is likely \ion{O}{5} $\lambda4124$. However, forming this feature in \package{SYNOW} requires the assumption of a higher specific excitation temperature value for this ion, leading to an overshoot in the strengths of the additional \ion{O}{5} absorption around 5040\,\AA\ and 5500\,\AA. This is not surprising given the limitations of the code; overall, the identification of \ion{O}{5} with multiple observed features seems secure. The third panel includes \ion{O}{4}, which serves mainly to explain the dip on the blue edge --- the strong \ion{O}{4} $\lambda\lambda$3726, 3729 lines (blueshifted by $\sim 4000$\,\kms).  \ion{O}{4} also contributes to the \ion{C}{4} feature around 4600\,\AA\ and creates additional dips that explain weak features in the spectrum, suggesting the likely existence of \ion{O}{4}. We note that the $\lambda\lambda$3726, 3729 lines are also associated with [\ion{O}{2}] transitions, but this interpretation is disfavored owing to the multiple other high-ionization features, and the low density associated with [\ion{O}{2}] transitions.
Also, while the feature at $\sim 4600$\,\AA\ is close to the \ion{He}{2} $\lambda4686$ line, associating the two would place the maximum absorption of the feature at $\sim 7000$\,\kms, which is inconsistent with the other features in the spectrum. This would not match the emission peak, missing it by $\sim 3000$\,\kms. Similarly, associating the 5800\,\AA\ feature with \ion{He}{1} {\secondref $\lambda5876$} requires an expansion velocity of 8000\,\kms, and it would place the peak emission $\sim 4500$\,\kms\ from the line rest wavelength. A C/O composition is favored, requiring a single expansion velocity and better matching the peak emission in all lines.

\begin{figure}[t]
\centering
\includegraphics[width=\columnwidth]{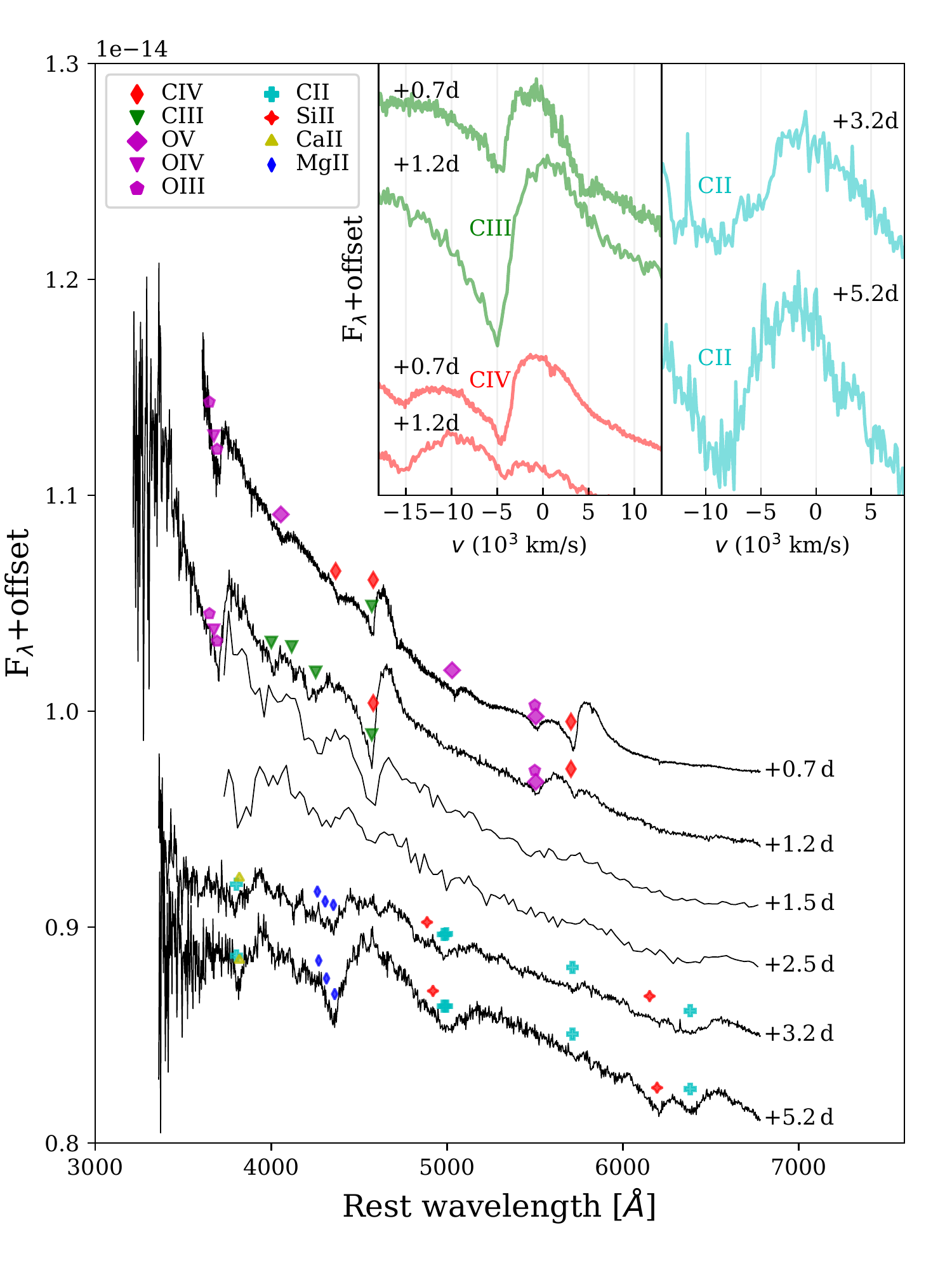} 
\caption{Early-time spectra of SN\,2022oqm. The main panel shows the spectral evolution in the first few days, and the inset shows a zoom-in view of the {\referee \ion{C}{4} $\lambda\lambda$5801, 5812 , \ion{C}{3} $\lambda\lambda$4647, 4650, and \ion{C}{2} $\lambda$6578} features. At $t<3$\,d, the spectrum is dominated by C/O high-ionization features with 4000--5000\,\kms\ velocities. After $t=3$\,d, the spectrum develops low-ionization absorption features with $\sim 10,000$\,\kms\ velocities.}
\label{fig:early_spectra}
\end{figure}

Figure ~\ref{fig:early_spectra} shows the early-time spectral evolution of \oqm\ at subsequent epochs. The high-ionization C/O features observed in the first spectrum evolve into lower-ionization features over the first 3\,days. In the second epoch, these features widen to a velocity of 5500\,\kms, measured from peak emission to absorption. By $t=3.2$\,days, all features broaden to a line velocity of $\sim 10,000$\,\kms. This evolution can be seen in the inset of Fig.~\ref{fig:early_spectra}.

While the absorption minima of the early-time spectra have velocities of 4000\,\kms\ and 5500\,\kms\ for the first and second epochs (respectively), the blue edge of the absorption reaches SN-ejecta-like velocities of $\sim 12,000$\,\kms\ in the first epoch and extends out to $\sim 15,000$\,\kms\ in the second epoch. It is well known for W-R stars that the asymptotic wind velocity is typically only measured in strong UV resonance lines. Indeed, \cite{Perley2022} show that in the early-time spectra of SN\,2021csp, a SN~Icn with a C/O expanding CSM {\referee (indicated by the narrow $\sim 2000$\,\kms\ features in its early spectrum)}, the blue edge of the optical \ion{C}{0} features is at lower velocities by a factor of 1.5 compared with those measured for UV \ion{C}{0} lines. Applying such a correction factor to our data would imply a velocity distribution extending to $\sim 20,000$\,\kms\ for the features in the early spectra. However, the emission maximum is much less extended, at $v=5000$\,\kms. 


During its photospheric phase, \oqm\ develops typical SN~Ic features ---  {\referee namely \ion{Si}{2} $\lambda6355$, \ion{O}{1} $\lambda7774$, \ion{Ca}{2} $\lambda\lambda$ 3934, 3969, \ion{Ca}{2} $\lambda\lambda\lambda$8498, 8542, 8662, and \ion{Mg}{2} $\lambda4481$, as well as a prominent \ion{Na}{1} $\lambda\lambda$5890, 5896}. As the evolution progresses, the spectrum develops Fe absorption features and unusually strong \ion{Ca}{2} and later also [\ion{Ca}{2}] emission. {\referee This suggests a Ca-rich SN~Ic classification is appropriate for \oqm.}

By the time it becomes {\referee partially} nebular at $t=60$\,days, the spectrum is dominated by the Ca emission features on the red side, with \ion{Fe}{2} absorption upon an elevated continuum on the blue side {\referee  as well as a \ion{Na}{1} $\lambda\lambda$5890, 5896 P~Cygni profile}. There is no detectable {\referee $\lambda\lambda$6300, 6364} [\ion{O}{1}] emission during the early nebular phase, indicating that \oqm\ falls into the category of ``Ca-rich SNe" \citep{Filippenko2003, Perets2010}.  To place an upper limit on the [\ion{O}{1}] emission, we assume it accounts for all the luminosity in a region surrounding the line with a similar velocity to the [\ion{Ca}{2}] feature, and find that $L_{[\rm O\ \rm I]} < 1.1 \times 10^{37}$\,erg\,s$^{-1}$ and that the flux ratio [\ion{Ca}{2}]/[\ion{O}{1}]$ > 4$. Since the [\ion{Ca}{2}]/[\ion{O}{1}] ratio can be time variable,  \cite{De2020} used a criterion of  [\ion{Ca}{2}]/[\ion{O}{1}]  $> 2$ for a single phase to ensure good separation of Ca-rich events at all phases. {\referee To extract the velocity of the  \ion{Ca}{2} $\lambda\lambda$7291, 7324 feature, we fit the velocity profile of \ion{Ca}{2} $\lambda\lambda$7291, 7324 with a Gaussian model. We adopt an average wavelength of 7307.5 for the reference wavelength, and fit two individual components with the same width, height, and offset. Our best-fit model had a full width at half-maximum intensity (FWHM) of 6900\,\kms\ (velocity of a single component) and a blueshift of $\Delta v = 1700$\,\kms. Thus,} in addition to its unusual strength, the [\ion{Ca}{2}] feature  has an FWHM at the high end of the SN~Ic distribution  \citep{Prentice2022}.

\begin{figure*}[t]
\centering
\includegraphics[width=\textwidth]{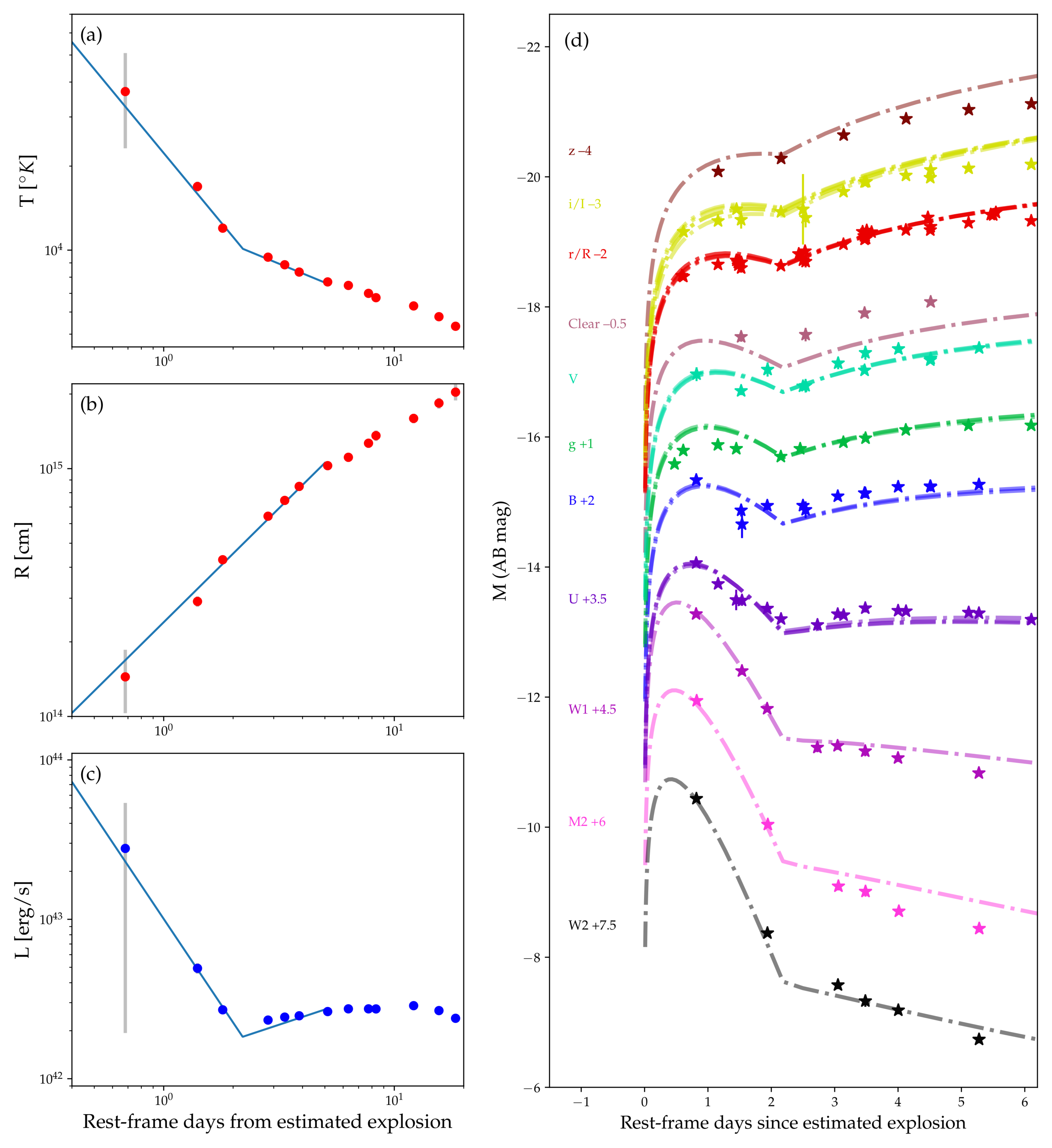} \\
\caption{(Blackbody evolution of \oqm. Data points are calculated by interpolating the SED at the times of UV photometry. The solid lines show the best-fit broken power law to the light curves. The blackbody evolution shows a dramatic transformation at $t \approx 2.5$\,d, characterized by a temperature break at $t_{\rm br}=2.2$\,days from a rapid $\sim t^{-1}$ temperature decline to a slower $\sim t^{-0.3}$ evolution (a), while maintaining a smooth rise in radius (b) with a corresponding impact on the bolometric luminosity (c). (d) shows the corresponding light-curve fits of SN\,2022oqm to a broken temperature and radius power-law evolution model. The break in temperature evolution at $t_{\rm br}=2.2$ naturally accounts for the rapid rise and early peak.}
\label{fig:comb_bb_lc_fits}
\end{figure*}

\subsection{Blackbody Evolution}
\label{subsec:blackbody}
We linearly interpolate the UV-optical light curves of \oqm\ to the times of UV observations and construct an SED. Using the \package{Scipy} \package{curve\_fit} package, we fit this SED to a Planck function and recover the evolution of the blackbody temperature, radius, and luminosity parameters $T_{\rm eff}$, $R_{\rm BB}$, and $L_{\rm BB}$, respectively. In order to have $\chi_{\nu}^{2}$ close to 1, we assume a $0.1$\,mag systematic error in addition to the statistical errors. This systematic error should account for both cross-calibration errors between different instruments and intrinsic deviations from a  perfect blackbody.  The fit results are shown in Fig.~\ref{fig:comb_bb_lc_fits} (a-c), and the SED fits are displayed in Fig.~\ref{fig:sed_fits}. In addition to the best-fit blackbody luminosity, we calculate a pseudobolometric luminosity: we perform a trapezoidal integration of the interpolated SED and extrapolate it to the UV and infrared (IR) using the blackbody parameters. Both estimates are consistent within the uncertainty for all times. However, as strong emission lines develop in the spectrum, {\referee the continuum contribution decreases}, and as the peak of the SED moves to the IR, the blackbody extrapolation is less reliable. This is likely more significant at $t>40$\,days when the spectrum is dominated by strong \ion{Ca}{2} lines and the directly observed luminosity accounts for only $30\%$ of the implied total luminosity. 

At the latest epoch ($t=66$\,days), we also include $JHK_s$ near-infrared (NIR) photometry in our fits. We find poor agreement between the full SED and a single blackbody. However, the $JHK_s$ bands alone are well fit with a blackbody at $\sim1650$\,K and a radius of $4.4\times10^{15}$\,cm, which is $\sim 80\%$ of the radius of freely expanding ejecta at 10,000\,\kms. We show the results of this fit in Fig.~\ref{fig:2comp_bb}. This NIR emission could be explained by the onset of dust formation within the ejecta around this time.  Alternatively, it could be a result of strong nebular lines forming in the IR. For this epoch, we extrapolate the pseudobolometric luminosities by fitting the $JHK_s$ bands, and extrapolating only to the IR. The blackbody fit parameters and pseudobolometric luminosities are given in Table \ref{tab:bb_fits}.

We find that the early-time light-curve behavior is fully explained by the blackbody evolution. During the first $2.5$\,days, the temperature cools rapidly, with a best-fitting power law of $\sim t^{-1}$. During this time the UV emission declines rapidly, with the $UVW2$ light curve falling by 1.5\,mag\,day$^{-1}$. After day 3, the temperature evolution slows down, and the $UVW2$ light-curve decline rate slows down by an order of magnitude. At early times, the photospheric radius is well described by an approximate free expansion, $R_{\rm BB} \propto vt$, with $v=20,000$\,\kms, which slows down significantly after $\sim 3.5$\,days. To check if the early light-curve behavior is fully explained by a cooling and expanding blackbody, we use an empirical light-curve model: we assume that $T_{\rm eff}$ and $R_{\rm BB}$ evolve according to

\begin{equation}
T_{\rm eff}=\begin{cases}
T_{0}(t-t_{0})^{\alpha} & (t-t_{0})<t_{\rm br}\, ,\\
T_{02}(t-t_{0})^{\alpha_{2}} & (t-t_{0})>t_{\rm br}\, ,
\end{cases}
\end{equation}

\begin{equation}
R_{\rm BB}=\begin{cases}
R_{0}(t-t_{0})^{\beta} & (t-t_{0})<t_{\rm br,2}\, ,\\
R_{02}(t-t_{0})^{\beta_{2}} & (t-t_{0})>t_{\rm br,2}\, .
\end{cases}
\end{equation}

\noindent
{\referee Here the variables are defined in a way similar to that of Eq. \ref{eq:bb_pl}}. This phenomenological model has 9 free parameters: $T_{0}$, $R_{0}$, $\alpha$, $\beta$, $\alpha_{2}$, $\beta_{2}$, $t_{\rm br}$, $t_{\rm br,2}$, and $t_{0}$, where $T_{02}$ and $R_{02}$ are calculated by demanding continuity at the power-law break. Given a set of parameters and the subsequent blackbody evolution, we generate light curves using
\begin{equation}
f_{\nu}\left(t\right)=4\pi^{2}R_{\rm BB}^{2}B_{\nu}\left(T_{\rm eff}\left(t\right)\right)\, 
\end{equation}
which we fit to the SN light curves by integrating the SED adopting each filter transmission curve. The fits are performed until $t=5$\,d, before significant features develop in the spectra. Our best-fit light-curve model is shown in Fig. \ref{fig:comb_bb_lc_fits}, and the corresponding blackbody power laws are plotted in Fig. \ref{fig:comb_bb_lc_fits}. We find that a cooling blackbody with $T_{\rm eff} =(22,000\,K)((t-t_0)/{\rm d})^{-1}$ and $R_{\rm BB} = (2.7 \times 10^{14}\,{\rm cm}) ((t-t_0)/{\rm d})^{0.9}$ can explain the full early light-curve behavior, up to $t=3$\,d. The break in the temperature evolution to $\alpha_{2} = -0.3$ naturally accounts for the first peak and the subsequent slowing in light-curve evolution in the blue bands. Table \ref{tab:pl_fits} shows the best-fit parameters and their respective uncertainties.\\


\begin{deluxetable*}{cccccccc}
\label{tab:bb_fits}

\centering

\tablecaption{Blackbody evolution of SN\,2022oqm}
\tablewidth{28pt} 

\tablehead{\colhead{JD} & \colhead{t [rest-frame days]} & \colhead{$\rm T_{eff}\ [^{\circ} \rm K]$} & \colhead{$\rm R_{\rm BB}\ [10^{14} \rm cm]$ } & \colhead{$\rm L_{\rm BB}\ [10^{42} \rm erg\ \rm s^{-1}]$} & \colhead{$\rm L_{pseudo}\ [10^{42} \rm erg\ \rm s^{-1}]$} & \colhead{$\rm L_{pseudo,extrap}\  [10^{42} \rm erg\ \rm s^{-1}]$} & \colhead{$\chi^{2}/dof$}} 

\tabletypesize{\scriptsize} 
\startdata
2459772.04 & 0.82 & $37000\pm13800$ & $1.44\pm0.41$ & $27.83\pm25.9$ & $3.02\pm3.02$ & $28.28\pm23.51$ & 10.6 \\
2459772.77 & 1.54 & $16900\pm700$ & $2.91\pm0.14$ & $4.93\pm0.37$ & $2.91\pm2.91$ & $4.88\pm0.15$ & 3.94 \\
2459773.18 & 1.94 & $12000\pm200$ & $4.29\pm0.11$ & $2.7\pm0.07$ & $2.19\pm2.19$ & $2.79\pm0.02$ & 1.71 \\
2459774.22 & 2.97 & $9400\pm200$ & $6.43\pm0.2$ & $2.33\pm0.06$ & $2.01\pm2.01$ & $2.44\pm0.01$ & 2.61 \\
2459774.74 & 3.48 & $8900\pm200$ & $7.44\pm0.27$ & $2.43\pm0.07$ & $2.13\pm2.13$ & $2.59\pm0.01$ & 3.57 \\
2459775.27 & 4.0 & $8300\pm200$ & $8.48\pm0.35$ & $2.48\pm0.08$ & $2.17\pm2.17$ & $2.66\pm0.02$ & 6.01 \\
2459776.55 & 5.28 & $7700\pm100$ & $10.28\pm0.41$ & $2.64\pm0.08$ & $2.23\pm2.23$ & $2.8\pm0.02$ & 5.8 \\
2459777.75 & 6.46 & $7500\pm100$ & $11.11\pm0.47$ & $2.74\pm0.08$ & $2.26\pm2.26$ & $2.89\pm0.02$ & 4.88 \\
2459779.17 & 7.87 & $7000\pm100$ & $12.66\pm0.53$ & $2.74\pm0.09$ & $2.15\pm2.15$ & $2.84\pm0.02$ & 5.23 \\
2459779.78 & 8.47 & $6800\pm100$ & $13.59\pm0.67$ & $2.74\pm0.11$ & $2.11\pm2.11$ & $2.84\pm0.03$ & 8.16 \\
2459783.62 & 12.27 & $6300\pm100$ & $15.95\pm0.7$ & $2.87\pm0.09$ & $2.13\pm2.13$ & $2.99\pm0.03$ & 6.11 \\
2459787.15 & 15.75 & $5800\pm100$ & $18.38\pm0.92$ & $2.68\pm0.1$ & $1.84\pm1.84$ & $2.78\pm0.03$ & 5.43 \\
2459790.0 & 18.57 & $5300\pm200$ & $20.35\pm1.51$ & $2.39\pm0.11$ & $1.43\pm1.43$ & $2.49\pm0.05$ & 8.23 \\
2459793.0 & 21.54 & $5200\pm200$ & $20.31\pm1.5$ & $2.07\pm0.09$ & $1.18\pm1.18$ & $2.13\pm0.04$ & 6.47 \\
2459796.0 & 24.51 & $5200\pm200$ & $18.09\pm1.23$ & $1.69\pm0.06$ & $0.94\pm0.94$ & $1.72\pm0.03$ & 4.04 \\
2459799.0 & 27.47 & $4900\pm100$ & $18.5\pm1.22$ & $1.4\pm0.05$ & $0.69\pm0.69$ & $1.45\pm0.03$ & 3.0 \\
2459802.0 & 30.44 & $4700\pm100$ & $18.88\pm1.25$ & $1.19\pm0.04$ & $0.54\pm0.54$ & $1.24\pm0.03$ & 2.83 \\
2459805.0 & 33.41 & $4600\pm100$ & $18.16\pm1.21$ & $1.02\pm0.04$ & $0.45\pm0.45$ & $1.05\pm0.02$ & 2.73 \\
2459807.0 & 35.38 & $4600\pm100$ & $17.01\pm0.98$ & $0.91\pm0.03$ & $0.39\pm0.39$ & $0.93\pm0.02$ & 2.06 \\
2459809.0 & 37.36 & $4600\pm100$ & $16.26\pm1.01$ & $0.83\pm0.03$ & $0.36\pm0.36$ & $0.85\pm0.02$ & 2.32 \\
2459813.0 & 41.32 & $4500\pm100$ & $15.01\pm1.07$ & $0.68\pm0.03$ & $0.28\pm0.28$ & $0.68\pm0.02$ & 2.5 \\
2459817.0 & 45.27 & $4600\pm200$ & $12.74\pm1.13$ & $0.52\pm0.03$ & $0.19\pm0.19$ & $0.53\pm0.02$ & 2.04 \\
2459820.0 & 48.24 & $4600\pm200$ & $11.26\pm1.04$ & $0.41\pm0.02$ & $0.16\pm0.16$ & $0.42\pm0.01$ & 1.99 \\
2459824.0 & 52.19 & $4600\pm200$ & $9.43\pm1.1$ & $0.29\pm0.02$ & $0.09\pm0.09$ & $0.29\pm0.01$ & 2.23 \\
2459837.5 & 65.54 & $1650\pm20$ & $44.4\pm1.3$ & $0.1\pm0.002$ & $0.09\pm0.002$ & $0.17\pm0.01$ & 0.2 \\
\hline
\enddata
\tablenotetext{a}{A 0.1 mag systematic error was assumed when performing the fits.}
\tablenotetext{b}{After $t=40$ days, we consider the blackbody fits and extrapolation to the IR and UV as unreliable, since the spectrum becomes line-dominated. We report the values here for completeness.}
\tablenotetext{c}{The last epoch is fit only to the $JHK$, but integrated using all observed bands, as discussed in the text.}
\end{deluxetable*}


\subsection{Light-Curve Evolution}
\label{subsec:evolution}
The early-time light-curve evolution of \oqm\ is characterized by a rapid decline in the UV (e.g., the $UVW2$ light curve drops by $1.5$\,mag\,day$^{-1}$), an early peak in the blue ($u$ and $g$ bands), and a rise in the red and IR bands. The UV decline slows after $t>3$\,days, as explained above.


At later times ($t>10$\,days), the light curve is well described by the radioactive decay of \Nifs\ diffusing from the inner part of the ejecta \citep{Arnett1982}. We fit the model of \citet{Inserra2013} to the bolometric light curves up to $t=40$\,days (after which we consider the bolometric luminosity unreliable), starting from the second peak in the bolometric light curve at $t \approx 12$\,days:
\begin{equation}
\label{eq:nifs_lum}
\frac{L_{\rm SN}\left(t\right)}{10^{43}\,\rm erg\,\rm s^{-1}}=e^{-\left(t/\tau_{m}\right)^{2}}\int\limits _{0}^{t/\tau_{m}}P\left(t'\right)2\left(\frac{t'}{\tau_{m}}\right)e^{\left(t'/\tau_{m}\right)^{2}}\frac{dt'}{\tau_{m}}\, ,
\end{equation}

\noindent
where $P\left(t\right)$ is the \Nifs\ decay energy and $\tau_{m}$ is the diffusion timescale parameter,

\begin{equation}
\tau_{m}=10.0\left(\frac{\kappa}{0.1\ {\rm cm^{2}\ g^{-1}}}\right)^{0.5}\left(\frac{M_{\rm ej}}{M_{\odot}}\right)^{\frac{3}{4}}\left(\frac{E_{\rm kin}}{10^{51}\,{\rm erg}}\right)^{-\frac{1}{4}}\, \rm day\, ,
\end{equation}
\noindent
{\referee where $\kappa$ is the ejecta opacity, $M_{\rm ej}$ is the ejected mass, and $E_{\rm kin}$ is the kinetic energy of the ejecta.} Here we adopt the following energy-deposition rate {\referee $Q_{\gamma}$} for $^{56}{\rm Ni}\rightarrow{}^{56}{\rm Co}\rightarrow{}^{56}{\rm Fe}$ decay \citep{Swartz1995,Junde1999} {\referee corresponding to a \Nifs\ mass $M_{\rm Ni}$}:
\begin{equation}
\frac{Q_{\gamma}\left(t\right)}{10^{43}\,{\rm erg\,s^{-1}}}=\frac{M_{\rm Ni}}{M_{\odot}}\left[1.38\,e^{\frac{-t}{111.4d}}+6.54\,e^{\frac{-t}{8.8{\rm d}}}\right]\, ,
\end{equation}
\begin{equation}
\frac{Q_{\rm pos}\left(t\right)}{10^{41}\,{\rm erg\,s^{-1}}}=\frac{M_{\rm Ni}}{M_{\odot}}4.64\left[\,e^{\frac{-t}{111.4d}}-e^{\frac{-t}{8.8{\rm d}}}\right]\, ,
\end{equation}
\begin{equation}
P=Q_{\gamma}f_{\rm dep}+Q_{\rm pos}\, ,
\end{equation}
where $f_{\rm dep}$ is the fraction of deposited energy due to $\gamma$-ray escape,
\begin{equation}
f_{\rm dep}=1-\exp\left(- t_{\gamma}^{2}/t^{2}\right)\, .
\end{equation}

Until $t=50$\,days, the bolometric light curve is well described {\referee ($\chi^2/\rm{dof} = 0.7$)} by a model with a \Nifs\ mass  of {\referee $M_{\rm Ni}=0.106\pm0.001$\,\msun}, a diffusion timescale $\tau_{m}=10\pm0.38$\,days, and a $\gamma$-ray escape time of  $t_{\gamma}=36.0\pm 0.8$\,days. We note that this fit accounts only for statistical uncertainties, and the errors on these parameters are therefore probably underestimated. After $t=50$\,days, the estimated bolometric luminosity declines sharply, but this is likely due to the underestimation of the IR flux owing to the lack of IR observations. This is illustrated during the last epoch at $t=66$\,days, where IR data have been obtained and the pseudobolometric luminosity is calculated using the $griJHK_{s}$ bands. An extrapolation based on the blackbody fit to the $JHK_s$ bands recovers 65\% of the missing luminosity compared to the Ni fit. Given the partial coverage of the SED, we consider it likely that the luminosity continues to follow the Ni model. 

\begin{figure}[t]
\centering
\includegraphics[width=\columnwidth]{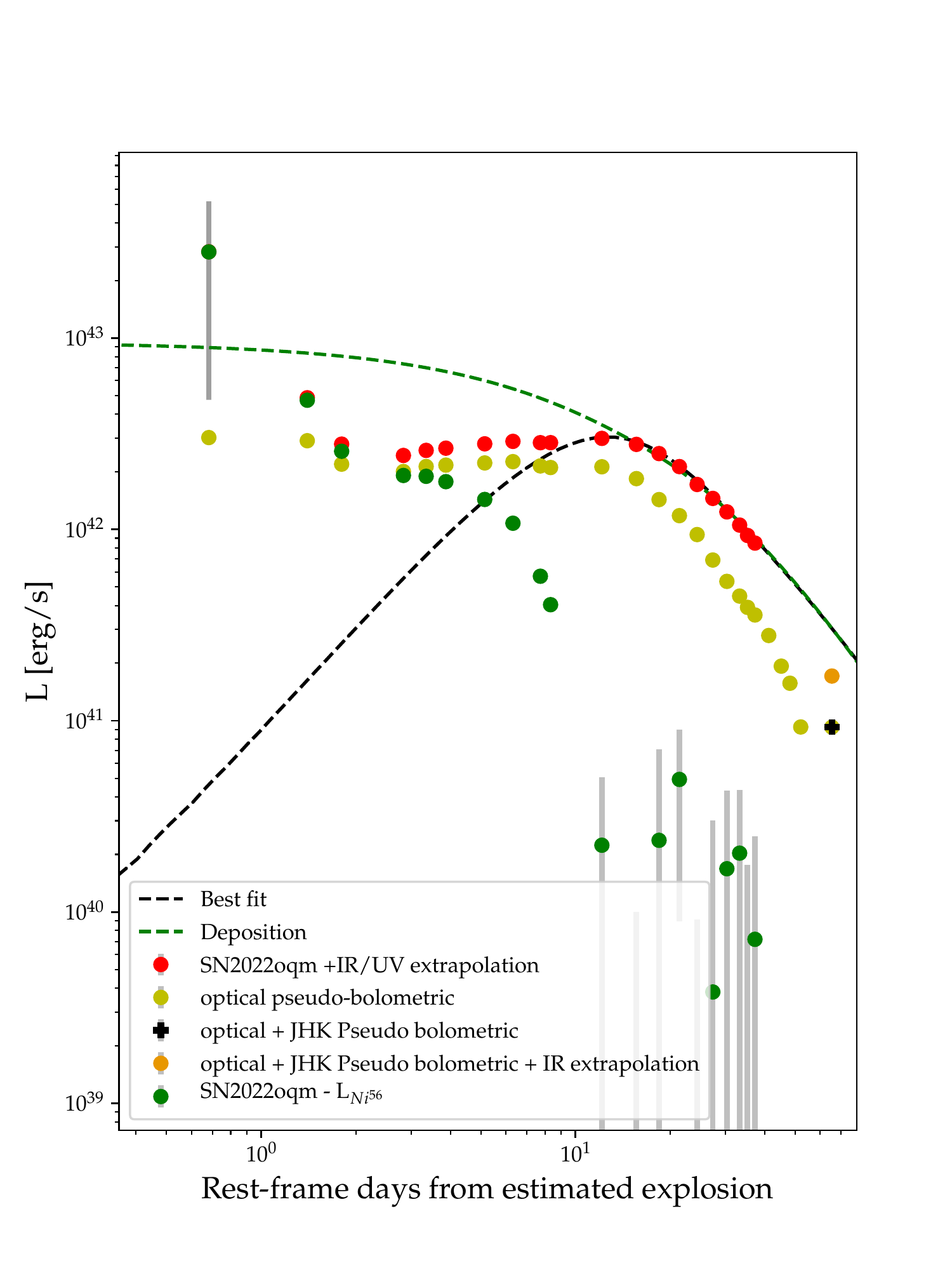} \\
\caption{A bolometric light-curve fit to an Arnett model. The best-fit model is the black dashed curve. The red data points are the integrated observed luminosity with blackbody extrapolation corrections for the UV and IR, and the green points are the difference between these two. The yellow points are the integrated luminosities with no UV/IR corrections (pseudobolometric). Note that these diverge from the bolometric data as the IR corrections become more important at late times. The black ``plus sign'' is the integrated luminosity including the late-time $griJHK_{s}$ bands, and the orange point is this luminosity including an IR extrapolation correction based on the $JHK_s$ blackbody fits, as described in the text. We do not fit an optical blackbody to this epoch for a UV component, since the optical SED is dominated by emission lines}
\label{fig:Ni}
\end{figure}

Assuming $v_{\rm ej}=10,000\,\rm km\,\rm s^{-1}$ (appropriate for the bulk of the mass) and $\kappa = 0.07\,\rm cm^{2}\,\rm g^{-1}$ as used by \cite{Barbarino2021},  we acquire from $\tau_{\rm m}$ an estimate of $M_{\rm ej}=1.1\pm0.04\,M_{\odot}$, and a kinetic energy of $E_{\rm kin}=6.6\times 10^{50}\,\rm erg$.\footnote{We note that assuming a higher typical ejecta velocity, as suggested by the early radius evolution and the nebular [\ion{Ca}{2}] FWHM, could result in a higher estimate for $M_{\rm ej}$. For example, if 15,000\,\kms\ is assumed, $M_{\rm ej}$ would be $1.7$\,\msun.} Fitting of the $t<50$\,days light curve using the methods of \citet{Sharon2020} yields $M_{\rm Ni}=0.113^{+0.002}_{-0.001}$\,\msun\ and $t_\gamma=36\pm2$\,days, in good agreement with the parameters derived using Eq.~\eqref{eq:nifs_lum}. Figure \ref{fig:Ni} shows the best-fit model to the late-time bolometric light curve. While $^{56}\rm Ni$ decay can account for the late-time behavior of the light curve, an additional powering mechanism is required to explain the early-time luminosity. The inferred \Nifs\ mass, kinetic energy, and ejected mass are within the distribution of values found for SNe~Ic by \cite{Barbarino2021} {\referee \cite{rodirigez2022}}, with the ejecta mass lying toward the low end of the mass distribution.  A rough order-of-magnitude estimate for the $\gamma$-ray escape time $t_{\gamma}$ and the diffusion time $\tau_{m}$ comes from demanding an optical depth of unity for $\gamma$-ray escape, and a dynamical timescale for the diffusion of order $c/v$. This implies a ratio of $\frac{t_{\gamma}}{t_{\rm diff}} \approx \sqrt{(\frac{c}{v})\frac{\kappa_{\gamma}}{\kappa_{\rm opt}}} \approx 3.3$, in good agreement with our findings. An order-of-magnitude estimate for the values of these timescales $t_\gamma \approx \sqrt{\frac{3\kappa_{\gamma}M}{4\pi v^{2}}} \approx 50$\,days and $t_{\rm diff}\approx \sqrt{\frac{3\kappa_{\rm opt}M}{4\pi vc}}\approx 14$\,days is also consistent with our fit results. {\referee The $\gamma$-ray escape time of \oqm\ is {\secondref short} for a typical SN~Ic, compared to the typical $t_{\gamma}\approx 100$\,days {\secondref found} by \citet{Sharon2020}. In their recent work, \cite{Sharon2023} measure the $\gamma$-ray deposition history for five Ca-rich SNe~Ib, and find that they have both low $^{56}\rm Ni$ masses (0.01--0.05\,$M_{\odot}$), and $t_{\gamma}$ in the 30--70\,day range. Compared with the Ca-rich SN~Ib population, \oqm\ has a higher $^{56}\rm Ni$ mass, but a similar $t_{\gamma}$, placing  it closer to the SN~Ia population in this parameter space.}


\begin{deluxetable*}{ccccccccc}
\label{tab:pl_fits}

\centering

\tablecaption{Power law fits for the early blackbody evolution SN\,2022oqm}
\tablewidth{24pt}

\tablehead{\colhead{$T_{0}\ [^{\circ} \rm K]$}&\colhead{$R_{0}\ [10^{14} \rm cm]$}&\colhead{$\alpha$}&\colhead{$\beta$}&\colhead{$t_{exp}\ [JD]$}&\colhead{$t_{\rm br}$ [rest days]} & \colhead{$\alpha_{2}$}} 
\tabletypesize{\scriptsize} 
\startdata
$22000\pm4000$ & ${ 2.7}^{+ 0.5}_{- 1.4}$  & ${-1.0}^{+ 0.1}_{- 0.2}$  & ${ 0.9}^{+ 0.3}_{-0.1}$ & ${ 2459771.2}^{+ 0.2}_{- 0.2} $ & ${ 2.2}^{+ 0.2}_{- 0.3} $  & $ {-0.3}^{+0.1}_{- 0.1} $\\
\hline
\enddata
\tablenotetext{a}{A 0.1 mag systematic error was assumed when performing the fits.}
\tablenotetext{b}{In our best fit, $\beta_{2}=\beta$. Thus, we do not report $\beta_{2}$ or $t_{\rm br,2}$}

\end{deluxetable*}

\subsection{Host-Galaxy Properties}
\label{subsec:host}

\oqm\ exploded at a distance of 16.6\,kpc (59\farcs3) from the center of the spiral galaxy NGC~5875 (Fig. \ref{fig:discovery}). To measure the galaxy properties, we retrieved science-ready stacked images from the \textit{Galaxy Evolution Explorer} (\galex) general release 6/7 \citep{Martin2005a}, the Sloan Digital Sky Survey data release 9 (SDSS DR9; \citealt{Ahn2012a}), the Panoramic Survey Telescope and Rapid Response System (Pan-STARRS, PS1) DR1 \citep{Chambers2016}, and \wise\ \citep{Wright2010a} images from the unWISE archive \citep{Lang2014a}.\footnote{\href{http://unwise.me}{http://unwise.me}} We measured the brightness of the host using LAMBDAR\footnote{\href{https://github.com/AngusWright/LAMBDAR}{https://github.com/AngusWright/LAMBDAR}} \citep[Lambda Adaptive Multi-Band Deblending Algorithm in R;][]{Wright2016a} and the methods described by \citet{Schulze2020}. {\referee In short, these involve the removal of contaminating foreground sources, identifying an appropriate aperture, and using it to extract photometry simultaneously from all available bands}. The photometry is summarized in Table \ref{tab:host_phot}. {\referee We find a half-light radius of $r_{50} = $21\farcs4 (6\,kpc) in the SDSS $r$ band, which places \oqm\ at an offset of $2.8\,r_{50}$ from the center of its host galaxy.}

\begin{table}
\caption{Photometry of the host galaxy of SN\,2022oqm}\label{tab:host_phot}
\centering
\begin{tabular}{llc}
\toprule
Survey & Filter    & Brightness [AB mag] \\
\midrule
\galex	    &$	FUV	$&$	15.79	\pm	0.03	$\\
\galex	    &$	NUV	$&$	15.25	\pm	0.02	$\\
SDSS	    &$	u	$&$	14.09	\pm	0.04	$\\
SDSS	    &$	g	$&$	12.90	\pm	0.04	$\\
SDSS	    &$	r	$&$	12.29	\pm	0.03	$\\
SDSS	    &$	i	$&$	11.96	\pm	0.04	$\\
SDSS	    &$	z	$&$	11.73	\pm	0.03	$\\
Pan-STARRS	&$	g	$&$	12.90	\pm	0.03	$\\
Pan-STARRS	&$	r	$&$	12.29	\pm	0.02	$\\
Pan-STARRS	&$	i	$&$	12.05	\pm	0.01	$\\
Pan-STARRS	&$	z	$&$	11.90	\pm	0.03	$\\
Pan-STARRS	&$	y	$&$	11.69	\pm	0.08	$\\
\wise	    &$	W1	$&$	11.99	\pm	0.01	$\\
\wise	    &$	W2	$&$	12.52	\pm	0.02	$\\
\bottomrule
\end{tabular}
\tablecomments{All measurements are reported in the AB system and not corrected for reddening. 
}
\end{table}
The SED was modelled with the software package \package{prospector} \citep{Johnson2021a}, as described in detail by \citet{Schulze2020}. We assumed a Chabrier initial mass function \citep[IMF;][]{Chabrier2003a} and approximated the star-formation history (SFH) by a linearly increasing SFH at early times followed by an exponential decline at late times (functional form $t \times \exp\left(-t/\tau\right)$). The model includes an extinction correction using the \citet{Calzetti2000a} model. We use the dynamic sampling package \package{dynesty} \citep{Speagle2020a} to sample the posterior probability distribution and extract the median host-galaxy properties. 

The host is a fairly massive ($\log_{10}(M/M_\odot) \approx 10.66^{+0.10}_{-0.31}$) star-forming galaxy (star-formation rate SFR $=3.52^{+1.17}_{-0.72}\,M_\odot\,{\rm yr}^{-1}$) with moderate extinction ($E(B-V)_{\rm star}=0.24^{+0.04}_{-0.03}$\,mag). The mass and the SFR are within the distributions measured for host galaxies of SNe~Ic from the PTF survey \citep{Schulze2020}. Although the SN is located in the outskirts of its host (Fig.~\ref{fig:discovery}), the location is not unusual for SNe~Ic exploding in galaxies of similar mass \citep{Schulze2020}. Our spectroscopic observations sampled different regions of the host galaxy. However, none of the slit alignments of our GMOS or NOT spectra showed any prominent \ion{H}{2} region emission at a distance $\lesssim 3$\,kpc along the slit, so we cannot constrain the metallicity or the SFR in the direct vicinity of SN\,2022oqm. In Fig. \ref{fig:PA_fig} we show the NOT and GMOS slit orientations, as well as the surroundings of the explosion site. {\referee Since our spectroscopic observations did not cover all nearby regions, we cannot rule out the presence of a nearby star-forming region. The most nearby well-defined star-forming region is 3.8\,kpc {\secondref southeast} of the SN explosion site. We measure the line fluxes of prominent emission lines ([\ion{O}{3}]\,$\lambda\lambda$4959, 5007, H$\alpha$, H$\beta$, and [\ion{N}{2}]\,$\lambda$6584), finding values of $1.4\pm0.2$, $1.6\pm0.2$, $5.6\pm0.6$, and $1.9\pm0.3$ in units of $10^{-15}\,\rm{erg\,s^{-1}\,cm^{-2}}$, respectively (calibrated to PS1 $r$-band photometry). Using the O3N2 and R3 strong-line metallicity indicators and the calibrations of \citet{Curti2017a}, we infer a metallicity of $0.88^{+0.10}_{-0.11}$ solar for this region. This value is near the average for the explosion-site metallicity in the sample of \cite{Galbany2018a}.}

A visual inspection of deep stacks from the Beijing-Arizona Sky Survey (BASS; \citealt{Dey2019}) and {\it GALEX} show no point source or elevated extended emission in the vicinity of the SN. BASS has a median point-source limit of $r=23.6$\,mag, implying $M_{r}\gtrsim -10$\,mag or an optical surface brightness limit of $\sim 23.5$\,mag\,arcsec$^{-2}$.  The {\it GALEX} all-sky survey has a limit of NUV $=20.5$\,mag, implying a region with $M_{\rm NUV} \gtrsim -13$\,mag or with a UV surface brightness of $\sim 24$\,mag\,arcsec$^{-2}$ can still exist in the vicinity of the SN, {\referee corresponding to a point-source SFR limit of $\Sigma_{\rm SFR} = 0.01\,M_{\odot}$\,yr$^{-1}$ \citep{Salim2007}.} {\secondref Since many \ion{H}{2} regions have a lower average SFR \citep{relano2009}, this does not rule out an \ion{H}{2} region below the {\it GALEX} detection limit.}

\section{Discussion}
\label{sec:discussion}
We presented extensive UV-optical observations of \oqm\ in $\S$~\ref{sec:observations}, as well as our X-ray limits and observations of the SN host galaxy. In $\S$~\ref{sec:analysis}, we analyzed these observations. We showed that the early-time spectra of \oqm\ are well explained by an expanding C/O shell moving at 4000\,\kms, with line velocities increasing to typical SN ejecta velocities by day 3. At the same time, the blackbody evolution transitions from a rapid cooling and a decline in the bolometric luminosity, to a slower evolution in both parameters. This transition is reflected by a double peak in the optical light curve, and a shift from a fast to slow decline in the UV bands. Following this transition, the spectrum evolves like those of spectroscopically normal {\referee (but relatively fast-rising)} SNe~Ic, until it becomes nebular at $t \approx 60$\,days. 

The nebular spectrum has strong [\ion{Ca}{2}] and \ion{Ca}{2} emission, with no detectable [\ion{O}{1}], indicating that the object is Ca-rich.  We fit the late-time post-peak ($t > 12$\,days) light curve to a \Nifs-decay model and find typical SN~Ic values of $M_{\rm ej}=1.1\,M_{\odot}$, $M_{\rm Ni}=0.12\,M_{\odot}$, and  $E_{\rm kin}=6.6\times 10^{50}\,\rm erg$ \citep[e.g.,][]{Barbarino2021}. We analyze the host-galaxy observations and find that it is a typical star-forming and massive spiral galaxy. However, the explosion site is more than 3\,kpc away from the nearest obvious star-forming region, and offset by 16\,kpc from the center of light of its host. In the following, we discuss the implications of our observations on the powering mechanism of the early-time light curve and on the progenitor star of \oqm. 

\subsection{The Early-Time Features}
At early times, the  spectra of SN\,2022oqm show high-ionization \ion{C}{0} and \ion{O}{0} features with absorption minima at velocities of $\sim 4000$--5500\,km\,s$^{-1}$, and a blue edge of 12,000--15,000\,\kms, which (as discussed in $\S$~\ref{subsec:spec_analysis}) could indicate a maximal expansion velocity of 18,000--22,000\,\kms\ as would have been measured in the UV. At the same time, we observe that the photospheric radius is expanding at $>20,000$\,km\,s$^{-1}$. Later in the evolution, the absorption minima and blue edge accelerate significantly, to absorption minima of 10,000\,\kms\ at $t=3.2$\,day. It is difficult to fully explain this evolution as being due to the ejecta alone, as it would require nonhomologous expansion (slow above fast), or with CSM alone, as the blue edge has ejecta-like high velocities. 

The absorption minimum at $\sim 4000$--5500\,km\,s$^{-1}$ implies that the photosphere is expanding behind an optically thin line-forming region, itself expanding at a lower velocity. The simplest interpretation is that the lines originate from an expanding shell of CSM surrounding the progenitor star, in addition to a weaker absorption component by the ejecta extending to the photospheric velocity.

An expansion velocity of 4000\,km\,s$^{-1}$ is consistent with a continuous wind around a W-R progenitor star \citep{Nugis2000}, with the escape velocity of a white dwarf (WD), or with a late-stage eruption resulting from a deposition of energy deep under the stellar surface \citep{Matsumoto2022}. An eruptive mass-loss episode occurring days to weeks before the explosion is often seen in other types of SNe. {\referee Such eruptions typically lack spectroscopic observations to constrain the ejected CSM velocity} \citep{Ofek2013,Ofek2014a,Strotjohann2015,strotjohann2021,JacobsonGalan2022b}. 

A distribution of expansion velocities in the CSM could explain the apparent line acceleration between the first ($t=0.7$\,days) and second ($t=1.2$\,days) spectra. First, the ejecta sweep up the slower CSM (at 4000\,\kms), and later they reach the faster material at 5500\,\kms, which accounts for the observed shift of the absorption minimum to higher velocities. As more and more material is accelerated to ejecta velocities, the blue edge of the absorption features becomes more pronounced and extends to higher velocities. 

{\referee The high velocity could also be explained by radiative acceleration of the optically thin material above the photosphere, by the free-streaming photons from the luminous underlying ejecta. The velocity gain of an optically thin shell of material above a source with integrated luminosity   $E_{\rm rad}(t)$ at radius $r_{\rm CSM}$ is given by
\begin{align}
        \frac{v_{\rm rad}\left(t\right)}{{\rm km\,s^{-1}}}=1000\left(\frac{E_{\rm rad}\left(t\right)}{10^{48}\,\text{{\rm erg}}}\right)\left(\frac{\kappa_{\rm fw}}{10\,\text{cm}^{2}\,\text{g}^{-1}}\right)\left(\frac{r_{\rm CSM}}{5\times10^{14}\,\text{{\rm cm}}}\right)^{-2}\, ,
\end{align}
where $\kappa_{\rm fw}\left(t\right)=\int\kappa_{\nu}f_{\nu}\left(t\right)d\nu / \int f_{\nu}\left(t\right)d\nu$ is the flux-weighted opacity (applicable at $\tau_{\rm diff}<1$), and $f_{\nu}$ is the spectral flux density. Scattering opacity alone ($\sim 0.2\,\rm{cm^{2}\,g^{-1}}$) is not sufficient to accelerate material to the observed velocities, or to explain the acceleration observed in the first few spectra. However, a high effective cross-section due to bound-free and bound-bound processes on the order of $\gtrsim10\,\rm{cm^{2}\,g^{-1}}$ is achievable with an illuminating blackbody spectrum at $T\gtrsim10,000$\,K, producing a large fraction of photons with energies $\gtrsim10$\,eV. It can also be achieved with a mild X-ray flux of $\sim 1$\% of the UV-optical luminosity, absorbed through photoionization in the CSM, and consistent with the highly ionized species observed during the first three days. We thus consider radiative acceleration as a plausible mechanism for explaining the initially high observed velocities and the acceleration between epochs. As we do not know the exact conditions in the CSM, we refrain from making an explicit calculation, leaving this for future work. }

Another explanation for the early emission is from an optically thick shell surrounding the ejecta. \cite{Soumagnac2019} show that breakout from an aspherical shell of CSM could form an increasing photospheric radius, with no actual expansion taking place. In this type of scenario, the expansion is unrelated to the ejecta velocity, but a result of breakout from an increasingly large region.  In $\S$~\ref{sec:csm_power} we show that the amount of mass required to make this material optically thick is inconsistent with the integrated luminosity, disfavoring this interpretation. {\referee In the absence of optically thick material which can facilitate a radiation-mediated shock, temporally-resolved acceleration of the CSM by the shock over a timescale of a few days can be ruled out. The shocks in such systems should be collisionless and would accelerate the material on very short ($\sim 1$\,m) length scales \citep{katz2012}, directly to the ejecta velocity seen at $t>3$\,days.}

Here we do not discuss other, more complicated asymmetric configurations. However, such a scenario would have to produce significant absorption at $\sim 4000$\,\kms. This is a challenge to line-of-sight-based interpretations, such as bipolar outflows, that can explain the slower components with material moving nearly perpendicular to our line of sight. Such models will have a hard time creating significant absorption at low velocities; the obscuring material needs to be placed in front of most of the emitting material. From this point, we assume that a spherical, slowly expanding CSM is the source of the 4000--5000\,\kms\ features.

\begin{figure}[t]
\centering
\includegraphics[width=\columnwidth]{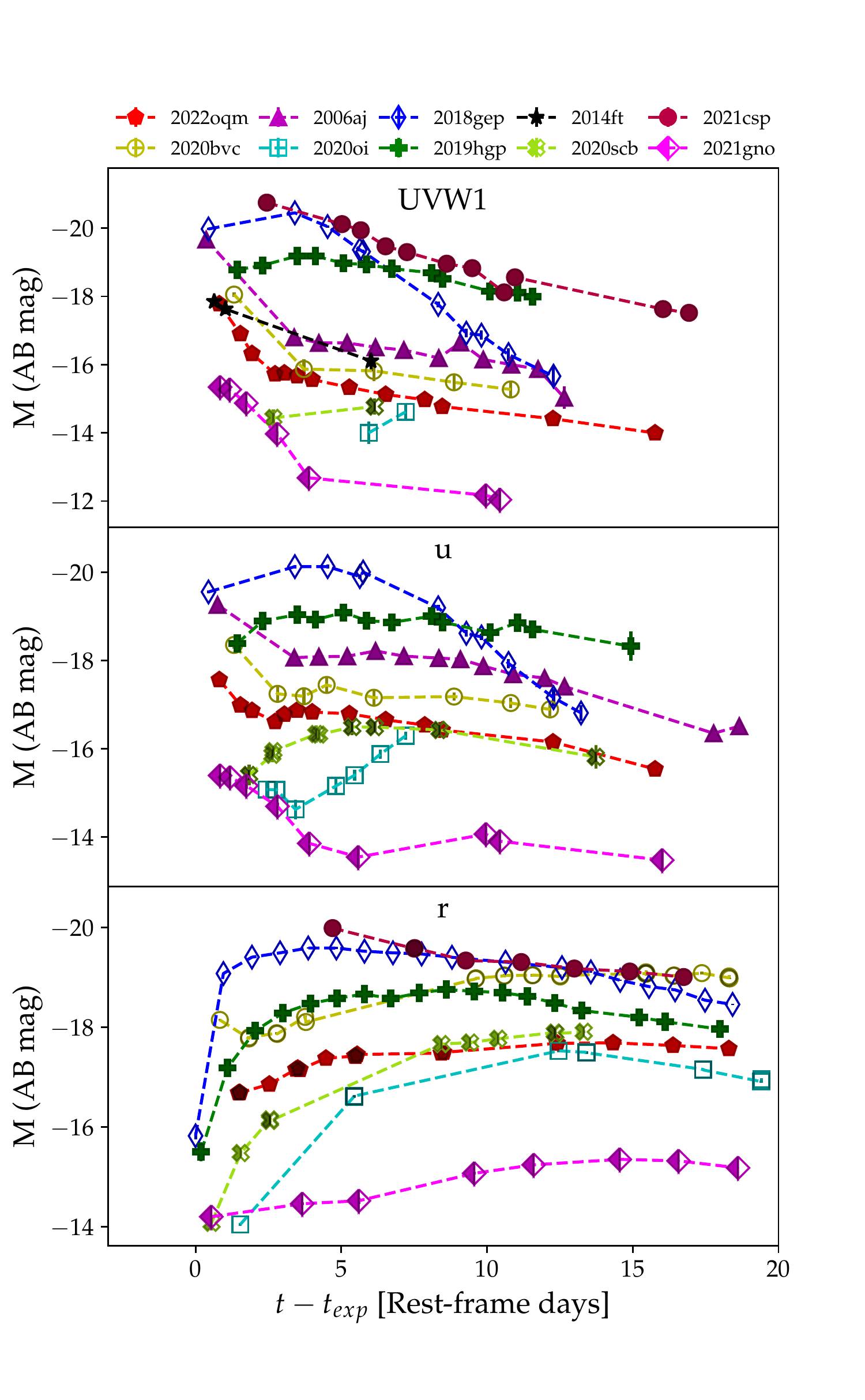} \\
\caption{The {\referee early} light-curve evolution of \oqm\ (filled pentagons) compared to that of several SNe~Ic, Ic-BL, and Icn in the $r$, $u$, and $UVW1$ bands. Several other SNe show a break in their UV light curves, combined with a fast rise in their optical light curve.}
\label{fig:lc_comp}
\end{figure}

\begin{figure}[t]
\centering
\includegraphics[width=\columnwidth]{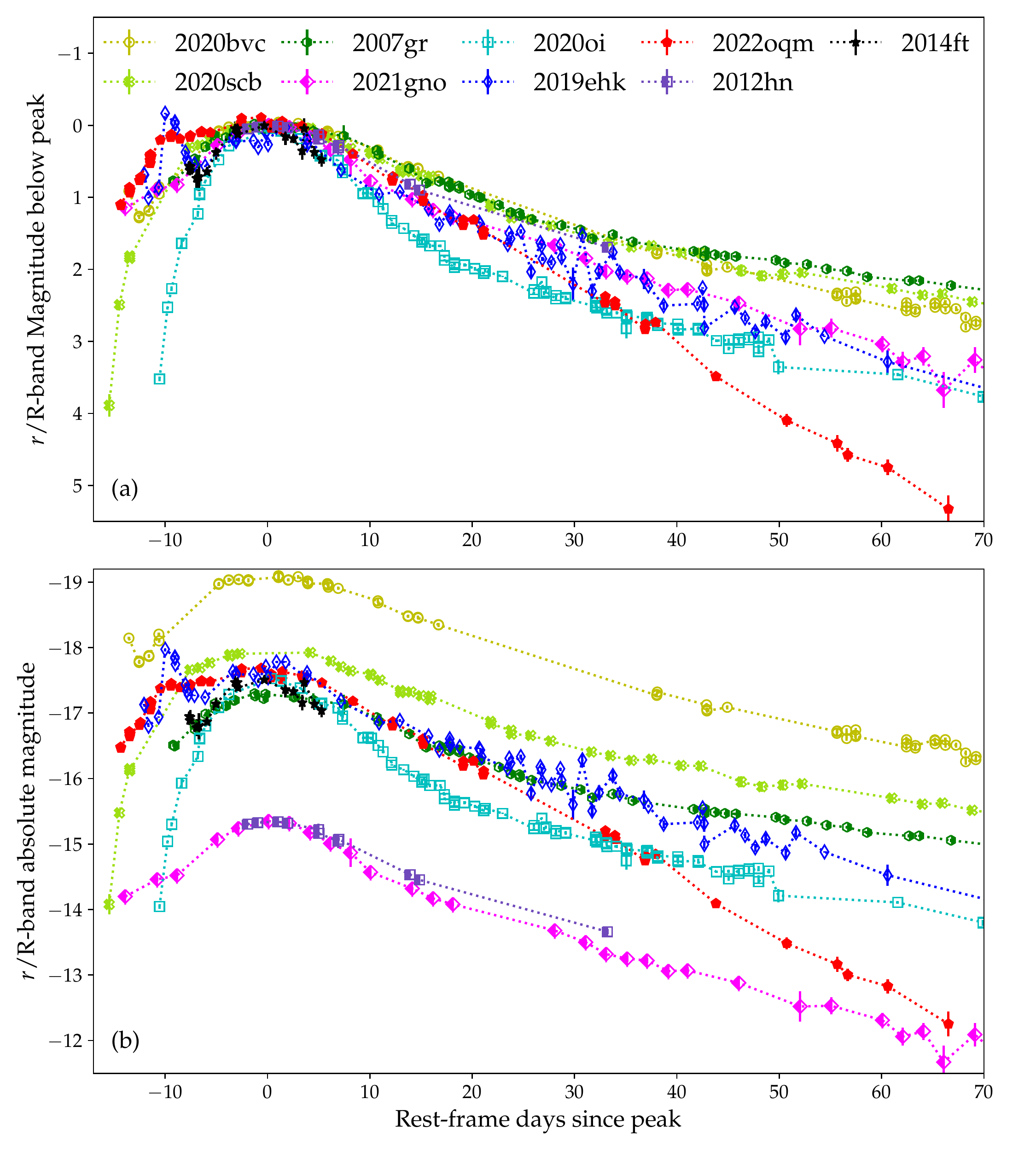} \\
\caption{{\referee A comparison of the $r$/$R$-band light curves of SN\,2022oqm to SN\,2007gr (Ic), SN\,2020oi (Ic), SN\,2020bvc (Ic-BL), SN\,2014ft (US-Ic), SN\,2012hn (Ca-Ic), SN\,2021gno (Ca-Ib), SN\,2019ehk (Ca-IIb), and SN\,2020scb (Ic). We show the light curve (a) relative to the $r$/$R$ peak and (b) in absolute magnitude. SN\,2022oqm shows an emission excess before peak, and then declines faster than other SNe~Ic.}}
\label{fig:full_lc_comp}
\end{figure}

\subsection{Comparison with Other Supernovae}
\label{sec:comp}
We compare the spectra, light curves, and blackbody evolution of \oqm\ with those of other SNe~Ic, Icn, and Ic-BL having extensive UV and optical observations at early times, and that either were suggested to have some amount of CSM around their progenitor star, or exhibit an early UV-optical peak. In {\referee order to contrast SN\,2022oqm with typical SNe Ic, which usually lack early UV observations}, we show a comparison with SN\,2020scb, a normal SN~Ic detected by ZTF with good constraints on its explosion time and early UV observations \citep{Dahiwale2020b}. For the sake of uniformity, UVOT and ZTF (if used) light curves were re-reduced using the methods described in $\S$~\ref{sec:observations}, and the blackbody fits are performed with the methods described in $\S$~\ref{sec:analysis}. 

In Fig.~\ref{fig:lc_comp}, we show a comparison of the {\referee early-time} $r$, $u/U$, and $UVW1$ light curves of these SNe with \oqm. While the diversity in absolute magnitude is large, \oqm\ is similar to SN\,2020bvc  \citep{Izzo2020,ho2020b}, SN\,2006aj \citep{Campana2006}, SN\,2014ft  \citep{De2018}, and SN\,2020oi \citep{Horesh2020,Rho_2021} in showing an early peak in the UV light curves and later rising to a second peak. In contrast to these, SN\,2020scb (this work), SN\,2018gep \citep{Ho2019}, SN\,2019hgp \citep{galyam2022}, and SN\,2021csp \citep{Perley2022} display a different behavior consistent with a single-peaked light curve. {\referee In Fig. \ref{fig:full_lc_comp}, we compare the $r$/$R$-band light curves of SN\,2022oqm to SN\,2007gr, SN\,2020oi, SN\,2020bvc, SN\,2014ft, SN\,2012hn, SN\,2019ehk, SN\,2021gno \citep{JacobsonGalan2022} and SN\,2020scb (a) relative to peak  and (b) in absolute magnitude. SN\,2022oqm shows a fast rise to peak, to an elevated emission unseen in other comparison objects. Initially it declines at a comparable rate to SN\,2021gno, but this changes at $t\approx35$\,days ($\sim20$\,days after peak), as the ejecta become transparent to gamma rays. }

Figure ~\ref{fig:bb_comp} illustrates a comparison of the blackbody evolution of these SNe with that of \oqm. 
In Fig.~\ref{fig:temp_pl}, we normalize the temperature evolution to an arbitrary time and temperature, selected to emphasise a transition in the temperature power-law slope (if such a transition exists). Similarly to \oqm, other SNe with an early peak in their light curve show a transition from a steep to a shallow power-law evolution. A steep temperature power law also provides a reasonable explanation for the fast rise, as the peak of the SED will move into the UV and optical bands faster than for a typical SN. 

Figure ~\ref{fig:spec_compare} shows a spectral comparison of selected objects with \oqm. In the upper panel, two spectra of SN\,2022oqm at +0.7\,day and +1.2\,day after the explosion are compared to the ultrastripped Type Ic SN\,2014ft \citep{De2018}, the broad-line Type Ic SN\,2018gep \citep{Ho2019}, and SN\,2020bvc \citep{Izzo2020,ho2020b}. Although the spectra of SN\,2014ft have lower signal-to-noise ratios, they closely resemble those of SN\,2022oqm; the prominent features match well with the C/O-dominated line profiles in SN\,2022oqm, suggesting a similar origin for the early-time spectroscopic features of SN\,2014ft. 

In the photospheric phase, the spectra of SN\,2022oqm look quite similar to typical SNe~Ic such as SN\,2007gr \citep{Valenti2008, Hunter2009}, as shown in the middle panel of Fig.~\ref{fig:spec_compare}. SN\,2007gr was a carbon-rich SN~Ic with \ion{C}{2} $\lambda\lambda$6580,7234\ clearly detected in the premaximum spectra \citep{Valenti2008}. The \ion{C}{2} $\lambda\lambda$6580,7234\ lines are likewise detected in SN\,2022oqm. 
As shown in the bottom panel, the (early) nebular-phase spectra of SN\,2022oqm exhibit both a strong Ca~II NIR triplet and the forbidden [\ion{Ca}{2}] $\lambda\lambda$7291, 7324, but no clear detection of [\ion{O}{1}] $\lambda\lambda$6300, 6363, similar to SN\,2014ft. 
The strong emission of [\ion{Ca}{2}] compared to [\ion{O}{1}] means that SN\,2022oqm belongs to the population of ``Ca-rich'' SNe such as SN\,2019ehk \citep{Jacobson-Galan2020,De2021} and SN\,2012hn \citep{Valenti2014}. We note that the [\ion{Ca}{2}] $\lambda\lambda7291,7234$ lines in the +60.1\,d spectrum of SN\,2022oqm are blueshifted {\referee by $\sim1700$\,\kms}, which was also shown for SN\,2014ft (+36.5\,d; {\referee $\sim 2500$\,\kms}) and SN\,2012hn (+31.0\,d; {\referee $\sim 1300$\,\kms}), indicating that those spectra may not be fully nebular in the red side of the spectrum. {\referee At this phase, the blue side of the spectrum shows an elevated continuum and a P~Cygni profile at $\sim5900$\,\AA. This feature can either be associated with the Na~I $\lambda\lambda$5890, 5896 doublet, or with He~I $\lambda5876$. The latter is disfavored owing to the lack of stronger features at 6678 and 7065\,\AA\ \citep{galyam2017}}.



As mentioned in $\S$~\ref{subsec:blackbody}, at $t=66$\,days, $\sim 75\%$ of the bolometric luminosity is observed in the NIR. This could be explained either by strong emission lines, or a blackbody component with 1650\,K and a radius of $4.4\times10^{15}$\,cm, in reasonable agreement with free expansion at 10,000\,\kms\ for the duration of the SN. The NIR $V-H$ color at this time ($V-H \approx 3$\,mag in the Vega system) is quite high compared to most of the 64 SESNe observed by \citet[][compare to their Fig. 13]{Bianco2014} during their entire evolution, and consistent with those of SN\,2006jc \citep{Pastorello2007,Foley2007} at a similar phase. One possible explanation would be dust formation, observed in some SNe~Ic as early as day 60 \citep{Rho_2021}. This would be consistent with the observed nebular \ion{Ca}{2} and [\ion{Ca}{2}] asymmetry toward the blue side, possibly due to the obscuration of the most redshifted parts of the ejecta. 



\subsection{Early-Time Powering Mechanism}
\label{sec:power}

While \Nifs\ provides a good mechanism for powering the second peak, it cannot explain the early-time contribution to the light curve. We integrate the difference between the bolometric light curve and the best-fit \Nifs\ model for all observed times and find $E=2.1\times 10^{48}\ \rm erg$ radiated by an early-time additional component. There are several possible origins for this component:\\
\begin{enumerate}
    \item shock cooling of a low-mass envelope; 
    \item CSM interaction; and
    \item shock breakout in {\referee extended} CSM.
    \end{enumerate} 
We examine each of these possibilities in light of the observed properties of \oqm.

\subsubsection{{\referee Shock Cooling at Early and Intermediate Times}}
\label{sec:SC}

The good agreement of the spectral energy distribution with a blackbody spectrum (Fig.~\ref{fig:sed_fits}) motivates the possibility of shock-cooling powering some or all of the early light curve, prior to \Nifs\ decay. During the first 3 days, as we show in $\S$~\ref{subsec:blackbody}, the temperature declines with a power law slope of $T\approx t^{-1}$, significantly more steeply than the expected $T\approx t^{-0.5}$ for shock-cooling \citep{Rabinak2011,Nakar2010,Piro2015}. It is possible to achieve a sharp temperature decline with existing models, assuming a low-mass envelope \citep{Piro2021}, when the luminosity is suppressed due to penetration of the diffusion depth deep into the envelope. While we acquire a good fit to the early-time light curve for a low-mass envelope of $M_{e}=0.05$\,\msun, $R_e=110\,R_{\odot}$, and $E=1.5\times 10^{50}$\,erg (see Fig. \ref{fig:piro_SC_fits}), we consider this fit to be unphysical. The fit implies the envelope is fully transparent by day 2.5, ($t_{\rm ph} = (0.08 \kappa M_{e}^{2}/E_{e})^{1/2}$\,s $=2.4$\,day). This implies a break in the photospheric radius to a receding $R_{\rm BB}$ should occur at roughly the same time, due to the same change creating the luminosity decline \citep[][ their Figs. 1 and 3]{Piro2021}.  However, this does not happen until much later in the evolution, around day 10. Furthermore, a power law of $R_{\rm BB}\approx t^{0.8}$ does not fit our data well even during the validity of the model (see Fig.~\ref{fig:piro_SC_fits_bb}), and so we disfavor this interpretation.\\

Following $t=2.2$ days, the temperature evolves with a power law slope of $\alpha_{2} = -0.3\pm{0.1}$, consistent with the predicted power-law slope for C/O or He/C/O composition \citep{Rabinak2011}. We fit a combined shock-cooling and \Nifs\ decay model (with the parameters found in $\S$~\ref{subsec:evolution}) to the light curve at $2<t<5$~days when \Nifs\ accounts for less than 50\% of the observed luminosity. We use the shock-cooling models of \cite{Morag2022}, calibrated to numerical grey simulations, and based on realistic opacities for a H-dominated composition. In the case of \oqm, a composition of C/O or He/C/O is appropriate, as some amount of He might be present even in the absence of He lines in the photospheric spectrum \citep{Hachinger2012, Teffs2020}  In order to account for a C/O or He/C/O composition of the ejecta, we chose a constant opacity of $\kappa=0.2 \ \rm cm^2$ g$^{-1}$, which we calculate to be appropriate for fully ionized He/C/O mixture (applicable to the early-time CSM) and for a wide He fraction range.\footnote{The choice of a constant opacity is in lieu of the approximate temperature-dependent opacity employed in the He/C/O model extensions in \citet{Rabinak2011}. The shock cooling luminosity is determined deep in the ejecta where the local temperature is higher than both the photosphere temperature and the observed emission temperature, and as a result the opacity in this regime is approximately constant, and higher than the opacity of $\kappa=0.07$\,cm$^2$\,g$^{-1}$ typically assumed for SNe~Ic. We defer a more detailed study of the effect of He/C/O composition on shock-cooling emission to later work.} The model is described in detail in $\S$~\ref{ap:SC_mod}. 

We use the nested-sampling \citep{Skilling2006} package \package{dynesty} \citep{Higson2019,Speagle2020a} to fit our likelihood function to the observed photometry. While we consider wide priors on all parameters, we limit ourselves to $M_{env}<1$ \msun, in order to remain consistent with our estimate for the ejected mass from $\S$~\ref{subsec:evolution}. The light-curve and blackbody evolution are well described by a model with $R=310_{-110}^{+30}\,R_{\odot}$, $M_{env}=0.23_{-0.07}^{+0.44}\,M_{\odot}$, and with a shock velocity parameter (related to the bulk velocity by $v_{\rm ej} \approx 5\times v_{s,*}$ \citealt{Morag2022}) of $v_{s,*}=1900_{-190}^{+850}\,{\rm km\,s^{-1}}$  which we show in Fig.~\ref{fig:SC_intermediate}, as well as the corresponding blackbody fits in Fig.~\ref{fig:SC_intermediate_bb}. In the \cite{sapir2016,Morag2022} framework, the fit is terminated at $t_{\rm tr}/2$, where $t_{\rm tr}=9.2$~days is the envelope transparency time for our best fit model, equivalent to $t_{\rm ph}$ in \cite{Piro2021}, and very close to $\tau_{m}$ in definition.\footnote{In the framework of \cite{sapir2016,Morag2022}, $t_{\rm tr} =  \sqrt{\frac{\kappa M_{\rm env}} {8 \pi c v_{\rm s \ast} }}\approx 0.9\tau_{m}$}, in excellent agreement with the diffusion time we get from the fit to the \Nifs\ peak, indicating our results are self-consistent.   
We conclude that while the early ($t<2.5$\,d) peak is unlikely to be powered by shock cooling, this process can explain the dominant emission seen between $2<t<5$\,d, until the Ni luminosity begins to dominate. 

\subsubsection{{\referee Ongoing CSM Interaction or Shock Breakout in a Wind?}}
\label{sec:CSM}

The presence of lines at velocities of $4000$\,\kms\ in both absorption and emission that disappear after 2.5 days provides compelling evidence for the presence of a CSM. We calculate here several estimates for the mass of this CSM lying above the photosphere when \oqm\ was first observed. Throughout this section, we assume a profile $\rho_{\rm CSM}=Ar^{-s}$, between $r_{in}=1.75\times 10^{14} \rm cm$ (the first observed photospheric radius) and $r_{out}=5\times 10^{14} \rm cm$ (the approximate location of the photosphere at the time when the early features disappear) with typical values between $s=0$, appropriate for a constant-density CSM shell and $s=2$, appropriate for a continuous wind \citep{Chevalier1989,Dwarkadas2011}. We assume the CSM opacity $\kappa$ is space independent. We can derive limits on $M_{\rm CSM}$ by using the CSM density profile and considering the optical depth $\tau$:
\begin{align}
        M_{\rm CSM}=\int\limits _{r_{\rm in}}^{r_{\rm out}}4\pi\rho r^{2}dr\ {\rm and}\ \tau =\int\limits _{r_{\rm in}}^{r_{\rm out}}\kappa\rho dr
\end{align}
 
 Since our earliest observations do not show {\referee direct} evidence for a wind shock breakout {\referee still ongoing at the time of detection} (in contrast, e.g., to SN 2006aj, \citealt{Waxman2007}), we can place an upper limit on the mass of the circumstellar material lying ahead of the photosphere during our first observations ($t\approx0.5$\,days).  At this time, the remaining CSM must have an optical depth $\tau \leq c/v$, so:
\begin{equation}
    \rho_{\rm in}\leq\frac{c}{v_{\rm ej}}\left|s-1\right|\kappa_{\rm cont}^{-1}r_{\rm in}^{-1}\left|\left(\frac{r_{\rm out}}{r_{\rm in}}\right)^{1-s}-1\right|^{-1}
\end{equation}
\noindent
which gives an upper limit of $\rho_{in}\leq\ 2\times 10^{-12} \, \rm g\ {\rm cm^{-3}}$ for $s=2$, and $\rho_{in}\leq\ 7\times10^{-13}\ \rm g\ {\rm cm^{-3}}$ for $s=0$ for $v_{\rm ej}=20,000\,  \rm km \, s^{-1}$. By integrating this density we can limit the CSM mass to  $M_{\rm CSM}\leq 0.06\,M_{\odot}$ for $s=2$, and $M_{\rm CSM}\leq 0.17\,M_{\odot}$ for $s=0$. 

We can also place a minimum bound on the mass and density of the CSM from the fact that line photons escape. We assume that the \ion{C}{3} and \ion{C}{4} lines in the unshocked CSM are emitted from a region $\tau_{l} = \int_{r}^{\infty} \rho \kappa_{\rm eff} dr =1$ (noting that the effective absorption opacity $\kappa_{\rm eff}\gg \kappa_{T}$ ), where $\kappa_{T}$ is the Thompson opacity. For Doppler-broadened lines that are resolved in frequency, $\kappa_{\rm eff}$ will be determined by the peak height of the frequency dependent opacity $\kappa_\nu$ for the broadened line \citep[for an in-depth discussion, see][]{Rabinak2011}. We therefore choose $\kappa_{\rm eff} \approx \frac{c}{v}\kappa_{l}$, where the intrinsic line opacity is $\kappa_l\equiv \frac{1}{\lambda_0} \int \kappa_\lambda d\lambda $, with the integral performed across the line, and $\lambda_0$ is the natural wavelength of the line. We get as a lower bound a mass of
\begin{equation}
M_{\rm CSM}=4\pi\frac{s-1}{s-3}\frac{r_{\rm out}^{3-s}-r_{\rm in}^{3-s}}{r_{\rm out}^{1-s}-r_{\rm in}^{1-s}}\kappa_{\rm eff}^{-1}\, .
\end{equation}

\begin{figure*}[t]
\centering
\includegraphics[width=\textwidth]{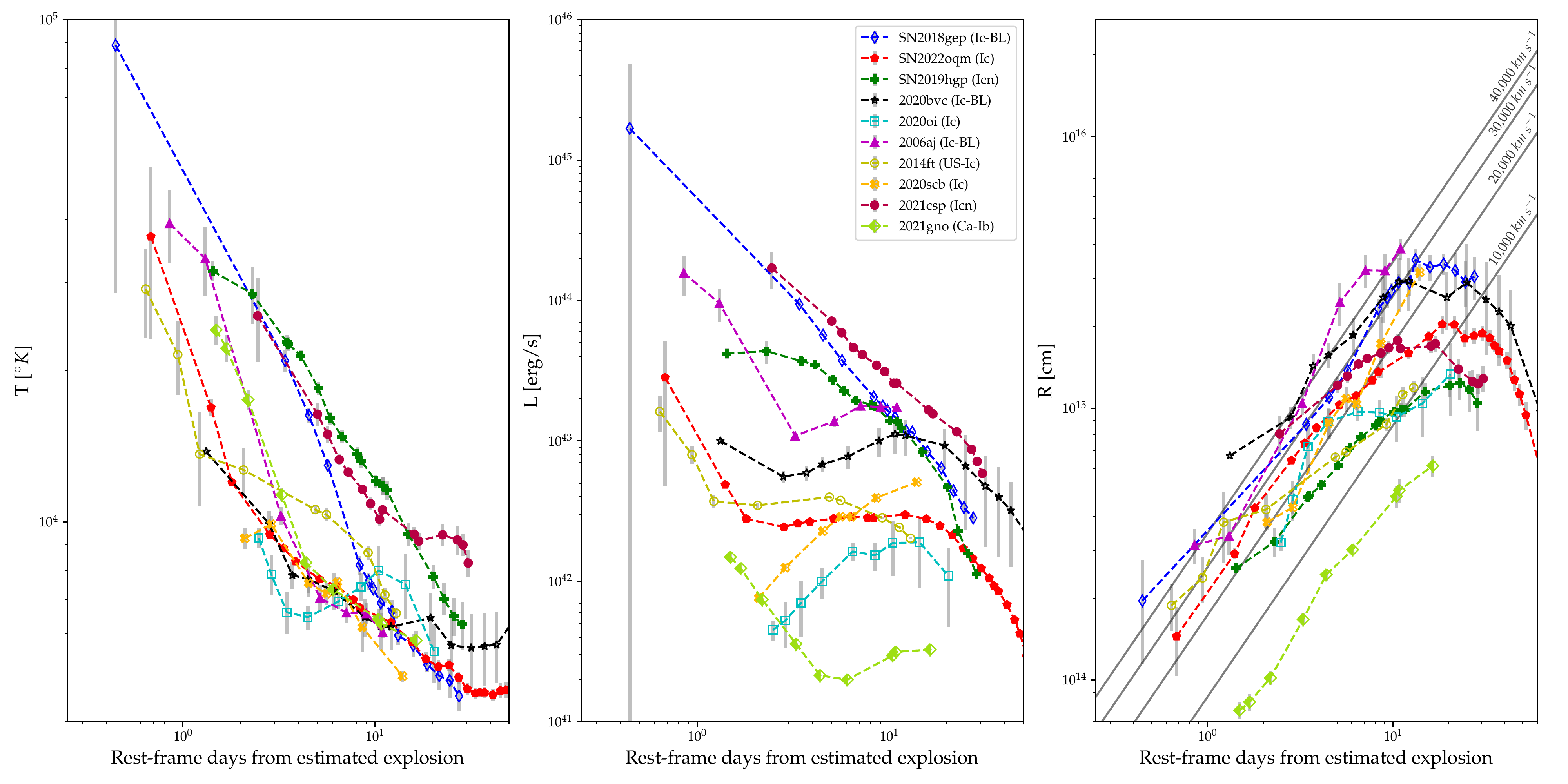} \\
\caption{Blackbody evolution of \oqm\ compared to that of several SNe Ic, Ic-BL, and Icn. Several other SNe Ic have early-time peaks in their luminosity, an initial rapid temperature decline, and high initial photospheric velocities. In particular, SN\,2020scb does not show this behavior, while SN\,2020oi and SN\,2006aj do.}
\label{fig:bb_comp}
\end{figure*}

\begin{figure}[t]
\centering
\includegraphics[width=\columnwidth]{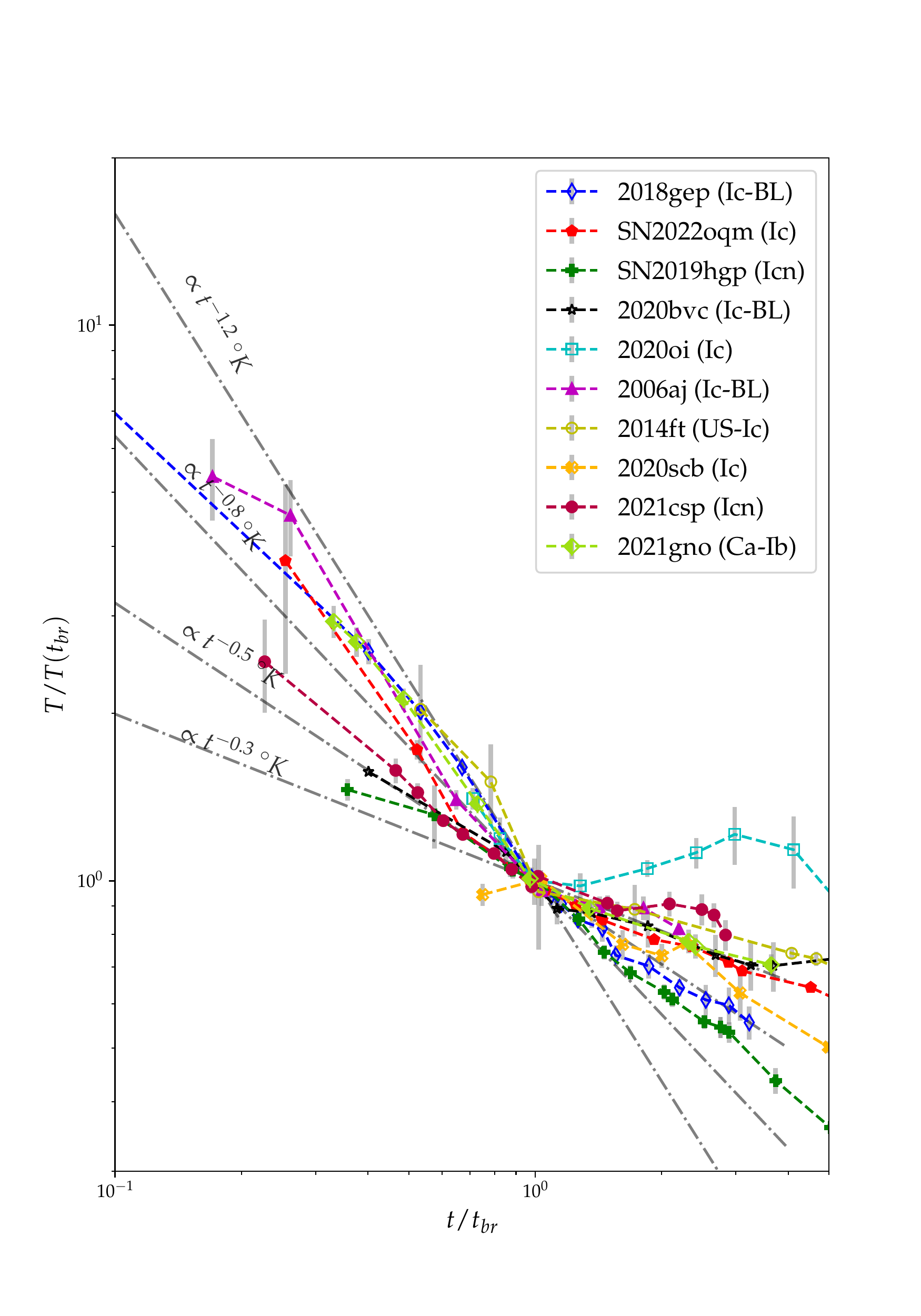} \\
\caption{Temperature evolution of \oqm\ compared with that of several SNe~Ic, Ic-BL, and Icn. The temperature has been  normalized to an arbitrary time where a break is observed in the power law evolution. Grey dashed lines represent various power laws. While the initial logarithmic slope of the temperature is diverse, fast-evolving SNe {\referee (SN\,2018gep, SN\,2019hgp, and SN\,2021csp)} exhibit a fast decline, and double-peaked SNe {\referee (SN\,2022oqm, SN\,2020bvc, SN\,2020oi, SN\,2021gno, and SN\,2006aj)} show a break in their evolution.}
\label{fig:temp_pl}
\end{figure}

\begin{figure*}[h]
\centering
\includegraphics[width=\textwidth]{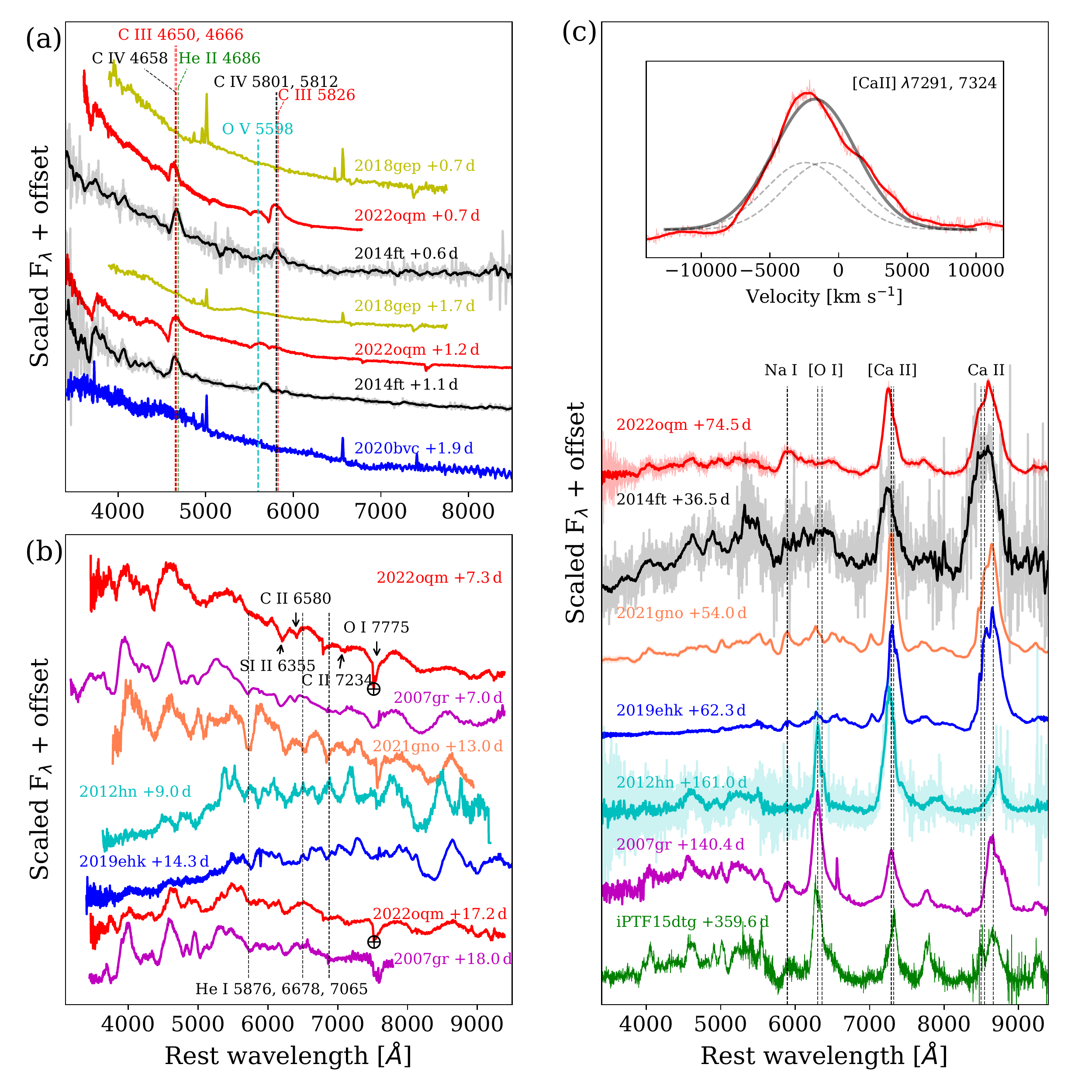} 
\caption{Spectral comparison between SN\,2022oqm and other SNe at different phases. {\referee All of the phases are given relative to the estimated explosion time.} {\referee (a)} Within two days after the estimated explosion time, {\referee (b)} spectra around the primary peak in optical bands, and {\referee (c)} spectra at $> 1$\,month after peak light. {\referee In panels (a) and (c), features of interest are marked at rest, and in panel (b) with a shift of 8000\,\kms.} At early times, \oqm\ is most similar to SN\,2014ft. During the photospheric phase, it is similar to typical SNe~Ic such as SN\,2007gr, and at late times it is characterized by a Ca-rich spectrum like SN\,2014ft and the Type IIb SN\,2019ehk, and unlike the double-peaked SN~Ic~iPTF15dtg. {\referee Some spectra are smoothed with a Savitzky–Golay filter \citep{Savitzky-Golay1964}.} {\referee The inset in panel (c) shows the velocity profile of the \ion{Ca}{2} $\lambda\lambda$7291, 7324, where the average wavelength of 7307.5\,\AA\ was adopted for the reference wavelength. The black solid line shows the best-fit Gaussian model (FWHM $=6900\,\rm{km\,s^{-1}}$, $\Delta v = 1700\,\rm {km\,s^{-1}}$), whereas the dashed lines show the individual emission components for which the fluxes were fixed to be equal in our fitting. }}
\label{fig:spec_compare}
\end{figure*}
\begin{figure}[h]
\centering
\includegraphics[width=\columnwidth]{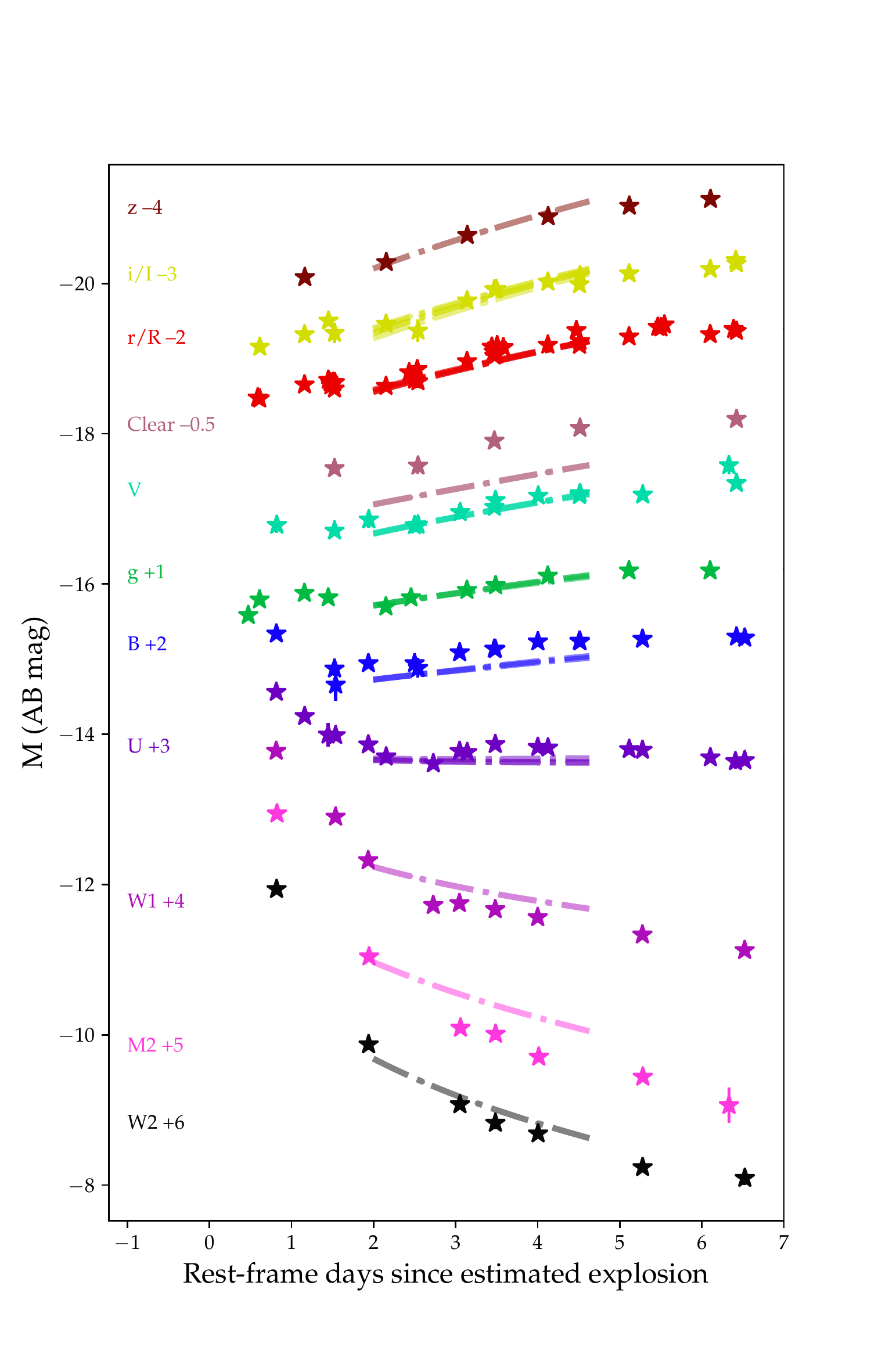} \\
\caption{Best fit shock-cooling \cite{Morag2022} model to the intermediate-time ($5>t>2$ days) light curves, along with the observations of \oqm\ at these times. The best fit model is plotted up to $t<t_{tr}/2 = 4.6$ days. The light-curve and blackbody evolution are well described by a  model with $R=310_{-110}^{+30}\,R_{\odot}$, $M_{env}=0.23_{-0.07}^{+0.44}\,M_{\odot}$, and by $v_{s,*}=1900_{-190}^{+850}\,{\rm km\,s^{-1}}\approx v_{\rm ej}/5$. We note $t_{tr}$ is in good agreement with the equivalent diffusion time $\tau_{m}$, acquired from the $t>12$\,days \Nifs\ fit.}
\label{fig:SC_intermediate}
\end{figure}

While we cannot infer an exact value for $\kappa_{l}$, as the density and temperature of the CSM at the line forming region are unknown, we can calculate it for a wide range of values and provide a limit. We calculate the opacity for the \ion{C}{3} $\lambda\lambda4647$, 4650 and for \ion{C}{4} $\lambda 4658$ features using the open-source opacity table described in \cite{Morag2022} and based on {\secondref \citet{kurucz_atomic_1995}} atomic line lists. We find an upper limit of $\kappa_{l} \lesssim 10^{-2}\ \rm cm^{2}\ {\rm g^{-1}}$. In Fig.~\ref{fig:CO_opac}, we show the line opacities near $4650$~\AA, for the density resulting in the highest line opacities. The opacity upper limit implies a lower mass limit of $M_{\rm CSM}\gtrsim 7 \times 10^{-4}\,M_{\odot}$ assuming $s=2$, and $M_{\rm CSM}\gtrsim 10^{-3}\,M_{\odot}$ assuming $s=0$.  

The upper mass limit from the continuum optical depth is quite robust. The lower mass limit is less strict: the line opacity depends on the occupation fraction of the electron states of the C ions, which is here determined by LTE, and can vary due to NLTE effects and due to the possible effect of ionizing X-ray photons absorbed in the material. Given these caveats we can proceed.  

\subsubsection{{\referee Self-Consistency of the Proposed Scenario}}
\label{sec:csm_power}
\cite{Murase2014} explore a simple framework for CSM interaction, where the ejecta collide with a CSM shell in a plastic collision. By demanding that the energy and momentum are conserved, the dissipated energy in the collision will provide an estimate for the interaction luminosity up to adiabatic losses. We consider a similar scenario, but modify it, considering only an external layer of the ejecta. The dissipated energy from a plastic collision between a shell of CSM and an ejecta layer with $M_{\rm ej,i}$ and $v_{\rm ej,i}$ is
\begin{equation}
    \Delta E=\frac{1}{2}\frac{M_{\rm ej,i}M_{\rm CSM}}{\left(M_{\rm ej,i}+M_{\rm CSM}\right)}\left(v_{\rm ej,i}-v_{\rm CSM}\right)^{2}
\end{equation}

During this collision, the ejecta creates a forward shock in the CSM, and the CSM will act as a piston on the ejecta, creating a reverse shock and decelerating it. The reverse shock is expected to dissipate when it sweeps up roughly an equal amount of ejecta mass to the CSM. Considering an external ejecta layer with mass $M_{\rm ej,i} = M_{\rm CSM}$ that is colliding with the CSM:

\begin{equation}
    \label{eq:dissipated_energy}
    \Delta E=\frac{1}{4}M_{\rm CSM}\left(v_{\rm ej}-v_{\rm CSM}\right)^{2}
\end{equation}

In an optically thin wind, all the dissipated energy (neglecting adiabatic losses) will be radiated within a light travel time $t\approx R/c$ for a spherical CSM. If the CSM is optically thick, the emission will occur on a dynamical time $t\approx R/v$, or a diffusion time $t = \sqrt{\frac{3\kappa M}{4\pi cv}}$ \citep{Ofek2010,Chevalier2011}. 

When integrating the bolometric light curve we find that $E_\text{rad}\approx1.5\times 10^{48}\,\text{erg}$ were emitted from 0.5 to 3~days after the explosion. Assuming that the spectroscopic line velocity we see in the first spectra $v_\text{CSM} = 4000\,\text{km}\,\text{s}^{-1}$ is that of the unshocked CSM that is then swept up by the SN ejecta and accelerated to $v_\text{ej}=20,000\,\text{km}\,\text{s}^{-1}$ that we deduce from the early-time blackbody radius evolution, we can estimate the CSM mass: 

\begin{equation}
\label{eq:CSMmass}
M_{\rm CSM}\thickapprox\frac{4E_{rad}}{\left(v_{\rm ej}-v_{\rm CSM}\right)^{2}}=6.6\times10^{-4}\,M_{\odot}
\end{equation}

The inferred CSM mass is much smaller than the ejecta mass, which is self-consistent with the assumption only the most outer ejecta layer is interacting with the CSM, and is in good agreement with our limits from the previous section. If we use Eq. \ref{eq:dissipated_energy} for the upper limit we derived on the CSM mass of 0.06 \msun, the condition that $\tau \leq \frac{c}{v}$, would result in $3 \times 10^{50}$ erg released as dissipated energy. Alternatively, a CSM mass of $7 \times 10^{-4}\,M_{\odot}$ implies an optical depth of $\tau = 0.4$ for $s=0$ or $\tau=0.1$ for $s=2$. As $\tau$ is much smaller than $\frac{c}{v}$ this argues against either a shock breakout from an optically thick CSM shell, or the interaction of the ejecta with an optically thick wind as the powering mechanism for the early light curve during the observed phase, and is consistent with our assumption of an optically thin wind.  

Next, we check if the presence of the reverse shock does not impact the intermediate-time shock-cooling emission, discussed in \S~\ref{sec:SC}. Since the reverse shock will dissipate when the swept-up CSM mass will match the shocked ejecta, we can estimate the fraction of the ejecta affected by the reverse shock. This estimate is relevant only if the amount of CSM mass above the photosphere is similar to the CSM mass already shocked when observations began. This is the case for $s=0-2$, but not for a steep density profile $s>2$.  We use \citet{Rabinak2011} Eq. 11 (recast in terms of $v_{\rm s*,8.5}$ using \citet{Morag2022} Eq. 3). Namely,
 \begin{equation}
 \label{deltaM_ph_RW}
     M_{\rm ph}/M_{\rm ej}\approx 4 \times 10^{-4} \frac{v_{\rm s*,8.5}^{1.6}}{(M_{\rm ej}/1\, M_{\odot})^{0.8}\kappa_{0.34}^{0.8}}t_{\rm days}^{1.5}
 \end{equation}
 where $M_{\rm ph}$ is the mass outside of the photosphere and $M_{\rm ej}$ is the ejecta mass. To estimate $M_{\rm ph}/M_{\rm ej}$, we choose the previously derived values $v_{s*}=1900$~\kms, $\kappa=0.2\ \rm cm^2 \,  \rm g^{-1}$ and $M_{\rm ej} \lesssim 1\,M_{\odot}$ for $t>2.5$\,days, we get $M_{\rm ph} \gtrsim 10^{-3}$, which is equal to or larger than our CSM estimate. We conclude the early CSM light-curve component does not significantly impact the use of shock-cooling models at later times, under the assumption the density profile of the CSM is not steep. 

Finally, we can check whether our non-detection of X-rays at early time is consistent with the optically thin CSM we find. A shock breakout in a stellar wind is expected to be accompanied by a forward-propagating collisionless shock that would harden the emitted spectrum and convert some of the thermal photons into hard X-rays \citep{Katz2011}, although the exact thermal and hard X-ray spectrum is currently unknown. X-ray radiation emitted by this mechanism is also likely to be absorbed by photo-ionization in the CSM, if it exists. For SN\,2006aj, where a wind breakout likely occurred \citep{Waxman2007} an X-ray flux of the same order of the optical flux was observed during the first day. Assuming this is also the case here, we check if the X-ray opacity of the CSM we deduce is high enough to bring the X-ray emission below our observed limit. 

We calculate the X-ray opacities of the CSM for a wide range of CSM temperature and densities and find that at $T\lesssim30$ eV the X-ray opacity is in the range $10^{2}-10^{5} \, \rm cm^2 \,  \rm g^{-1}$, so that $\kappa > 100 \rm \, cm^2 \, g^{-1}$. In Fig. \ref{fig:xray_opac}, we show a representative example for the effect of temperature on the bound-free absorption in the CSM.  We calculate the optical depth of X-rays given this lower limit and find $\tau > 130$. Alternatively, less than $\sim 7\times 10^{-5}$\,\msun\ are sufficient to make the CSM optically thick to X-rays. While our calculation does not  include NLTE effects, and taking into account the breakout flash (as opposed to the X-rays from the collisionless shock), this analysis shows that for a C/O composition, unless the CSM is almost fully ionized, a small amount of matter is sufficient to totally absorb the initial X-ray radiation. Since both \ion{O}{4} and \ion{C}{4} features are identified in the first spectrum, and the exponential dependence of the ionization fraction on temperature, a large fraction of the CSM being fully ionized is strongly disfavored, and high X-ray suppression is likely. 

\subsection{Implications of the Lack of Pre-SN Emission}

The early emission lines disappear at day three after the explosion which might indicate that the ejecta have swept up the entire CSM at this time. For an ejecta velocity of $v_\text{ej} = 20,000\,\text{km}\,\text{s}^{-1}$ as we measure at early times, this would imply that the CSM is located at a distance of $R_\text{CSM} = 5\times10^{14}\,\text{cm}$ and that the CSM was ejected $\sim 15$\,days before the SN explosion, given a CSM velocity of $\sim4000$\,\kms. {\referee We note that if the velocity difference observed between the first two epochs is due to acceleration, the CSM might have been ejected earlier.}

We here estimate the energy that is required to unbind $7\times 10^{-4}\,M_\odot$ of material from a massive compact progenitor. We assume a W-R progenitor star with a radius of $R_\text{star}=1\,R_\odot$ and a mass of $M_\text{star}=10\,M_\odot$ \citep{Nugis2000}, and find that unbinding the CSM from the stellar surface requires 


\begin{equation}
\frac{E_{\rm pot}}{10^{46}\,{\rm erg}}=3.8\left(\frac{M_{\rm star}}{10\,M_{\odot}}\right)\left(\frac{M_{\rm CSM}}{10^{-3}\,M_{\odot}}\right)\left(\frac{R_{\rm star}}{R_{\odot}}\right)^{-1}
\end{equation}\, ,
\noindent
which is negligible compared to the CSM kinetic energy given by

\begin{equation}
\frac{E_{\rm CSM,kin}}{10^{47}\,{\rm erg}}=1.6\left(\frac{M_{\rm CSM}}{10^{-3}\,M_{\odot}}\right)\left(\frac{v_{\rm CSM}}{4000\,{\rm km\,s^{-1}}}\right)^{2}\, . 
\end{equation}


As shown in \S~\ref{subsec:precursor}, we can rule out precursors that are brighter than $-14\,\text{mag}$ in the $i$ band and last for at least two weeks in the last 100 days before the SN explosion. The precursor luminosity depends on its duration and is given as

\begin{equation}
L_\text{prec} = \epsilon  \frac{E_\text{CSM,kin}}{\Delta t} = 2\times10^{41}{\rm erg\,s^{-1}}\left(\frac{\Delta t}{{\rm week}}\right)^{-1}
\end{equation}

The fact that no precursor was detected allows us to constrain the radiative efficiency $\epsilon$, the fraction of CSM kinetic energy converted to optical radiation e.g., by collision with pre-existing CSM. 

While the progenitors of {\referee most} SNe~IIn are likely surrounded by material ejected during earlier eruptions, the immediate surroundings of the progenitor of \oqm\ could have had a low matter density at the time of the outburst. This could significantly reduce the efficiency of a pre-explosion outburst. Furthermore, CSM interaction in an optically thin environment would likely result in radiation outside the optical bands. For example, in the first UV observation of \oqm\ only a small fraction ($<10$\%) of the total observed luminosity is radiated in the optical bands. 

We require that the precursor is fainter than $-14$. For a week-long precursor, this constrains the radiative efficiency to $\epsilon<0.45$, or $\epsilon<0.2$ for a 3-day-long outburst, both of which are not constraining limits, and indicate that the CSM could have been ejected in an outburst below our detection threshold in the observed bands. 

\subsection{{\referee Searching for $>100$\,keV Breakout Emission}}
\label{subsec:SB_wind}
When a massive progenitor explodes, a radiation-mediated shock will travel down the density profile of the star, until the optical depth of the material above the shock region drops below $c/v$ \citep{Weaver1976}. If it is  sufficiently dense, this process will occur in the CSM. Depending on the breakout radius $R_{\rm br}$, an early UV-optical flash might be observed, lasting for a time equal to $\sim R_{\rm br}/v$ \citep{Ofek2010,Chevalier2011,Svirski2012}. As mentioned in $\S$~\ref{sec:csm_power}, the optical depth of the CSM above the photosphere is smaller than unity at the time we first start observing. Thus, a CSM breakout would have occurred before observations began. This is consistent with the observations of SN\,2006aj a SN accompanied by a low-luminosity GRB lasting $\sim 10^{4}$\,s, interpreted as the CSM breakout \citep{Waxman2007}. In that case, the optical and UV bands rose to peak over a day timescale, resulting in an early UV-optical peak similar to that observed for \oqm, as shown in Fig. \ref{fig:lc_comp} and \ref{fig:bb_comp}. 

While we most likely did not observe the breakout flash, a considerable amount of CSM (compared to $\sim 10^{-3}$\,\msun\ we infer) might have been shocked prior to our observations if the CSM has a steep density profile ($s>2$). In this case, a large amount of shocked material originating from both the CSM and the ejecta might still be cooling up to $t=3$\,d, as described by Eq. \ref{deltaM_ph_RW}.  The post-breakout cooling of this material might account for some of the early radiation, and could possibly account for the shock-cooling-like temperature and radius evolution at $t>2$ days. \cite{Chevalier2011} show that the shocked CSM can be approximated with a self-similar evolution with \textbf{$n=7$}, compared to $n=10$--12 for stellar envelopes \citep{Matzner_1999}, so the density profile at the photosphere can be steep. If the velocity profile is similar to the stellar case, one might expect a shock-cooling-like blackbody evolution.\\ 

As was observed for SN\,2006aj, a CSM breakout around a SN Ib/c progenitor is expected to peak in the 100\,keV -- MeV range \citep{Waxman2007,Katz2011,Granot2018,Margalit2022}, resulting in a low-luminosity GRB. We search for a coincident GRB in the {\it Fermi}/GBM and {\it Swift}/BAT instruments. No onboard, or sub-threshold trigger, was found during the putative breakout window of $2459771.2 \pm 0.5$ JD, that is also consistent with the location of SN 2022oqm. During this time period SN 2022oqm was visible to {\it Fermi} and {\it Swift} (above the Earth limb) $\sim70\%$ of the time. Using the {\it Fermi}/GBM trigger sensitivity, we rule out the existence of a GRB with peak flux greater than $\sim1\times10^{-7}\,\mathrm{erg}\,\mathrm{s}^{-1}\,\mathrm{cm}^{-2},$ (50--300\,keV) within this window. However, the sensitivity to a GRB 060218-like transient with {\it Fermi}/GBM is degraded due its relatively slow evolution, with variability timescales comparable to the background variability experienced by {\it Fermi}/GBM in Low-Earth Orbit.  A search using data from Konus-Wind could likely rule out a CSM breakout over the entire time window, but to shallower depths of $\sim5\times10^{-7}\,\mathrm{erg}\,\mathrm{s}^{-1}\,\mathrm{cm}^{-2}$ (20\,keV -- 10\,MeV) \citep{konusupperlims}. Neither of these limits are sensitive enough to constrain a GRB 060218-like transient at the distance of SN 2022oqm, which would peak at a flux of $\sim3\times10^{-8}\,\mathrm{erg}\,\mathrm{s}^{-1}\,\mathrm{cm}^{-2}$ (15--150\,keV).

In the next few years, {\it The Ultraviolet Transient Astronomy Satellite} ({\it ULTRASAT}) will begin a 200 deg$^2$ high-cadence UV survey {\referee \citep{Shvartzvald2023}}, and is expected to detect the early UV emission of hundreds of CCSNe, of which a fraction will be SNe Ib/c \citep{ganot2016}. An early UV flash observed with {\it ULTRASAT} will not only provide information about CSM emission, but will also enable early X-ray observations and a systematic study of coincident GRBs. Finding coincident low-luminosity GRBs for a large fraction of SNe Ib/c with an early UV peak will demonstrate these are the result of spherical CSM breakouts, while having meaningful limits on coincident GRBs will favor a beamed interpretation. We encourage sub-threshold searches for similar future discoveries.  

\subsection{A Population of Explosions?}
{\referee Although its bulk properties such as peak time and luminosity are similar to those of the general SN~Ic population, }\oqm\ shows several peculiarities separating it from {\referee spectroscopically regular SNe~Ic}. It has an early peak only seen in a few other SNe~Ib/c and SNe~Ic-BL. {\referee The SN interacts with a compact distribution of C/O-dominated CSM, directly observed} so far only in rare cases such as SNe~Icn \citep[e.g.,][]{Benami2014,galyam2022,Perley2022,Gagliano2022,Pellegrino2022}, and indirectly implied in (for example) SN\,2018gep through its precursor emission \citep{Ho2019}. {\referee Compared with Ca-rich transients, its Ca-dominated nebular spectrum, rise to peak luminosity by 12\,days, and offset location are consistent with the Ca-rich population \citep{Perets2010,Kasliwal2012,De2018}. However, few examples of Ca-rich SNe~Ic (rather than Ib) have been previously observed. \oqm\ is significantly more luminous, with a higher $^{56} \rm Ni$ mass, and more rapidly declining than most Ca-rich SNe~Ib \citep{De2018,Sharon2023}. Notably, this also holds with respect to the Ca-rich SN~Ic SN\,2012hn. In at least two Ca-rich SNe~Ib (SN\,2021gno, SN\,2021inl), a short-lived and blue peak similar to that of \oqm\ has been observed. In the case of SN\,2019ehk, an early blue peak has been accompanied by short lived and narrow H and He emission lines from a compact CSM. While the early behavior is similar, the different composition challenges a similar progenitor or explosion mechanism as that of \oqm.} 

{\referee Though different in its total radiated luminosity, a notably} similar SN to \oqm\ is the Ca-rich Ic SN\,2014ft \citep{De2018}. It has an early peak, a fast drop in temperature, {\referee a similar peak magnitude,} a Ca-dominated nebular spectrum, and is extremely offset ($\sim50$\,kpc) from the nearest massive galaxy with the same redshift. The common features between the early-time spectrum of SN\,2014ft and \oqm\ {\referee (Fig. \ref{fig:spec_compare})} suggests a similar CSM composition. {\referee We propose} that lines of \ion{C}{4} and \ion{C}{3} dominate the early spectra, rather than \ion{He}{2} as originally inferred from the {\referee early-time} spectrum of SN\,2014ft, {\referee as it better matches the peak emission, and owing to the presence of other highly ionized C lines.} However, \oqm\ has different bulk properties. An order of magnitude more mass was ejected and \Nifs\ synthesized in the explosion compared to SN\,2014ft, as evident by the slower evolution of \oqm.  It remains to be seen if future SNe show common similarities to these two objects in CSM, location, and nebular-phase spectra, supporting a common origin. {\referee Since Ca-rich transients, as well as SNe~Ic, are diverse in their properties, and might originate from different channels, it is unclear if SN\,2022oqm and SN\,2014ft are the extreme end of a distribution of the SN~Ic population, the Ca-rich transient population, or represent their own unique group. Any single explosion mechanism or progenitor channel investigated in future studies would need to account for an order of magnitude difference in ejected mass and $^{56} \rm Ni$ mass between the two.} 

\section{Interpretation}
\label{sec:Interpretation}

\subsection{Option 1: A Massive Progenitor, Embedded in CSM Ejected during its Final Weeks}
Many of the properties of \oqm\ are consistent with the general properties of SNe~Ic, favoring a massive star origin. The presence of C/O CSM can be explained {\referee naturally} in a massive-star scenario by an eruptive ejection of material shortly before the terminal explosion. {\referee or by the radiative acceleration of a shell of pre-existing dense CSM}. Such a pre-SN eruption is expected to eject material in the last stages of the evolution {\referee of massive stars} \citep{Smith2014b,Fields_2021,Varma2021,Yoshida_2021,Matsumoto2022}. In terms of location, while remote, the offset of \oqm\ is consistent with the general offset distribution of SNe~Ic \citep{Schulze2020}, and cannot exclude a massive star origin, and while the [\ion{O}{1}] nebular luminosity has been connected to progenitor mass \citep{Jerkstrand2015}, the general SN Ic population shows no correlation between $M_{\rm ej}$ and [\ion{O}{1}] nebular luminosity \citep{Prentice2022}.

Shock-cooling models of an extended $R\approx300\,R_{\odot}$ progenitor describe the behavior of the light curves at $2<t<5$ days, and produce a diffusion timescale which is consistent with the one acquired from the \Nifs\ fits at $t>12$ days. As discussed in $\S$~\ref{subsec:SB_wind}, the shock cooling following a CSM breakout might also produce a similar behavior, for which we do not have a numerically calibrated model allowing parameter estimation. If we have observed the cooling of the stellar envelope, the progenitor would have to be a stripped star with an inflated envelope. If we are observing CSM cooling, a CSM originating in a W-R star could explain the observations. {\referee A massive-star origin has been previously suggested for several Ca-rich transients \citep{Jacobson-Galan2020,De2021}. \cite{Sharon2023} test various explosion models from the literature and find that most are inconsistent with the observed $M_{\rm Ni}$--$t_{\gamma}$ distribution of He-shell detonations and core collapse of ultrastripped stars, but are consistent with some SNe~Ia and core collapse of stripped-star models. Our inferred values for \oqm\ of $M_{\rm Ni}=0.106\,M_{\odot}$, $M_{\rm ej}=1.1\,M_{\odot}$, and $t_{\gamma}=36$\,d place \oqm\ in the region broadly consistent with stripped-envelope SN explosions of \cite{Dessart2016,Woosley2021}, with an ejected mass in the 0.5--3\,$M_{\odot}$ range. }

\subsection{Option 2: A White Dwarf Progenitor Disrupting a C/O Companion}
Since the ejected mass of the explosion is within the mass range of WDs, we consider a system containing such a star as a possible progenitor for the explosion. The velocity of the features in the first few spectra are around 4000~\kms\ 
{\referee (P~Cygni minimum), and up to 15,000\,\kms.}  Such a velocity is of the order of the escape velocity from the surface of a WD, {\referee projected on the line of sight}. A WD progenitor for \oqm\ is consistent with the lack of detectable nebular [\ion{O}{1}] emission, which correlates with progenitor mass in nebular spectral modeling of core-collapse SNe \citep{Jerkstrand2015}{\referee , and is thus expected for a massive star progenitor}. The strong Ca emission in the nebular phase, marking \oqm\ as Ca-rich, connects it with a population of transients which is associated with non-star-forming locations and with a thermonuclear origin \citep{Perets2010, Kasliwal2012, Lunnan2017,De2020}. {\referee However, this preference of Ca-rich transients is, to the best of our knowledge, not demonstrated for Ca-rich spectral subtypes independently.  While \cite{De2020} explain all H-poor Ca-rich events within a framework consisting of double detonation of He shells on WDs, some Ca-rich transients have been suggested to have a massive star origin, such as the Ca-rich SNe IIb iPTF 15eqv \citep{Milisavljevic2017} and SN\,2019ehk \citep{Jacobson-Galan2020,De2021}.}

A non-massive-star origin would be consistent with the location of the explosion in the outskirts of its host, and $>3$ kpc from any luminous UV source. As massive stars have short lifetimes ($<10$\,Myr), a progenitor star would have to travel at more than 300\,\kms\ for 10\,Myr to cover such a distance.  While CCSNe (and specifically, SNe~Ic) do occur occasionally in offset regions or regions with low star-formation \citep{Hosseinzadeh2019,Irani2019b,Irani2022}, a population which preferentially explodes in non-star-forming regions cannot originate from  massive stars. If SN\,2014ft and \oqm\ are part of the same population, and other SNe with similar properties will be found in similar sites, this would imply a non-massive-star origin for these events. Our limits on an underlying point source do not exclude the presence of globular clusters \citep{Richtler2003} and ultra-compact dwarf galaxies \citep{Bruns2012} where the environment is dense and close binary interactions between compact objects are more likely. However, \cite{De2020} demonstrates that the offset distribution of Ca-rich SNe in general is inconsistent with the globular cluster offset distribution - arguing against their association.

Models predicting an early flux excess for SNe Ia that arise from WD systems involve companion interaction through Roche-lobe overflow \citep{Kasen2010,Magee2021}, CSM interaction \citep{Kromer2016,Piro_2016}, and clumpy \Nifs\ distribution in the ejecta \citep{Dimitriadis_2018,Shappee_2018,Magee2020b}. An early flux excess above the expected \Nifs-powered light curves has been found to occur in a significant fraction of SNe Ia \citep{Magee2020a,Deckers2022}. 

{\referee In the comparisons of \cite{Sharon2023}, the location of \oqm\ in the $M_{\rm Ni}$--$t_{\gamma}$ parameter space and its $M_{\rm ej}$ are consistent with low-luminosity thermonuclear WD sub-Chandra detonations of \cite{Kushnir2020} or the WD collisions of \cite{Kushnir2013}. However the sub-Chandra models of \cite{Kushnir2020} required to produce $M_{\rm Ni}=0.106\,M_{\odot}$ and $t_{\gamma}=36$\,d have a progenitor mass of $M_{\rm prog}=0.85\,M_{\odot}$, in tension with the observed $M_{\rm ej}=1.1\,M_{\odot}$ (on top of the remnant mass).  }

We propose a scenario where a
C/O WD is disrupted by a heavier WD companion. The disruption deposits the CSM we see, while continuous accretion eventually triggers the explosion of the heavier primary. This could satisfy many of the observed properties of \oqm. The relatively low amount of \Nifs\ synthesized compared to SNe~Ia \citep[e.g.][]{Stritzinger2006,Scalzo2014} in combination with the high velocities in the early and nebular phase and the lack of strong \ion{Si}{0} absorption set this event apart from regular WD explosions as SNe Ia. 


\section{Summary}
\label{sec:summary}
\begin{itemize}
    \item \oqm\ is a SN Ic detected $< 1$\,day after the explosion, with early UV-optical photometric coverage and a spectrum within 0.6\,days of explosion.
    \item The early spectra of \oqm\ show high ionization C/O features, with a mean velocity of 4000--5500\,\kms, with extended blue-edge velocities of 12,000--15,000\,\kms\ which disappear after 2--3\,days. We interpret these lines as a result of combined emission from an optically thin CSM, and the underlying ejecta. 
    \item We infer these lines arise from a CSM with a mass of $\gtrsim 7\times 10^{-4}$\,\msun, which is sufficient to drive the luminosity during the first days. 
    \item We find no significant X-ray emission, expected from interaction shocks, or sub-threshold $\gamma$-ray emission. This is consistent with absorption by the CSM mass we estimate, and the expected optical depth in the X-ray band. The $\gamma$-ray limits cannot rule out a GRB 060218-like burst, associated with the CSM-breakout of SN\,2006aj. 
    \item \oqm\ rose rapidly to peak in the optical bands, rising more than 2.6\,mag\,day$^{-1}$, while rapidly declining in the UV. After 3 days, the light curve evolution slows, and the optical light curves rise to a second peak after 15\,days.  
    \item During the first 2--3\,days, the blackbody temperature and luminosity decline quickly, while the radius expands at 20,000\,\kms. This behavior changes roughly at the same time the absorption lines evolve to lower ionization C/O expanding at $\sim 10,000$\,\kms. The break in the blackbody evolution naturally explains the double-peaked light curve structure.
    \item Up to the second peak, the luminosity and temperature evolve as expected from shock-cooling. 
    \item About the main peak, \oqm\ is similar to a typical SN Ic, with a light curve powered by $0.12$ \msun\ of \Nifs, $M_{\rm ej} = 1.1$ \msun, and $t_{\gamma} = 36$ days, and displays a typical spectrum.
    \item The SN becomes nebular by $t=60$\,days, developing strong NIR \ion{Ca}{2} and [\ion{Ca}{2}] emission, with a high FWHM compared to other SNe~Ic, and with no detectable [\ion{O}{1}]. This marks \oqm\ as Ca-rich. 
    \item The explosion site is located at the outskirts of a massive star-forming galaxy. While its global properties and the offset are consistent with the general SN~Ic population, the combination of no elevated galaxy emission at the SN site and no nearby H~II regions challenges a massive star origin. 
    \item \oqm\ is similar to several other SNe Ic {\referee and Ca-rich transients}. SN\,2014ft has a similar early and nebular spectrum. Notably, SN\,2006aj (a GRB-SN associated with a wind-breakout), SN\,2020oi  (a regular SN Ic associated with CSM through radio emission), {\referee and SN\,2019ehk (a double-peaked Ca-rich SN~IIb with He/H narrow features)} have an early UV peak, and show similar early declining temperature profiles. 
    
    \item The upcoming ULTRASAT survey will be able to detect stripped-envelope SNe in their first hours, characterize how common an early UV-optical peak is, and determine its origin.
\end{itemize}

\section{Data Availability}
All photometric and spectroscopic data of \oqm\ and SN2020\,scb used in this paper are made available via WISeREP\footnote{\href{https://www.wiserep.org}{https://www.wiserep.org}} \citep{yaron2012}. The blackbody fits reported in \S~\ref{sec:analysis} will be made available through the journal website in a machine-readable format. The code used for the fitting of the early-time light curve to a power law will be released to \hyperlink{https://github.com/idoirani}{https://github.com/idoirani} upon publication.

\section{Acknowledgements }
We thank Doron Kushnir, Eran Ofek, and Eli Waxman  for their insights on the analysis. U.C. Berkeley undergraduate students Raphael Baer-Way, Kate Bostow, Victoria Brendel, Asia deGraw, Kingsley Ehrich, Connor Jennings, Gabrielle Stewart, and Edgar Vidal  are acknowledged for their effort in taking Lick/Nickel data. We are grateful to the staff at the various observatories where data were obtained.
This work made use of data supplied by the UK Swift Science Data Centre at the University of Leicester.

A.G-Y.’s research is supported by the EU via ERC grant 725161, the ISF GW excellence center, an IMOS space infrastructure grant and BSF/Transformative and GIF grants, as well as the André Deloro Institute for Advanced Research in Space and Optics, The Helen Kimmel Center for Planetary Science, the Schwartz/Reisman Collaborative Science Program and the Norman E Alexander Family Foundation ULTRASAT Data Center Fund, Minerva and Yeda-Sela;  A.G.-Y. is the incumbent of the Arlyn Imberman Professorial Chair.
S. Schulze acknowledges support from the G.R.E.A.T research environment, funded by {\em Vetenskapsr\aa det},  the Swedish Research Council, project 2016-06012.
N.L.S. is funded by the Deutsche Forschungsgemeinschaft (DFG, German Research Foundation) via the Walter Benjamin program -- 461903330.
A.V.F.'s supernova group at U.C. Berkeley has been supported by Steven Nelson, Alan Eustace, Landon Noll, Sunil Nagaraj, Sandy Otellini, Gary and Cynthia Bengier, Clark and Sharon Winslow, Sanford Robertson, Briggs and Kathleen Wood, the Christopher R. Redlich Fund, the Miller Institute for Basic Research in Science (in which A.V.F. was a Miller Senior Fellow), and numerous individual donors.

Based in part on observations obtained with the Samuel Oschin Telescope 48-inch and the 60-inch Telescope at the Palomar Observatory as part of the Zwicky Transient Facility (ZTF) project. ZTF is supported by the National Science Foundation (NSF) under grants AST-1440341 and AST-2034437, and a collaboration including current partners Caltech, IPAC, the Weizmann Institute of Science, the Oskar Klein Center at Stockholm University, the University of Maryland, Deutsches Elektronen-Synchrotron and Humboldt University, the TANGO Consortium of Taiwan, the University of Wisconsin at Milwaukee, Trinity College Dublin, Lawrence Livermore National Laboratories, IN2P3, University of Warwick, Ruhr University Bochum, Northwestern University, and former partners the University of Washington, Los Alamos National Laboratories, and Lawrence Berkeley National Laboratories. Operations are conducted by COO, IPAC, and UW.
The ZTF forced-photometry service was funded under the Heising-Simons Foundation grant \#12540303 (PI M. J. Graham). 
The SED Machine at Palomar Observatory is based upon work supported by the NSF under grant 1106171. 

A major upgrade of the Kast spectrograph on the Shane 3\,m telescope at Lick Observatory, led by Brad Holden, was made possible through gifts from the Heising-Simons Foundation, William and Marina Kast, and the University of California Observatories.
KAIT and its ongoing operation were made possible by donations from Sun Microsystems, Inc., the Hewlett-Packard Company, AutoScope  Corporation, Lick Observatory, the NSF, the University of California, the Sylvia \& Jim Katzman Foundation, and the TABASGO Foundation. 
Research at Lick Observatory is partially supported by a generous gift from Google.
{\secondref Some of the data presented herein were obtained at the W. M. Keck Observatory, which is operated as a scientific partnership among the California Institute of Technology, the University of California, and NASA; the observatory was made possible by the generous financial support of the W. M. Keck Foundation.}
The Liverpool Telescope is operated on the island of La Palma by Liverpool John Moores University in the Spanish Observatorio del Roque de los Muchachos of the Instituto de Astrof\'isica de Canarias with financial support from the UK Science and Technology Facilities Council. Partly based on observations made with the Nordic Optical Telescope, operated at the Observatorio del Roque de los Muchachos.
\\


{\secondref \textit{Facilities}: P48, \swift (UVOT, XRT), P60 (RC, SEDM), Liverpool telescope (IO:O, SPRAT), Gemini-North, Keck I (LRIS), Shane (KAST), NOT (ALFOSC), P200 (DBSP)}

\textit{Software}: \package{Astropy} \citep{Astropy2013,Astropy2018}, \package{IPython} \citep{Perez2007}, \package{Matpotlib} \citep{Hunter2007}, \package{Numpy} \citep{Oliphant2006}, \package{Scipy} \citep{Virtanen2020}, \package{exctinction} \citep{Barbary2016}, \package{FSPS} \citep{Conroy2009a,ForemanMackey2014a}, \package{prospector} V1.1 \citep{Johnson2021a},
 \package{dynesty} \citep{Skilling2004,Skilling2006,Feroz2009,Higson2019,Speagle2020a}
 
\bibliographystyle{apj},
\bibliography{bibliograph.bib}

%

\appendix

\section{}
\subsection{SN\,2020scb}
In this paper, we publish the light curves of SN\,2020scb (ZTF20abwxywy), a SN Ic detected by ZTF on UT August 26.38, 2020, and classified as a SN Ic shortly after \citep{Prentice2020}. SN\,2020scb exploded in the face-on spiral CGCG 456-055, at a redshift of $z=0.017429$, for which we adopt a distance estimate of 76.1\,Mpc,  corrected for Virgo, Great Attractor and Shapley supercluster infall as discussed in $\S$~\ref{discovery}.  We acquired ZTF, LT/IO:O, and {\it Swift}/UVOT  photometry of the SN using the methods described in Sec \ref{subsec:opt-photometry}, and correct these for a Galactic extinction value of 0.052 mag. We also infer a host-galaxy extinction of $E(B-V) = 0.022$ mag using the $g-r$ color 10 days after maximum light \citep{stritzinger2018b}, as discussed in $\S$~\ref{extinction}. We recover a pre-discovery detection of $r=20.53\pm0.16$ mag on August 25.36, following a non-detection 0.9 day prior. Our high cadence light curve and rapid UVOT triggering allowed us to acquire UV photometry by August 26.668, only 2.2 days after the non-detection, and 1.3 days after the first detection - making SN\,2020scb one of the earliest observed SNe Ic in the UV.  We fit the early $t-t_{first}<5$ days light curve to a power-law evolution  in the radius and temperature according to Eq. \ref{eq:bb_pl}, and find a good fit for $t_0= \rm JD\ 2459086.3\pm0.3$ days. The spectral data for this object will be published together with the rest of the ZTF SNe Ic (Yang et al., in prep.).

\begin{figure}[t]
\centering

\includegraphics[width=\columnwidth]{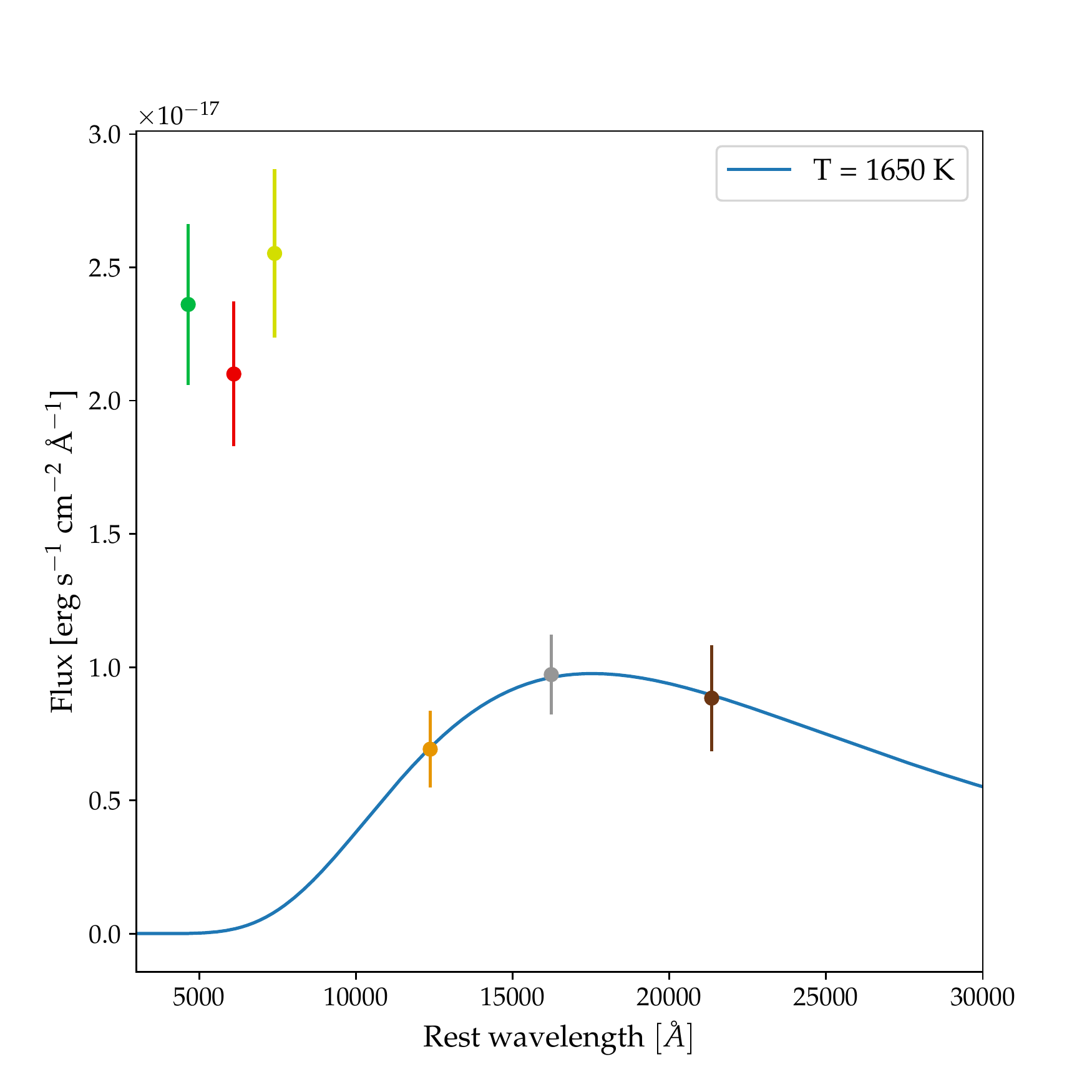} \\
\caption{Blackbody fit to the SED of \oqm\ at $t=66$\,days,  fit separately for $JHK_{s}$ bands. The best-fit blackbody has a temperature of 1650\,K and a radius of $4.4\times10^{15}$\,cm, consistent with the size of the system at $t=66$\,days. We did not fit a blackbody to the optical component, since the spectrum is line dominated.}
\label{fig:2comp_bb}
\end{figure}

\subsection{Shock-Cooling Models}
\label{ap:SC_mod}
For $\S$~\ref{sec:discussion}, we fit intermediate-time observations to the shock-cooling model of \cite{Morag2022}. This model describes the blackbody evolution of a cooling envelope until recombination or sufficient transparency of the envelope, using a set of four free parameters: (1) $R_{13}$, the radius of the progenitor star in units of $10^{13}$\,cm, (2) $f_{\rho}M_{0}$, where $f_{\rho}$ describes the structure of the density near the edge of the stellar envelope and $M_{0}$ is the progenitor mass prior to the SN in units of \msun, (3) $v_{s*,8.5}$, the shock-velocity parameter in units of $10^{8.5}\,{\rm cm\,s^{-1}}$, which roughly corresponds to $v_{\rm ej}/5$, and (4) $M_{\rm env}$, the envelope mass; also, $\kappa_{0.34}$ is the opacity in units of $0.34\,\rm cm^2$\,g$^{-1}$, and $t_{\rm d}/t_{\rm hr}$ is the time since explosion in units of days or hours (respectively). Following their notation, $L$ and $T$ evolve according to
\begin{equation}
\label{eq:AL_sum}
    L_{\rm SC}=L_{\rm planar}+0.9\exp\left[-\left(\frac{2t}{t_{\rm tr}}\right)^{0.5}\right]\,L_{\rm RW}\, ,
\end{equation}
\begin{equation}
\label{eq:AT_col_MSW_2}
    T_{\rm col}=1.1\min\left[T_{\rm ph,planar}\,,\,T_{\rm ph,RW}\right]\, ,
\end{equation}
which are valid during 
\begin{equation}
\label{eq:Ats}
   3 R / c = 17\,R_{13} \, {\rm min} < t < \min[t_{\rm 0.7 eV}, t_{\rm tr}/2].
\end{equation}

\begin{figure}[t]
\centering
\includegraphics[width=\columnwidth]{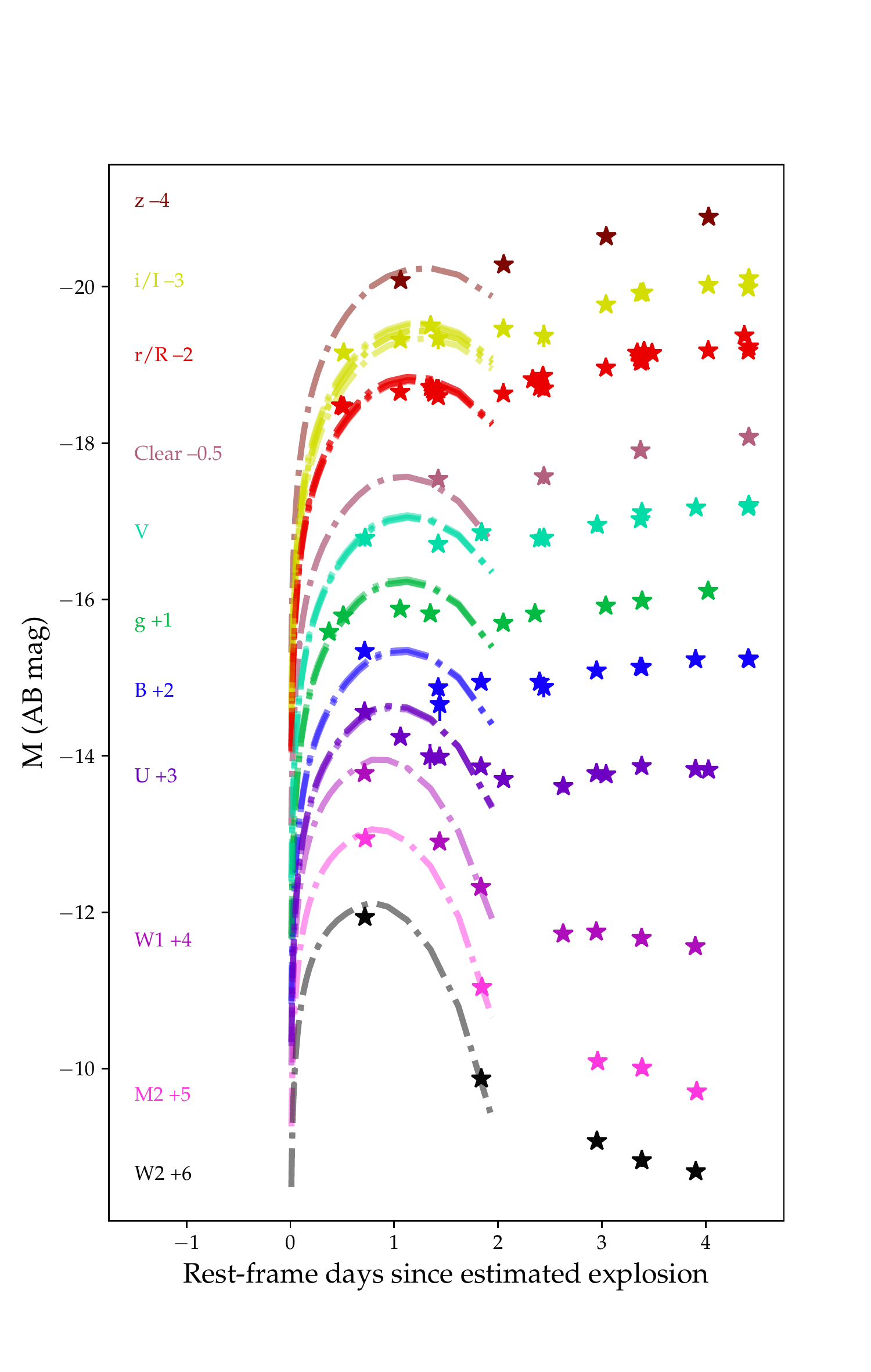} \\

\caption{Best \cite{Piro2021} shock-cooling fits to the light curves of \oqm, for a model with $M_{e}=0.05$\,\msun, $R_e=110\,R_{\odot}$, and $E=1.5\times 10^{50}$\,erg.}
\label{fig:piro_SC_fits}
\end{figure}

\noindent
The terms in Eqs.~(\ref{eq:AL_sum}--\ref{eq:Ats}) are
\begin{align}
\label{eq:AL_planar_powerlaw}
    \frac{L_{{\rm planar}}}{10^{42}\,{\rm erg\,s^{-1}}}=3.01\,R_{13}^{2.46}v_{{\rm s*,8.5}}^{0.60}(f_{\rho}M_{0})^{-0.06}t_{{\rm hr}}^{-4/3}\kappa_{0.34}^{-1.06},
\end{align}
\begin{align}
\label{eq:AT_planar_powerlaw}
    \frac{T_{{\rm ph,planar}}}{\,{\rm eV\,}}=6.94\,R_{13}^{0.12}v_{{\rm s*,8.5}}^{0.15}(f_{\rho}M_{0})^{-0.02}\kappa_{0.34}^{-0.27}t_{{\rm hr}}^{-1/3},
\end{align}
\begin{align}
\label{eq:ALRW}
   \frac{L_{{\rm RW}}}{2.08\times10^{42}\,{\rm erg\,s}^{-1}}=\,R_{13}v_{{\rm s*,8.5}}^{1.91}(f_{\rho}M_{0})^{0.09}\kappa_{0.34}^{-0.91}t_{{\rm d}}^{-0.17},
\end{align}
\begin{align}
\label{eq:ATphRW}
\frac{T_{{\rm ph,RW}}}{\,{\rm \,eV}}=1.66\,R_{13}^{1/4}v_{{\rm s*,8.5}}^{0.07}(f_{\rho}M_{0})^{-0.03}\kappa_{0.34}^{-0.28}t_{{\rm d}}^{-0.45},
\end{align}
\begin{equation} \label{eq:At_0_7eV}
    t_{0.7 \rm eV} = 6.86 \, R_{13}^{0.56} v_{\rm s*,8.5}^{0.16} \kappa_{0.34}^{-0.61} (f_{\rho}M_0)^{-0.06} \rm  \, days \, ,
\end{equation}
\begin{equation}
\label{eq:At_transp}
    \begin{split}
        \begin{aligned}
            {\rm and}~t_{\rm tr} &= 19.5 \, \sqrt{\frac{\kappa_{0.34}M_{\rm env,0}} {v_{\rm s*,8.5}}}\, \text{days}.
        \end{aligned}
    \end{split}
\end{equation}
In addition to the luminosity set by the shock-cooling component, we assume a \Nifs\ decay component, such that the temperature is simply $T_{\rm col}$ and the total luminosity is a sum of Eqs. \ref{eq:AL_sum} and \ref{eq:nifs_lum}:
\begin{equation}
        \begin{aligned}
            L=L_{^{56} \rm Ni}+L_{SC},\ T=T_{\rm col}\, .
        \end{aligned}
\end{equation}
Since the validity of this model is dependent on the model parameters, a $\chi^2$ minimization is not applicable. Instead, we fit this model with a likelihood function adapted for a variable validity domain, as discussed in detail by \cite{Soumagnac2020}:
\begin{equation}
   \mathcal{L}={\rm PDF}\left(\chi^{2},{\rm dof}\right);\,\,\chi^{2}=\sum_{i}\frac{f_{i}-m_{i}}{\sigma_{i}^{2}}\, ,
\end{equation}
where PDF is the $\chi^2$ distribution given the number of degrees of freedom, $f_i$ are the observed fluxes, $\sigma_i$ are the observational uncertainties including a 10\% systematic error, and  $m_i$ are the integrated synthetic fluxes for the model. We do not treat deviations from a blackbody spectrum in our fitting process.

\begin{figure}[t]
\centering

\includegraphics[width=\columnwidth]{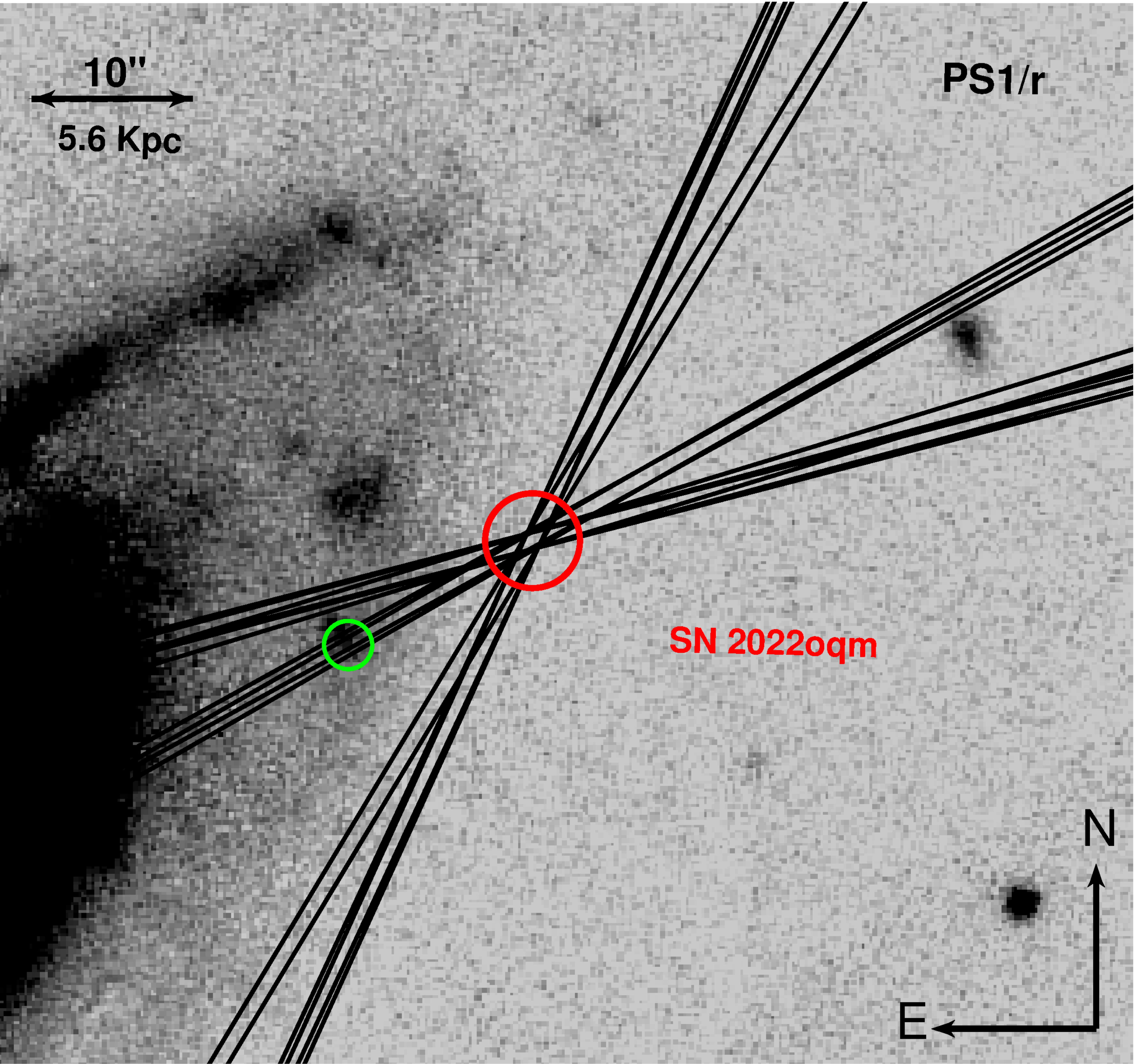} \\
\caption{The host environment of \oqm, displayed using PS1 $r$-band images overlaid with the 8 
NOT and GMOS spectra slit orientations. Slits are drawn with a representative width of 1{\arcsec}. {\secondref The nearest (3.8 kpc) \ion{H}{2} region we identify is marked with a green circle.}}
\label{fig:PA_fig}
\end{figure}

\begin{figure*}[t]
\centering
\includegraphics[width=\textwidth]{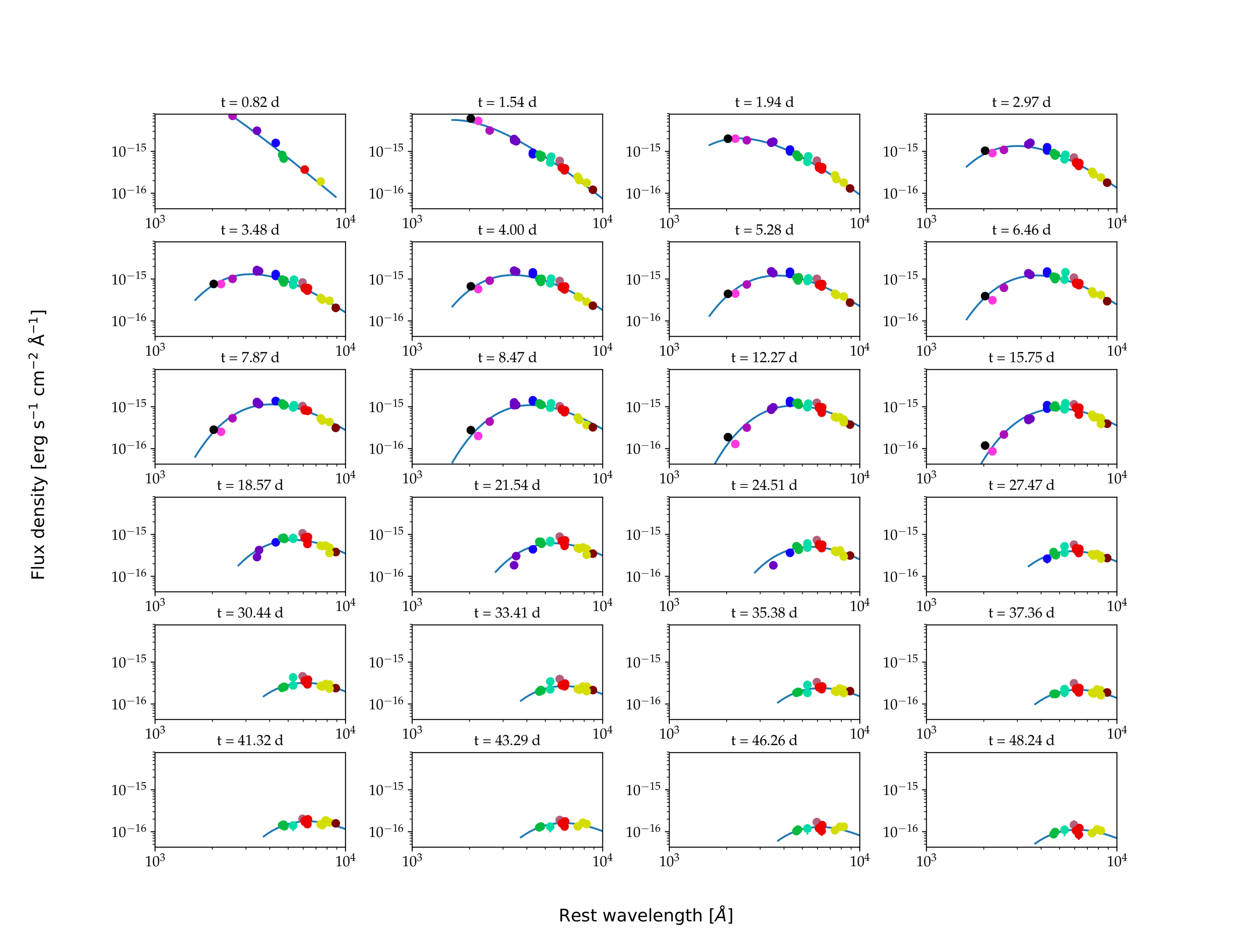} \\
\caption{Blackbody fits to the photometry of SN\,2022oqm.}
\label{fig:sed_fits}
\end{figure*}



\begin{figure*}[t]
\centering

\includegraphics[width=\textwidth]{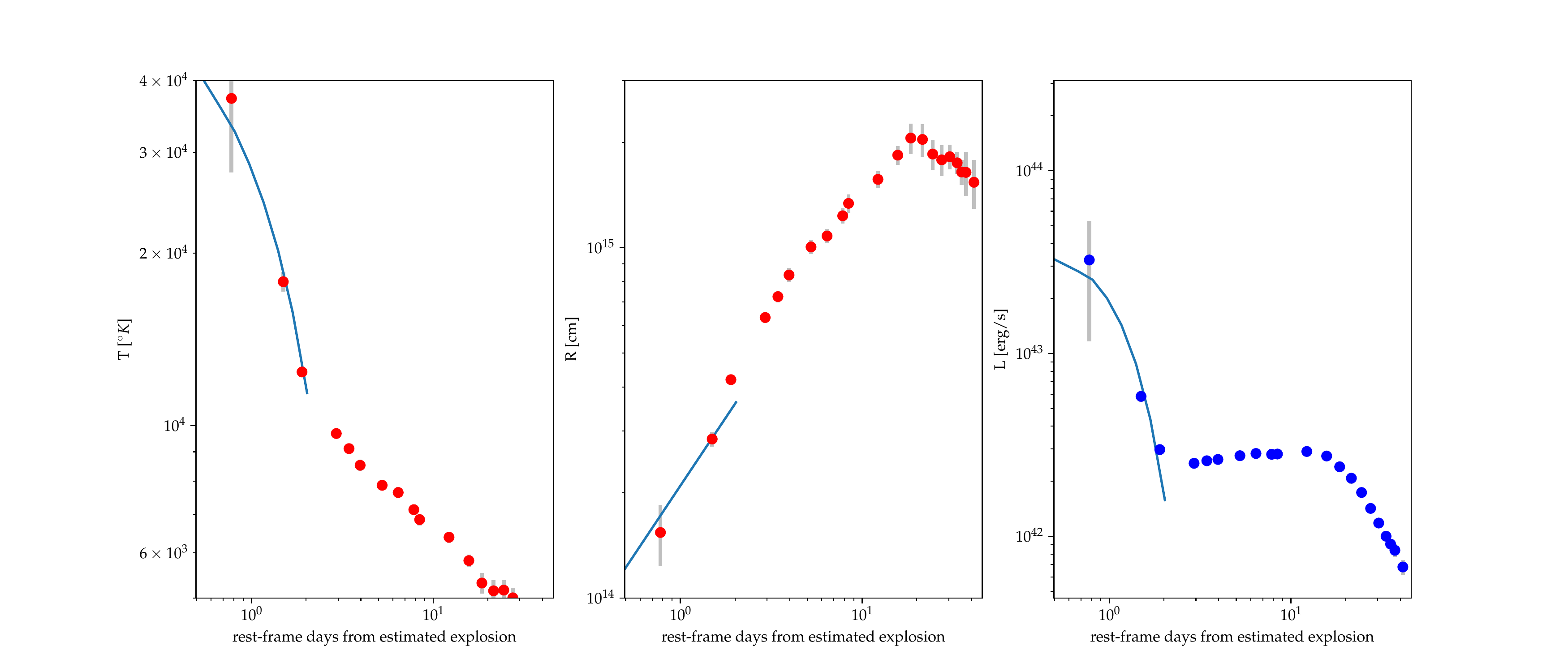} \\
\caption{The blackbody evolution of the best-fitting shock-cooling \cite{Piro2021} model with the blackbody evolution of \oqm. The models are plotted up to $t=t_{\rm ph}$, when the envelope becomes fully transparent. }
\label{fig:piro_SC_fits_bb}
\end{figure*}

\begin{figure*}[t]
\centering

\includegraphics[width=\textwidth]{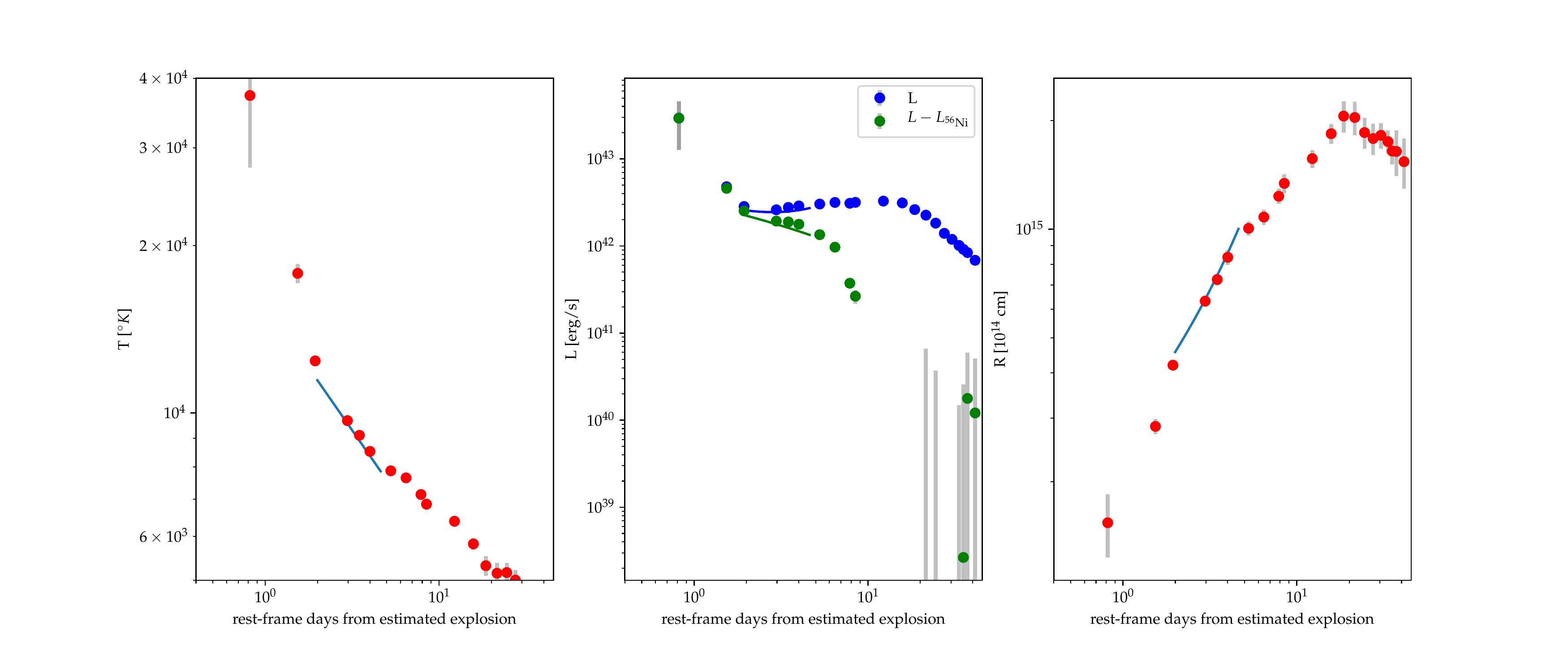} \\
\caption{The blackbody evolution of the best-fit shock-cooling model to the intermediate-time ($5>t>2$\,days) light curve, along with the blackbody of \oqm\ at these times. In the middle panel, we show both the combined \Nifs\ and shock-cooling luminosity (blue line) with the bolometric luminosity of \oqm\ (blue points), and the shock-cooling fit alone (green curve), with the residual bolometric luminosity from the \Nifs\ fit (green curve). The best-fit model is plotted at $t<t_{tr}/2 = 4.6$\,days.}
\label{fig:SC_intermediate_bb}
\end{figure*}

\begin{figure*}[b]
\centering

\includegraphics[width=\textwidth]{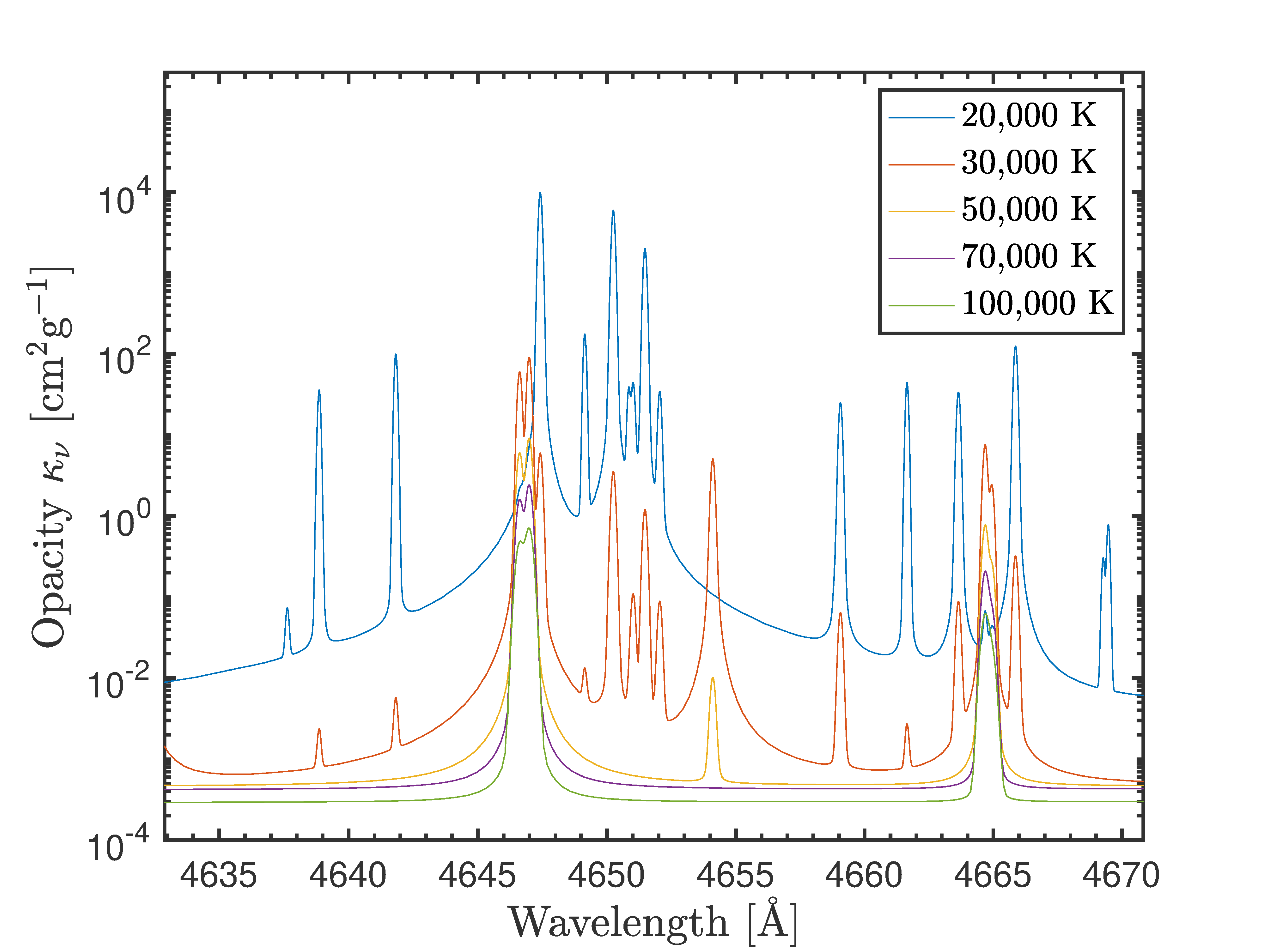} \\
\caption{Line opacities, produced using the code of \cite{Morag2022} for an equal C/O composition near 4650\,\AA. For illustration purposes, we show the most constraining (i.e., the highest) set of opacities we acquired for a density of $\rho = 10^{-12}\,{\rm g\,cm^{-3}}$ and at various CSM temperatures. }
\label{fig:CO_opac}
\end{figure*}

\begin{figure*}[b]
\centering

\includegraphics[width=\textwidth]{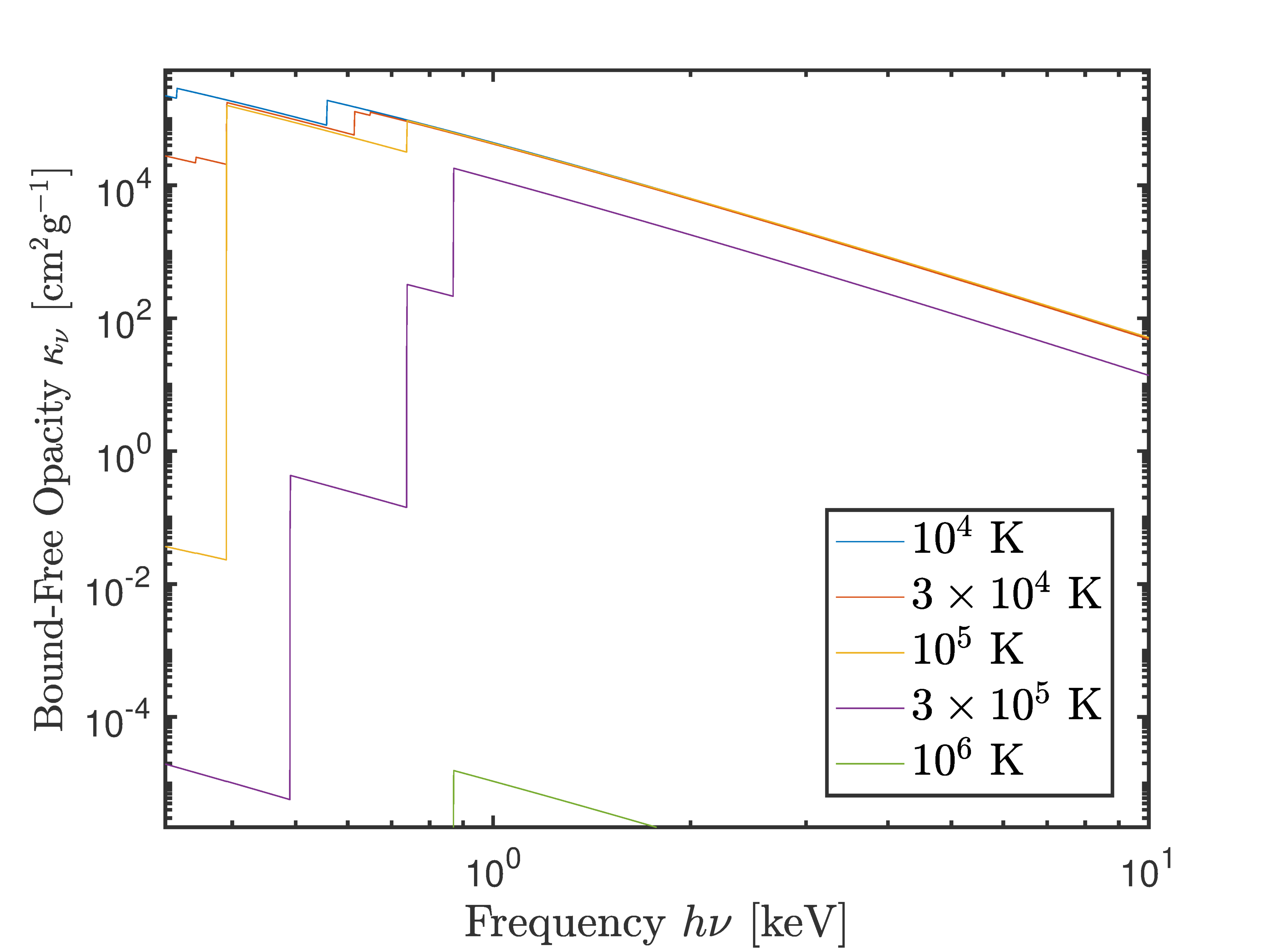} \\
\caption{Bound-free opacities in the {\it Swift}/XRT band (0.3--10\,keV), produced using the code of \cite{Morag2022} for an equal He/C/O composition. For illustration purposes, we show the most constraining (i.e., the lowest) set of opacities we acquired for a density of $\rho = 10^{-12}\,{\rm g\,cm^{-3}}$ and at various CSM temperatures. These opacities are representative of various fractions of He in the composition.}
\label{fig:xray_opac}
\end{figure*}

\end{document}